\documentclass[english,aps, prx, reprint, amsmath, amssymb, longbibliography]{revtex4-2}
\pdfoutput=1
\usepackage[T1]{fontenc}
\usepackage[latin9]{inputenc}
\usepackage{mathtools}
\usepackage{amsmath}
\usepackage{graphicx}
\usepackage{esint}
\usepackage{babel}
\begin{document}
\title{Real-Time Motion of Open Quantum Systems: Structure of Entanglement,
Renormalization Group, and Trajectories}
\author{Evgeny A. Polyakov}
\address{Russian Quantum Center, Skolkovo IC, Bolshoi Bulvar 30, bld 1, Moscow,
Russia 121205}
\email{evgenii.poliakoff@gmail.com}

\begin{abstract}
In this work we provide a complete description of the lifecycle of
entanglement during the real-time motion of open quantum systems.
The quantum environment can have arbitrary (e.g. structured) spectral
density. The entanglement can be seen constructively as a Lego: its
bricks are the modes of the environment. These bricks are connected
to each other via operator transforms. The central result is that
each infinitesimal time interval one new (\textsl{incoming}) mode
of the environment gets coupled (\textit{entangled}) to the open system,
and one new (\textit{outgoing}) mode gets irreversibly decoupled (\textit{disentangled}
from future). Moreover, each moment of time, only a few \textit{relevant}
modes (3 - 4 in the considered cases) are non-negligibly coupled to
the \textit{future} quantum motion. These relevant mode change (\textit{flow},
or\textit{ renormalize}) with time. As a result, the\textit{ temporal}
entanglement has the structure of a matrix-product operator. This
allows us to pose a number of questions and to answer them in this
work: what is the intrinsic quantum complexity of a real time motion;
does this complexity saturate with time, or grows without bounds;
how to do the real-time renormalization group in a justified way;
how the classical Brownian stochastic trajectories emerge from the
quantum evolution; how to construct the few-mode representations of
non-Markovian environments. We provide illustrative simulations of
the spin-boson model for various spectral densities of the environment:
semicircle, subohmic, Ohmic, and superohmic.
\end{abstract}
\maketitle

\section{Introduction}

A major innovation brought by quantum mechanics is that a physical
system can be in a superposition of its classical configurations.
While the classical state is a point in phase space, the quantum state
is a wave in the space of classical configurations \citep{Feynman2010}.
All current experimental evidence suggests that the quantum theory
works. Thereofore, the superposition principle is expected to apply
at all scales: from a single microscopic degree of freedom to one
mole of any substance with $N_{A}\propto10^{23}$ degrees of freedom.
However if we try to imagine the quantum state as a wave in the $10^{23}$-dimensional
space, then the quantum reality appears before us as an object of
tremendous complexity. This complexity is reflected in the notion
of a Hilbert space \citep{Dirac1982}.

We believe it is obvious that the overwhelming part of this\textit{
potential} complexity is not realized in the natural observable processes
\citep{Poulin2011,Brandao2019,Lin2021}. It is enough to review those
tremendous difficulties that the experimenters face in their attemps
to build a quantum computer \citep{Zhou2020,Franca2020}. This convinces
us that there are rather effective mechanisms of how the quantum complexity
decays during the natural processes. 

We can put it differently: it is expected that only a tiny fraction
of the Hilbert space is relevant for the observable motion of quantum
systems. Therefore, in order to reduce the complexity, we need some
tools to identify this relevant subspace. We can call this the problem
of efficient Hilbert space decimation. The motivation is two-fold:
conceptual and pragmatic. Conceptually, we want to understand the
stucture of quantum states and their evolution; we want to answer
the basic questions like the characterization of the boundary between
the quantum and the classical reality. Pragmatically, we want to be
able to calculate any property we are interested in. 

The main idea of this work is that the structure of entanglement can
provide answers to all these questions. The whole point is to look
at the enanglement constructively. We consider it like a Lego. The
bricks of this Lego are the degrees of freedom (the modes) of the
environment in some trivial (separable) state. The entanglement is
built by connecting these bricks via some (non)unitary operators.
If we know how the entanglement is built in real time from its bricks,
we can keep track of them and rigorously consider such questions as:
what is the quantum complexity of real-time motion; how this complexity
behaves in time: whether it saturates or grows; how to construct a
few-mode approximation of the non-Markovian environment. What is even
more important, if we know how the entanglement is built, then we
know how to compute the properties of quantum state by \textit{disassembling}
the entanglement brick-by-brick. This is the renormalization group
\citep{Vidal2007,Vidal2008,Acoleyen2020,Haegeman2013,White1992,Schollwock2011}
in one of its most advanced forms. 

As a specific example, we consider in this work the model of a finite
quantum system coupled to an infinite environment. This model plays
important role in quantum physics. Its practical significance is due
to the large number of covered situations: behavior of mesoscopic
degrees of freedom in physical chemistry and condensed matter \citep{Vega2017};
models of quantum measurement and control \citep{Brandes2010,Kiesslich2012,Gough2014,Brandes2016,Luo2016,Wagner2016}.
At the same time, the unique combination of simplicity and capability
to exhibit many-body physics \citep{Kondo1984,Wilson1975,Bulla2003,Bulla2008}
makes it a paradigmatic model which is used to challenge and improve
our perspective on quantum phenomena.

In this work we use the model of open quantum system to study the
lifecycle of entanglement during the real-time evolution \citep{Lazarou2012}.
Suppose we couple the open system to the environment. Then it starts
to emit quanta into the environment. Some of them will be reabsorbed,
but a significant number of quanta will fly further and further into
the environment, therefore entangling increasingly more modes of the
environment. This leads to a combinatorial growth of the dimension
of the joint wavefunction. Such a combinatorial growth of complexity
is called the entanglement barrier \citep{Rams2020} and it presents
a significant obstactle both to interpretation and real-time computational
methods. 

\begin{figure}
\includegraphics[scale=1.2]{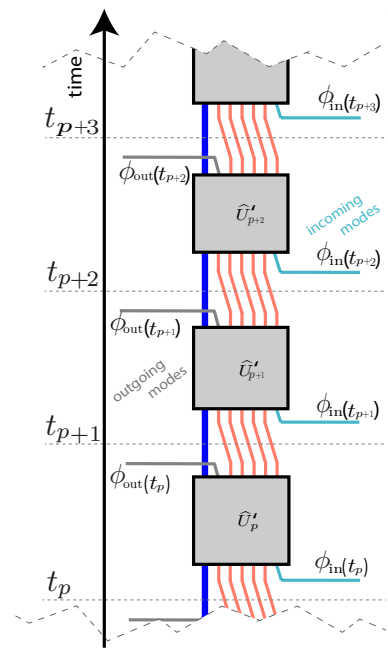}

\caption{\label{fig:entanglement_structure}The temporal structure of entanglement
which follows from the existence of the streams of incoming and outgoing
modes. The blue line depicts the quantum numbers of the open system.
The evolution $\widehat{U}_{p}^{\prime}$ , which occurs during each
infinitesimal time interval $\left[t_{p},t_{p+1}\right]$, couples
one additional incoming mode $\phi_{\textrm{in}}\left(t_{p}\right)$.
The corresponding quantum numbers are denoted by the turquoise line.
After that, one additional ougoing mode $\phi_{\textrm{out}}\left(t_{p+1}\right)$
becomes irreversibly decoupled from the future evolution. Its quantum
numbers are denoted by the gray line.}
\end{figure}

In an attempt to make sense of the entanglement barrier let us ask
the question: what is the ultimate fate of the emitted quantum field?
First of all, we expect that the emitted field becomes asymptotically
decoupled from the \textit{future} motion of open system. For concreteness,
let us assume that at some time moment $t^{*}$ a certain mode $\phi_{\textrm{out}}$
of the environment becomes effectively and irreversibly decoupled.
Then it is important how this mode is entangled to the rest of the
environment. The mode $\phi_{\textrm{out}}$ can be entangled to those
modes of the environment which were coupled to open system before
$t^{*}$. We say figuratively that $\phi_{\textrm{out}}$ is entangled
to the past. However, $\phi_{\textrm{out}}$ can not be significantly
entangled to any mode $\phi_{\textrm{in}}$ which couples to open
system after $t^{*}$. This is because the entanglement can be only
created via unitary evolution under coupling, which is effectively
absent for $\phi_{\textrm{out}}$ after $t^{*}$. Therefore, we say
figuratively that $\phi_{\textrm{out}}$ is not entangled to the future.
Now suppose that we have succeed to arrange the following two streams
of modes: (i) the stream of incoming modes such that the mode $\phi_{\textrm{in}}\left(t\right)$
assigned to the time moment $t$ couples to open system only after
$t$; (ii) the stream of outgoing modes such that the mode $\phi_{\textrm{out}}\left(t\right)$
assigned to the time moment $t$ is irreversibly decoupled from the
future motion after $t$. This system of modes leads to the entanglement
structure depicted in Fig. \ref{fig:entanglement_structure}.

From this picture one sees that our entanglement Lego contains the
two types of bricks: the incoming modes $\phi_{\textrm{in}}\left(t_{p}\right)$
and the outgoing modes $\phi_{\textrm{out}}\left(t_{p}\right)$. If
we look at how the entanglement builds forward in time, we can observe
that these two types of bricks play different roles. The incoming
modes $\phi_{\textrm{in}}\left(t_{p}\right)$ are sequentially attached
to the structure via the evolutions $\widehat{U}_{p}^{\prime}$ thus
increasing its complexity. At the same time, the outgoing modes $\phi_{\textrm{out}}\left(t_{p}\right)$
do not participate in the construction. Therefore, we can remove them
by tracing out them as soon as they emerge. This is how we can disassemble
the entanglement by continuously removing all the emerging such ``idle''
bricks. The competition between the two processes, the attachment
of $\phi_{\textrm{in}}\left(t_{p}\right)$ and the removal of $\phi_{\textrm{out}}\left(t_{p}\right)$,
determines the final complexity of the structure. In Fig. \ref{fig:entanglement_structure}
we have assumed that $\phi_{\textrm{in}}\left(t_{p}\right)$ and $\phi_{\textrm{out}}\left(t_{p}\right)$
occur with equal rates in time. Then, each moment of time only a finite
number of bricks (in Fig. \ref{fig:entanglement_structure} they are
the vertical salmon-colored lines connecting the adjacent evolutions)
are attached to the structure. The quantum complexity of real-time
motion only survives in these bricks. The quantum complexity will
saturate and become bounded with time if the occupation of the coupled
modes is bounded. In other words, the flux $j_{\textrm{in}}\left(t_{p}\right)$
of the quanta emitted by the open system should balance the flux $j_{\textrm{out}}\left(t_{p}\right)$
of the quanta leaking out to the outgoing modes. 

The structure in Fig. \ref{fig:entanglement_structure} yields a few-mode
representation of the bath. If we consider the reduced joint state
of open system and of the coupled modes, then all the effects of the
environment on the system will be taken into account. In this work
we verify this idea by considering the model of open quantum system
in an environment with subohmic, Ohmic, superohmic, and semicircle
(waveguide) spectral density.

The structure in Fig. \ref{fig:entanglement_structure} indicates
that the time plays the role of a flow parameter of some renormalization
group: the movement in time is accompanied with the emergence of degrees
of freedom which are unentangled to the future flow. In the language
of RG they are called the irrelevant degrees of freedom. Such degrees
of freedom are iteratively traced out while computing the flow. Here
we mean the modern varinants of RG, like the entanglement renormalization
(ER) \citep{Vidal2007,Vidal2008,Acoleyen2020,Haegeman2013}. In this
work we propose a numerical implementation of the real-time RG which
follows from the Fig. \ref{fig:entanglement_structure}. We verify
the idea by a calculation for the environment with a semicircle spectral
density.

The second thing we expect to eventually happen is that the emitted
field becomes observed. For example, one can imagine that the environment
is filled with observers at an exteremy low density. So that when
the emitted field hits the observer, it has already effectively decoupled
from the open system. These observers can percieve e.g. the signal
of displaced vacuum of the environment (i.e. they observe some global
classical background of the environment). Such a signal is expected
to be a superposition of the stochastic vacuum noise and some deterministic
motion due to interaction with the open system \citep{Polyakov2019}.
Because of the entanglement structure in Fig. \ref{fig:entanglement_structure},
such a measurement does not disturbs the future quantum evolution:
the trace over the outgoing modes can be represented as a stochastic
average over the ensemble of their measurement signals. In Fig. \ref{fig:entanglement_structure}
this corresponds to replacing each occuring gray line with some classical
noise. Then we immediately conclude that there exist two kinds of
entanglement: (i) the entanglement which is coupled to the future
quantum motion. It appears to be a genuinely quantum phenomenon. (ii)
The \textit{outgoing} entanglement which is irreversibly decoupled
from the future quantum motion (and which is carried by the outgoing
modes $\phi_{\textrm{out}}\left(t_{p}\right)$). The latter kind of
entanglement is equivalent to the stochastic ensemble of classical
signals. This fact has both pragmatic and conceptual consequences.
Pragmatically, we can efficiently simulate/analyze the outgoing entanglement
via the Monte Carlo sampling of the signal realizations. The resulting
procedure, the renormalization group for quantum trajectories, is
presented in this work. Conceptually, we conjecture that the outgoing
entanglement is the carrier of classical reality. We verify the idea
by a calculation for the environment with a subohmic spectral density.

In the following sections we develop this picture up to numerical
schemes by proposing a \textit{measure of average coupling to the
future}. 

This work is structured as follows. We place our ideas into the current
research context in sec. \ref{sec:RELATION-TO-OTHER}. The model of
open quantum system and its environment are introduced in sec. \ref{sec:MODEL-OF-OPEN}.
Then in sec. \ref{sec:HOW-THE-QUANTUM} we study how the emitted field
decouples from its source. We encourage the reader to pay special
attention to the Figs. \ref{fig:When-the-real-time} and \ref{fig:emergence_of_outgoing_mode}
which describe the physical mechanism of how the stream of the incoming
modes and of the irreversibly decoupled (outgoing) modes emerge in
real time. We verify our ideas by providing the calculations for the
subohmic/Ohmic/superohmic and semicircle environments. The reader
who is interested in the justification of the real-time RG should
pay attention to sec. \ref{subsec:Implications-for-the} and to the
discussuon of Figs. \ref{fig:mode_10_example-1} and \ref{fig:Plot-of-the}.
The Schrodinger equation for the model of open quantum system is reformulated
in terms of the incoming and outgoing modes in sec. \ref{sec:EQUATIONS-OF-MOTION}.
As a result, in Fig. \ref{fig:QSA_production} we obtain the entanglement
structure of Fig. \ref{fig:entanglement_structure}. In sec. \ref{sec:TRACING-OUT-THE}
the outgoing modes are traced out in real time as soon as they occur:
this is the renormalization group for density matrix. We verify this
idea by a calculation for the semicircle spectral density (waveguide
environment). In sec. \ref{subsec:Balance-of-complexity}, Fig. \ref{fig:The-outgoing-current},
we present a calculation which supports our conjecture that there
is a balance between the fluxes of incoming and outgoing quanta, i.e.
the complexity of real-time motion is expected to saturate. In section
\ref{sec:MARKOVIAN-LIMIT} we consider the Markovian limit of our
RG procedure. In sec. \ref{sec:STOCHASTIC-RENORMALIZATION-GROUP}
we implement another alternative for the RG procedure: to measure
the irrelevant modes in real time as soon as they occur. We verify
this by a calculation for the environemnt with a subohmic spectral
density. Our approach is compared with a state-of-the-art tensor network
method (TEMO \citep{Strathearn2018}). In section \ref{sec:ENTANGLEMENT-IN-THE}
we introduce the measure of temporal entanglement. We conclude in
sec. \ref{sec:CONCLUSION}. There are also four appendices where we
provide some implementation details of our RG procedures. 

\section{\label{sec:RELATION-TO-OTHER}RELATION TO OTHER APPROACHES}

Ultimately our approach is to find such a quantum circuit (Fig. \ref{fig:entanglement_structure})
which shows how the entanglement is built along the flow (which is
the real time in our case) from some set of uncoupled degrees of freedom.
This promotes the viewpoint of entanglement as some kind of Lego.
Its bricks are the uncoupled degrees of freedom. These bricks are
sticked together by applying (non)unitary operations. The merit of
this viewpoint is that the observable properties of the resulting
complex many-body states can be computed by gradually disassembling
this Lego brick-by-brick. 

Actually this viewpoint is close in spirit to the ideas of the entanglement
renormalization (ER) \citep{Vidal2007,Vidal2008,Acoleyen2020,Haegeman2013}.
However, we extend ER in the following two aspects.

First, ER studies how the entanglement is arranged over the increasing
length scales. In other words, for ER the relevant degrees of freedom
are those which contribute to the low-energy ('infrared') properties.
While this is appropriate for the description of low-temperature equilibrium
properties, it becomes unjustified in the case of real-time evolution.
During the real-time motion, the widely-separated energy scales may
become coupled e.g. after a sudden quench of parameters. We avoid
this problem since in our approach the relevant degrees of freedom
are exactly those which have a non-negligible coupling to the future
evolution. 

The second difference is how our circuit model is derived. Below we
introduce the metric $I_{2}\left[\phi;\tau\right]$ which measures
how much the given mode $\phi$ at a given time moment $\tau$ is
coupled to the future motion. This way we rigorously identify the
modes $\phi_{\textrm{out}}$ with are irreversibly irrelevant (decoupled)
because their metric $I_{2}\left[\phi_{\textrm{out}};\tau\right]$
is below a certain treshold. Thefore, we prove our entanglement model.
This should be contrasted to the reasoning behind ER which is more
heuristic, e.g. by matching the structure of a tensor network against
the expected entanglement scaling laws \citep{Eisert2010}. Main justification
of ER comes from the corpus of the numerical calculation results.

In literature there are many other formal ways to describe the evolution
of open systems. They can be divided into the following four groups.
The first group are the methods which do not rely on the structure
of entanglement. They are the master equations \citep{Vega2017,Breuer2011}
and their stochastic unravelings \citep{Percival1999,Wiseman1996,Wiseman1993,Daley2014},
path integrals with influence functionals \citep{Segal2010,Makri1995,Makri1995a,Makri1996,Makarov1994,Richter2017},
equivalent chain representations \citep{Woods2014}, pseudomodes \citep{Mazzola2009,Lambert2019}
and Markovian embeddings \citep{Tamascelli2018}. 

The second group of methods consists of the straightforward extensions
of the Wilson renormalization group ideas (e.g. the logarithmic discretization
of energy scales, the focus on the infrared limit) to the real time
case: impurity numerical renormalization group \citep{Bulla2008}.
These methods make the (often unjustified) assumption that the real-time
structure of entanglement is similar to the equilibirium case.

The third group of methods rely on the tensor network models of quantum
states and evolution \citep{Strathearn2018,Jorgensen2019,Luchnikov2019,Somoza2019},
usually within the framework of the influence functional formalism.
The language of tensor networks is suitable for the description of
the structure of entanglement \citep{Zhu2020}. However, in these
methods the main focus is to employ the linear algebra methods (e.g.
SVD) to compress the numerical representation of quantum states. They
neither reveal nor exploit the lifecycle of entanglement in real time. 

The final group consists of methods which are derived from different
principles, but which, in our opinion, turn out to exploit implicitly
some aspects of the structure described in Fig. \ref{fig:entanglement_structure}.
These are the non-Markovian quantum state diffusion \citep{Polyakov2019,Diosi1998,Diosi1997}
and stochastic Schrodinger equations \citep{Shao2008,Yan2016,Shao2004,Han2019};
various variants of the approach of hierarchical equations of motion
\citep{Suess2014,Hartmann2017,Moix2013,Strumpfer2012,Ishizaki2009,Duan2017,Rahman2019}. 

\global\long\def\ps#1#2#3{\prescript{#1}{#2}{#3}}%

\section{\label{sec:MODEL-OF-OPEN}MODEL OF OPEN QUANTUM SYSTEM}

Here we describe the model for which we implement our ideas in this
paper: the open quantum system. The latter is defined by a Hamiltonian
$\widehat{H}_{\textrm{s}}$. The system is coupled to bosonic environment
via some operator $\widehat{s}$. The environment is harmonic with
a Hamiltonian $\widehat{H}_{\textrm{b}}$,
\begin{equation}
\widehat{H}_{\textrm{b}}=\intop_{0}^{+\infty}d\omega\widehat{a}^{\dagger}\left(\omega\right)\widehat{a}\left(\omega\right).\label{eq:bath_in_sp}
\end{equation}
Here the environment's degrees of freedom are $\left[\widehat{a}\left(\omega\right),\widehat{a}^{\dagger}\left(\omega^{\prime}\right)\right]=\delta\left(\omega-\omega^{\prime}\right)$.
The environment is coupled to the open system via some site operator
\begin{equation}
\widehat{b}=\intop_{0}^{+\infty}d\omega c\left(\omega\right)\widehat{a}\left(\omega\right).
\end{equation}
The coupling constant is related to the spectral density $J\left(\omega\right)=\pi\left|c\left(\omega\right)\right|^{2}$.
The total Hamiltonian of the joint system is 
\begin{equation}
\widehat{H}_{\textrm{sb}}=\widehat{H}_{\textrm{s}}+\widehat{s}^{\dagger}\widehat{b}+\widehat{s}\widehat{b}^{\dagger}+\widehat{H}_{\textrm{b}}.\label{eq:open_quantum_system_model}
\end{equation}
 In the interaction picture wrt the free environment we have 
\begin{equation}
\widehat{H}_{\textrm{sb}}\left(t\right)=\widehat{H}_{\textrm{s}}+\widehat{s}\widehat{b}^{\dagger}\left(t\right)+\widehat{s}^{\dagger}\widehat{b}\left(t\right),\label{eq:model_Hamiltonian}
\end{equation}
where
\begin{equation}
\widehat{b}\left(t\right)=\intop_{0}^{+\infty}d\omega c\left(\omega\right)\widehat{a}\left(\omega\right)e^{-i\omega t}.\label{eq:coupling_site}
\end{equation}
For convenience we assume that initially at $t=0$ the environment
is in its vacuum state $\left|0\right\rangle _{\textrm{b}}$, so that
the initial joint state is
\begin{equation}
\left|\Phi\left(0\right)\right\rangle _{\textrm{sb}}=\left|\phi_{0}\right\rangle _{\textrm{s}}\otimes\left|0\right\rangle _{\textrm{b}},\label{eq:initial_condition}
\end{equation}
 where $\left|\phi_{0}\right\rangle _{\textrm{s}}$ is the initial
state of the system. The vacuum initial condition can be easily generalized
to a finite temperature state of the environment \citep{Polyakov2019,Polyakov2017a}.
Here the subscripts 'b', 's', and 'sb' designate the belonging to
the Hilbert space of the open system $\mathcal{H}_{\textrm{s}}$,
the Fock space $\mathcal{F}_{\textrm{b}}$ of environment, and the
joint space $\mathcal{H}_{\textrm{s}}\otimes\mathcal{F}_{\textrm{b}}$
of the open system and environment correspondingly. 

\section{\label{sec:HOW-THE-QUANTUM}HOW THE QUANTUM FIELD GRADUALLY DECOUPLES
FROM ITS SOURCE}

In this sections we find what are the incoming and outoing modes from
Fig. \ref{fig:entanglement_structure}. The entanglement is generated
by coupling. Therefore, the analysis of the entanglement structure
is essentially the analysis of the decoupling mechanism.  In this
section we consider the superspositions of single-quantum states which
are emitted by the open system, and describe the mechanism behind
their irreversible decoupling which was proposed in \citep{Polyakov2018b,Polyakov2019}.
Later in the next section we apply this mechanism to the full many-quanta
case.

\subsection{Physical mechanism of decoupling}

During its evolution the open system emits and absorbs quanta in the
environment. There are several important asymmetries between the emission
and absorption processes which we discuss and exploit here. 

\begin{figure*}
\includegraphics{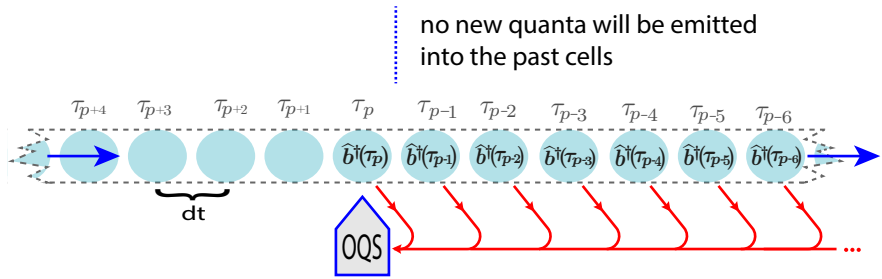}

\caption{\label{fig:time_as_a_tape}We may think of time as a stream of infinitesimally
close time moments $\tau_{0},\tau_{1},\ldots,\tau_{p},\tau_{p+1},\ldots$.
The time moves with constant speed in one direction (the blue arrows)
like a tape inside a tape recorder (or a \textit{Turing machine}).
Each time moment $\tau_{p}$ corresponds to a cell of the tape (the
light turquoise circles). Then the open system acts like a write/erase
head. At the time moment $\tau_{p}$, the head is located in front
of the corresponding cell. The head writes new quanta strictly to
the cell $\tau_{p}$ (which is denoted by $\widehat{b}^{\dagger}\left(\tau_{p}\right)$):
it is a local operation. At the same time, the erasing of quanta is
a causal but long-range operation: the quanta can be erased from an
arbitrarily distant past cell $\tau_{r}$, with a slowly decaying
amplitude $M\left(\tau_{p}-\tau_{q}\right)$ (red arrows). Erasing
is the only way for the past cells to interact with the future motion.
What is important is that in this representation the compexity of
the emitted quantum field grows gradually. For example, after one
infinitesimal time step the Fock space of the emitted field is spanned
by only one mode $\widehat{b}^{\dagger}\left(\tau_{0}\right)$.}
\end{figure*}

\subsubsection{Parametrization of the emitted quantum field}

The Fock space of the environment (\ref{eq:bath_in_sp}) is infinite.
The important nontrivial fact is that the Fock space of the quantum
field which can be emitted by open system is dramatically smaller.
Indeed, since the open system sits at one local site $\widehat{b}^{\dagger}$
of the environment, clearly it cannot resolve the full Fock space
of the environment. The spatiotemporal resolution of the emitted field
is determined by the spectral properties of $\widehat{b}^{\dagger}\left(t\right)$.

Formally, from the Hamiltonian (\ref{eq:model_Hamiltonian}) it is
seen that at a time moment $t$ the quanta are emitted via $\widehat{b}^{\dagger}\left(t\right)$.
Let us assume that the evolution happens during the time interval
$\left[0,t\right]$, starting from the initial state $\left|\Phi\left(0\right)\right\rangle _{\textrm{sb}}$,
eq. (\ref{eq:initial_condition}). One can imagine that this time
interval consists of a sequence of $N$ time moments $\tau_{0},\tau_{1}\ldots,\tau_{p},\tau_{p+1},\ldots$,
$0\leq\tau_{p}\leq t$, $p=0\ldots N$, which are equally spaced at
an infinitesimal interval $dt$. Then at each time moment $\tau_{p}$
the new quanta are created via $\widehat{b}^{\dagger}\left(\tau_{p}\right)$.
This means that the wavefucntion $\left|\Phi\left(t\right)\right\rangle _{\textrm{sb}}$
can only depend on $\widehat{b}^{\dagger}\left(\tau_{0}\right),\ldots\widehat{b}^{\dagger}\left(\tau_{p}\right),\ldots\widehat{b}^{\dagger}\left(\tau_{N}\right)$.
We may write it in the following form 
\begin{multline}
\left|\Phi\left(t\right)\right\rangle _{\textrm{sb}}=\sum_{n_{0}=0}^{\infty}\ldots\sum_{n_{N}=0}^{\infty}\frac{1}{\sqrt{n_{0}!\ldots n_{N}!}}\\
\times\left|\phi\left(n_{0}\ldots n_{N}\right)\right\rangle _{\textrm{s}}\otimes\widehat{b}^{\dagger n_{0}}\left(\tau_{0}\right)\ldots\widehat{b}^{\dagger n_{N}}\left(\tau_{N}\right)\left|0\right\rangle _{\textrm{b}}.\label{eq:interaction_wavefunctiob}
\end{multline}
We can visualize the resulting evolution by imagining it as a tape
recorder, Fig. \ref{fig:time_as_a_tape}: the time is like a tape
which moves with a constant speed \citep{Diosi2012}. This tape is
divided into discrete cells, one cell for each time moment $\tau_{p}$.
Then the open system is like a read/erase head. At the time moment
$\tau_{p}$ the head writes new quanta into the $\tau_{p}$th cell
of the tape. At the same time, the head can erase quanta from any
past cell $\tau_{q}$, $\tau_{q}\leq\tau_{p}$, with the amplitude
\begin{multline}
\left[\widehat{b}\left(\tau_{p}\right),\widehat{b}^{\dagger}\left(\tau_{q}\right)\right]=\frac{1}{\pi}\intop_{0}^{+\infty}d\omega J\left(\omega\right)e^{-i\omega\left(\tau_{p}-\tau_{q}\right)}\\
=M\left(\tau_{p}-\tau_{q}\right),\label{eq:temporal_commutation_relation}
\end{multline}
where $M\left(\tau_{p}-\tau_{q}\right)$ is the so-called memory function
of the environment. This equation follows from eq. (\ref{eq:coupling_site})
and from the commutation relation $\left[\widehat{a}\left(\omega\right),\widehat{a}^{\dagger}\left(\omega^{\prime}\right)\right]=\delta\left(\omega-\omega^{\prime}\right)$.
The past cells may be entangled with each other and with the head
(open system), as it follows from (\ref{eq:interaction_wavefunctiob}).

We see that in this picture both the creation and the absorption of
quanta are causal processes: the future cells with $\tau_{q}>\tau_{p}$
are empty and uncoupled from the open system. They couple to the evolution
one-by-one as the tape moves in front of the head. We call them the\textit{
stream of incoming modes}. They are the bricks of the Lego from which
the entanglement is being built in real time. 

The first important asymmetry is that the new quanta can be written
in the present and future cells, with $\tau_{q}\geq\tau_{p}$, whereas
no new quanta are created in the past cells with $\tau_{q}<\tau_{p}$.
The past cells are coupled to the future motion only via the annihilation
of quanta. 

The second important asymmetry is that the creation of new quanta
is a local operation, whereas the annihilation of quanta is a long-range
operation, since the memory function $M\left(t\right)$ always has
the inverse-power law tails \citep{Polyakov2018b,Polyakov2019}. Indeed,
since the physical spectral function $J\left(\omega\right)$ cannot
have the negative-frequency components, it should go to zero as a
certain power of $\omega$: $\omega^{p_{0}}\theta\left(\omega\right)$,
where $\theta$ is a Heaviside function, and $p_{0}>0$. This leads
to a tail $\propto\left(\pm it\right)^{-p_{0}-1}$ in the large-$t$
asymptotic behaviour of $M\left(t\right)$. Moreover, every frequency
$\omega_{k}$, where $J\left(\omega_{k}+\delta\omega\right)$ has
a discontinous part $\delta\omega^{p_{k}}\theta\left(\pm\delta\omega\right)$,
leads to an additional tail $\propto e^{-i\omega_{k}t}\left(\pm it\right)^{-p_{k}-1}$
in $M\left(t\right)$. These discontinuities can be e.g. the band
edges. Such tails were called the memory channels in \citep{Polyakov2019,Polyakov2018b}.

For illustration, let us consider the widely used model of the environment
with a spectral density 
\begin{equation}
J\left(\omega\right)=\frac{\alpha\omega_{\textrm{c}}}{2}\left[\frac{\omega}{\omega_{\textrm{c}}}\right]^{s}\exp\left(-\frac{\omega}{\omega_{\textrm{c}}}\right).\label{eq:power_density}
\end{equation}
It covers many types of environment by varying the parameter $s$:
the subohmic ($0\leq s<1$), Ohmic ($s=1$) and superohmic ($s>1$).
This spectral density corresponds to the memory function 
\begin{equation}
M\left(t\right)=\frac{\alpha\omega_{\textrm{c}}^{2}}{2\pi}\frac{\Gamma\left(s+1\right)}{\left(1+it\omega_{\textrm{c}}\right)^{s+1}},\label{eq:power_memory}
\end{equation}
which has a tail $\propto\tau^{-s-1}$ due to the discontinuity at
the frequency $\omega=0$. Another example is the waveguide environment
(in the Schrodinger picture)
\begin{equation}
\widehat{H}_{\textrm{b}}=\sum_{j=1}^{\infty}\left\{ h\widehat{a}_{j+1}^{\dagger}\widehat{a}_{j}+h\widehat{a}_{j}^{\dagger}\widehat{a}_{j+1}+\varepsilon\widehat{a}_{j}^{\dagger}\widehat{a}_{j}\right\} ,\label{eq:waveguide_system}
\end{equation}
with the open system being coupled to the first site $\widehat{a}_{1}$.
In the interaction pictire with respect to $\widehat{H}_{\textrm{b}}$,
we get $\widehat{b}\left(t\right)\equiv\widehat{a}_{1}\left(t\right)=\exp\left(it\widehat{H}_{\textrm{b}}\right)\widehat{a}_{1}\exp\left(-it\widehat{H}_{\textrm{b}}\right)$.
Such an environment has a finite band of energies $\left[\varepsilon-2h,\varepsilon+2h\right]$,
and the memory function is
\begin{equation}
M\left(t\right)=e^{-i\varepsilon t}\frac{J_{1}\left(2ht\right)}{ht}\propto t^{-\frac{3}{2}}e^{-i\left(\varepsilon-2h\right)t}+t^{-\frac{3}{2}}e^{-i\left(\varepsilon+2h\right)t},\label{eq:waveguide_memory_function}
\end{equation}
where now we have two tails, one tail per band edge at the frequencies
$\omega_{1}=\varepsilon-2h$ and $\omega_{2}=\varepsilon+2h$. 

This long-range character of $M\left(t\right)$ is a problem because
it leads to a large number of past cells which are entangled and significantly
coupled to the future motion. As a result, the complexity of real-time
motion accumulates combinatorially-fast and becomes prohibitive. 

In order to solve this problem, let us cosider the amplitude of the
annihilation process. Suppose that by the time moment $\tau_{p}$
one quantum was emitted. Its state $\left|\phi\right\rangle _{\textrm{b}}$
is a superposition of the cell-write events: 
\begin{equation}
\left|\phi\right\rangle _{\textrm{b}}=\left\{ \phi_{0}\widehat{b}^{\dagger}\left(\tau_{0}\right)+\ldots+\phi_{p-1}\widehat{b}^{\dagger}\left(\tau_{p-1}\right)\right\} \left|0\right\rangle _{\textrm{b}}.\label{eq:tape_single_quantum_state}
\end{equation}
Here for the moment we neglect the degrees of freedom of the open
system (in the next section \ref{sec:EQUATIONS-OF-MOTION} we derive
the full equations of motion). Now suppose that in the future the
open system points to some cell $\tau_{p^{\prime}}$, with $\tau_{p^{\prime}}>\tau_{p}$.
Then it will erase this quantum state (\ref{eq:tape_single_quantum_state})
with the amplitude 
\begin{equation}
\mathcal{A}_{p^{\prime}p}\left[\phi\right]=\sum_{r=0}^{p-1}M\left(\left(p^{\prime}-r\right)dt\right)\phi_{r}.\label{eq:annihilation_as_convolution}
\end{equation}
We see that the quanta in the past cells are coupled to the future
via the convolution with $M\left(t\right)$. Now we recall that for
the distant past the memory function $M\left(t\right)$ behaves as
a superposition of its tails $M_{k}\left(t\right)=e^{-i\omega_{k}t}\left(\pm it\right)^{-p_{k}-1}$.
Every such tail has the property that its local spectrum gradually
becomes more and more narrowed to the frequency $\omega_{k}$ as the
time argument $t$ is increased. Equivalently, one may say that as
$t$ is increased, there is an increasingly large time scale $\Delta\left(t\right)$
over which the tail $M_{k}\left(t\right)$ can be considered as being
effectively constant. Therefore, starting from $t$ every spectral
component of $\phi$ outside the frequency range $\approx\left[\omega_{k}-\Delta\left(t\right)^{-1},\omega_{k}+\Delta\left(t\right)^{-1}\right]$
will be averaged to zero by convolving with $M\left(t\right)$ in
eq. (\ref{eq:annihilation_as_convolution}). Physically this means
that in the remote past, the spectral content of the emitted quanta
is completely decoupled from the future motion, except a progressively
small vicinity of these singular frequencies $\omega_{k}$,
\begin{figure}
\includegraphics{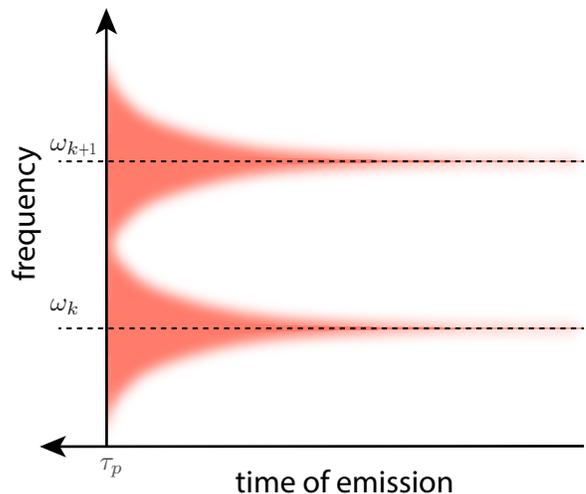}

\caption{\label{fig:The-spectral-content}The emitted quantum field is coupled
to the future motion only via convolution with the memory function,
eq. (\ref{eq:annihilation_as_convolution}). As a result, the spectral
content of the emitted quanta is completely decoupled from the future
motion, except a progressively small vicinity (salmon-colored area)
of the singular frequencies $\omega_{k},\omega_{k+1},\ldots$ where
the spectral function $J\left(\omega\right)$ has discontinuities.
These singular frequencies may be e.g. the boundaries of energy bands
or other sharp features of the spectral density.}

\end{figure}
 see Fig. \ref{fig:The-spectral-content}.

The analysiz above convinces us that as the time proceeds, new modes
of the environment should continously emerge which are effectively
decoupled from the future motion, 
\begin{figure}
\includegraphics{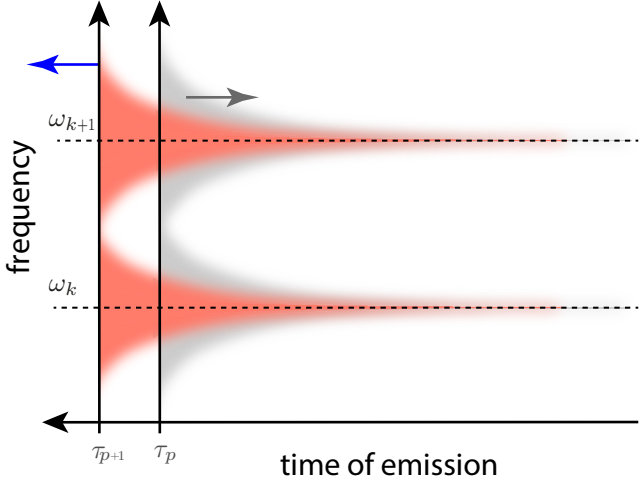}\caption{\label{fig:When-the-real-time}When the real-time evolution is propagated
forward one time step (blue arrow), from $\tau_{p}$ to $\tau_{p+1}$,
the emitted quantum field is shifted one step further into the past
(gray arrow). Then the frequency scales $\Delta\left(t\right)^{-1}$
slightly shrink. As a result, each propagation step yields new spectral
content (gray area) which is \textit{irreversibly} decoupled from
the future motion. The gray area should correspond to some continuously
emerging \textit{outgoing} modes of the environment.}

\end{figure}
see Fig. \ref{fig:When-the-real-time}. We call these the \textit{stream
of outgoing modes.} We will construct them formally in the following
section. 

\begin{figure}
\includegraphics[scale=1.7]{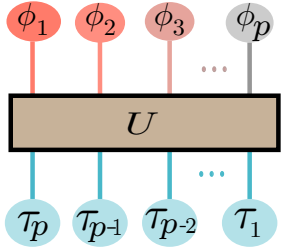}

\caption{\label{fig:If-we-find}If we find the degrees of freedom (the modes)
which correspond to the coupled and decoupled spectral components
of the emitted quantum field, then we can switch to the basis $\phi_{1},\phi_{2},\ldots$
of such modes by applying a suitable unitary transform $U$ to the
past cell degrees $\tau_{1},\tau_{2},\ldots,\tau_{p}$ of freedom
of the time tape. The new cells $\phi_{k}$ are numbered in the decreasing
order of their significance for the future evolution. Their color
match the color of the spectral areas in Fig. \ref{fig:When-the-real-time}. }

\end{figure}

Now suppose we succeeded in constructing the degrees of freedom which
represent the different spectral areas in Fig. \ref{fig:When-the-real-time}.
Namely, the cells from Fig. \ref{fig:time_as_a_tape} are treated
as independent quantum degrees of freedom. Then the frame is changed
(via a unitary transform, Fig. \ref{fig:If-we-find}) 
so that instead of the past cells we have: (i) the cells which are
significantly coupled to the future (\textit{the relevant modes},
from the salmon-colored area in Fig. \ref{fig:When-the-real-time})
and (ii) the cells which are effectively and irreversibly decoupled
from the future (the outgoing modes, from the gray area in Fig. \ref{fig:When-the-real-time}),
\begin{figure*}
\includegraphics{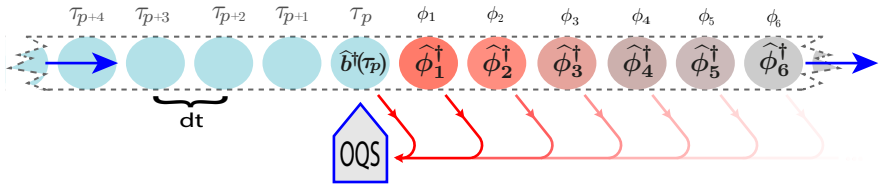}

\caption{\label{fig:In-the-basis}In the basis of the modes $\phi_{k}$ whose
significance for future motion rapidly decreases, the coupling to
the past is expected to become short-ranged.}
\end{figure*}
see Fig. \ref{fig:In-the-basis}. We expect that this way the coupling
to the past will become short-ranged, and the complexity of the real-time
motion will become bounded. This is because the decoupled past modes
can be traced out as we demonstrate below in this work.

\subsection{\label{subsec:How-to-find}How to find the outgoing modes}

In this section we implement the intuition of Fig \ref{fig:The-spectral-content}:
we find the modes which are significantly coupled to the future motion
(lie inside the salmon-colored area) and those which are irreversibly
decoupled (lie outside the samlon-colored area).

Let us again assume that by the time moment $\tau_{p}$ there is a
single-quantum state of the tape (\ref{eq:tape_single_quantum_state}).
Recall that $\mathcal{A}_{p^{\prime}p}\left[\phi\right]$ from eq.
(\ref{eq:annihilation_as_convolution}) is the amplitude to annihilate
this quantum at a later time moment $\tau_{p^{\prime}}$. We introduce
the short-hand bra-ket notation for the equation (\ref{eq:annihilation_as_convolution}):
\begin{equation}
\mathcal{A}_{p^{\prime}p}\left[\phi\right]=\left(p^{\prime}\left|M\right|\phi\right),\label{eq:annihilation_amplitude_2}
\end{equation}
where the matrix $M$ is the discretized memory function $M_{lr}=M\left(\left(l-r\right)dt\right)$,
with $l,r\geq0$. The set of amplitudes $\phi_{0},\ldots,\phi_{p-1}$
is considered as a semiinfinite vector $\phi_{r}$, $r\geq0$, with
$\phi_{r}=0$ for $r\geq p$. This vector is denoted in the bra-ket
notation as $\left|\phi\right)$. The notation $\left(p\left|\phi\right.\right)$
means the amplitude on the cell $\tau_{p}$: $\left(p\left|\phi\right.\right)=\phi_{p}$. 

In order to estimate the significance of $\phi$ for the future quantum
motion, we can evaluate the average intensity of annihilation processes
over all the future time moments: 
\begin{equation}
I_{q}\left[\phi;\tau_{p}\right]=\sum_{p^{\prime}=p}^{\infty}\left(\left|\left(p^{\prime}\left|M\right|\phi\right)\right|^{q}\right)^{\frac{1}{q}}/\left(\phi\left|\phi\right.\right)^{\frac{1}{2}}\label{eq:mean_intensity}
\end{equation}
with some parameter $q>0$. The quantity $I_{q}\left[\phi;\tau_{p}\right]$
is the power mean of $\mathcal{A}_{p^{\prime}p}\left[\phi\right]$
over the time interval $\left[\tau_{p},\infty\right)$. Here the denominator
in (\ref{eq:mean_intensity}) removes the trivial dependence on normalization
of $\phi$. It is natural to expect that the mode $\phi_{1}$ which
makes the largest average contribution to the future annihilation
processes is the most important for the future motion. Such a mode
can be found by maximizing $I_{q}\left[\phi;\tau_{p}\right]$. The
second most imortant mode $\phi_{2}$ can be found by maximizing $I_{q}\left[\phi;\tau_{p}\right]$
subject to the orthogonality constraint $\left(\phi_{1}\left|\phi_{2}\right.\right)=0$.
Repeating iteratively this process by maximizing $I_{q}\left[\phi;\tau_{p}\right]$
subject to the constraint of orthogonality to all the previously found
modes, we find the \textit{fastest decoupling basis} of modes $\phi_{1}\ldots\phi_{p}$.

Let us recall the intuitive picture of Fig. \ref{fig:The-spectral-content}.
The annihilation amplitude (\ref{eq:annihilation_amplitude_2}) has
the convolutional form. The intensity $I_{q}\left[\phi;\tau_{p}\right]$
is the average magnitude of such convolutions which are shifted to
future times. Therefore, we expect that the modes which maximize $I_{q}\left[\phi;\tau_{p}\right]$
correspond to the salmon-colored spectral area in Fig. \ref{fig:The-spectral-content}.
The modes which yield the least significant contribution to $I_{q}\left[\phi;\tau_{p}\right]$
lie ouside the salmon-colored spectral area of Fig. \ref{fig:The-spectral-content}. 

In this paper we consider the case of a root mean square intensity
$I_{2}\left[\phi;\tau_{p}\right]$. In this case the intensity assumes
the form 
\begin{equation}
I_{2}^{2}\left[\phi;\tau_{p}\right]=\left(\phi\left|K\left(p\right)\right|\phi\right),
\end{equation}
where the matrix $K\left(p\right)$ is
\begin{equation}
K\left(p\right)_{r^{\prime}r}=\sum_{p^{\prime}=p}^{\infty}M^{*}\left(\left(p^{\prime}-r^{\prime}\right)dt\right)M\left(\left(p^{\prime}-r\right)dt\right),\label{eq:K_matrix_element}
\end{equation}
$0\leq r,r^{\prime}<p$. Then the fastest decoupling basis $\phi_{1}\ldots\phi_{p}$
is given by the eigenvectors of $K\left(p\right)$, 
\begin{equation}
K\left(p\right)\left|\phi_{k}\right)=\lambda_{k}\left|\phi_{k}\right),\label{eq:FDMS}
\end{equation}
where $k=1,2,\ldots$, and we sort the eigenvalues $\lambda_{k}\left(p\right)$
in the descending order. 

\subsubsection{Example calculations: some tests of the ideas}

\begin{figure}
\includegraphics[scale=0.5]{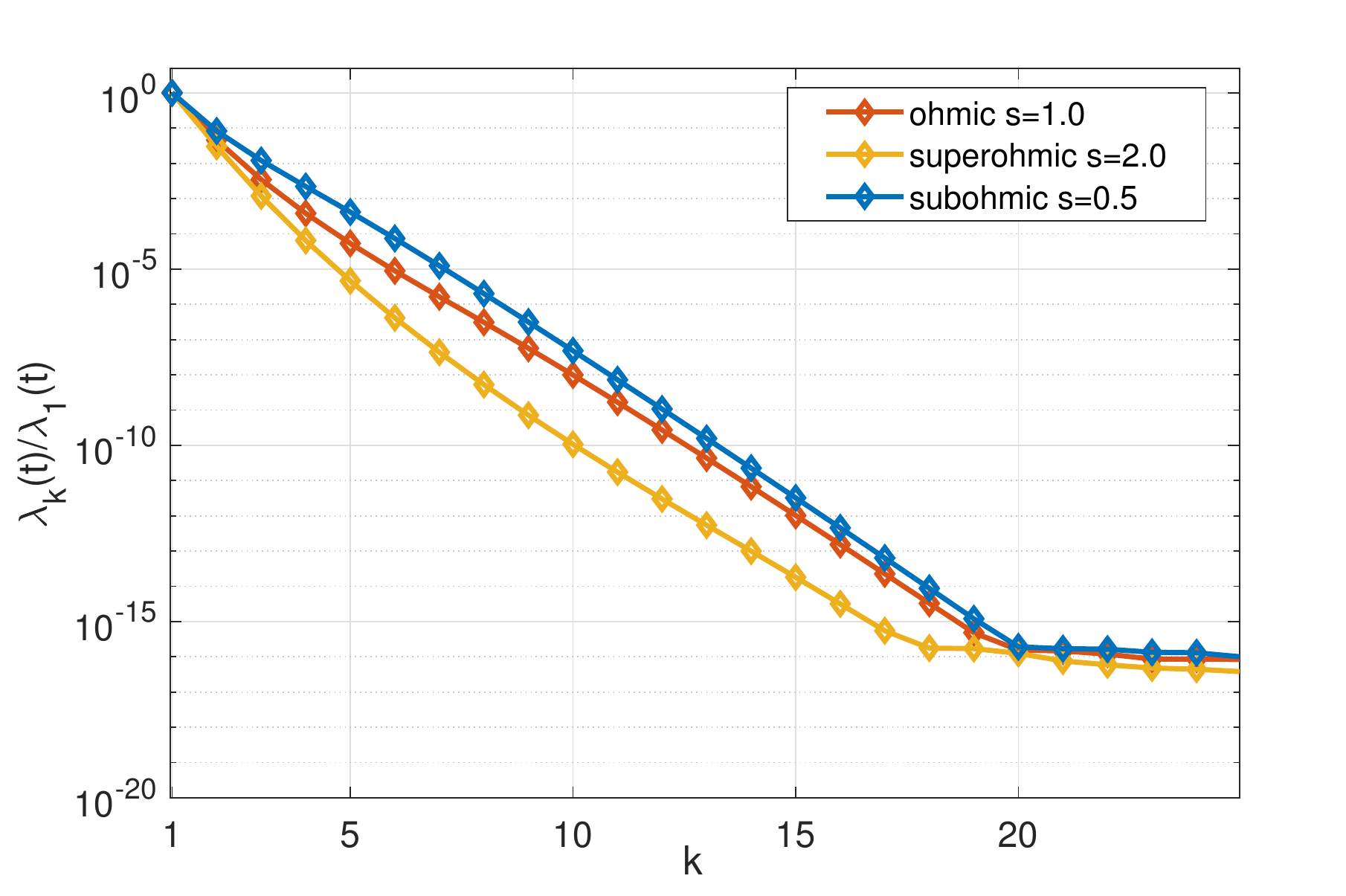}

\caption{\label{fig:decay_of_eigenvals-1}The average contribution of the modes
$\phi_{k}$ of the fastest decoupling basis to the future evolution
decays exponentially fast. Here the cases of subohmic ($s=0.5$),
ohmic ($s=1.0$), and superhomic ($s=2.0$) environments are considered
for the time moment $t_{p}=pdt=100$ and $dt=0.01$. The saturation
on the level of $10^{-16}$ is due to roundoff errors.}
\end{figure}

As illustration, we resent in Fig. \ref{fig:decay_of_eigenvals-1}
the plot of normalized eigenvalues $\lambda_{k}\left(p\right)/\lambda_{1}\left(p\right)$
for the cases of subohmic ($s=0.5$), ohmic ($s=1.0$) and superohmic
($s=2.0$) environments, eq. (\ref{eq:power_memory}), for the time
moment $t_{p}=pdt=100$. 
One observes that the average coupling of these eigenvectors to the
future evolution decays exponentially fast.

\begin{figure}
\includegraphics[scale=0.5]{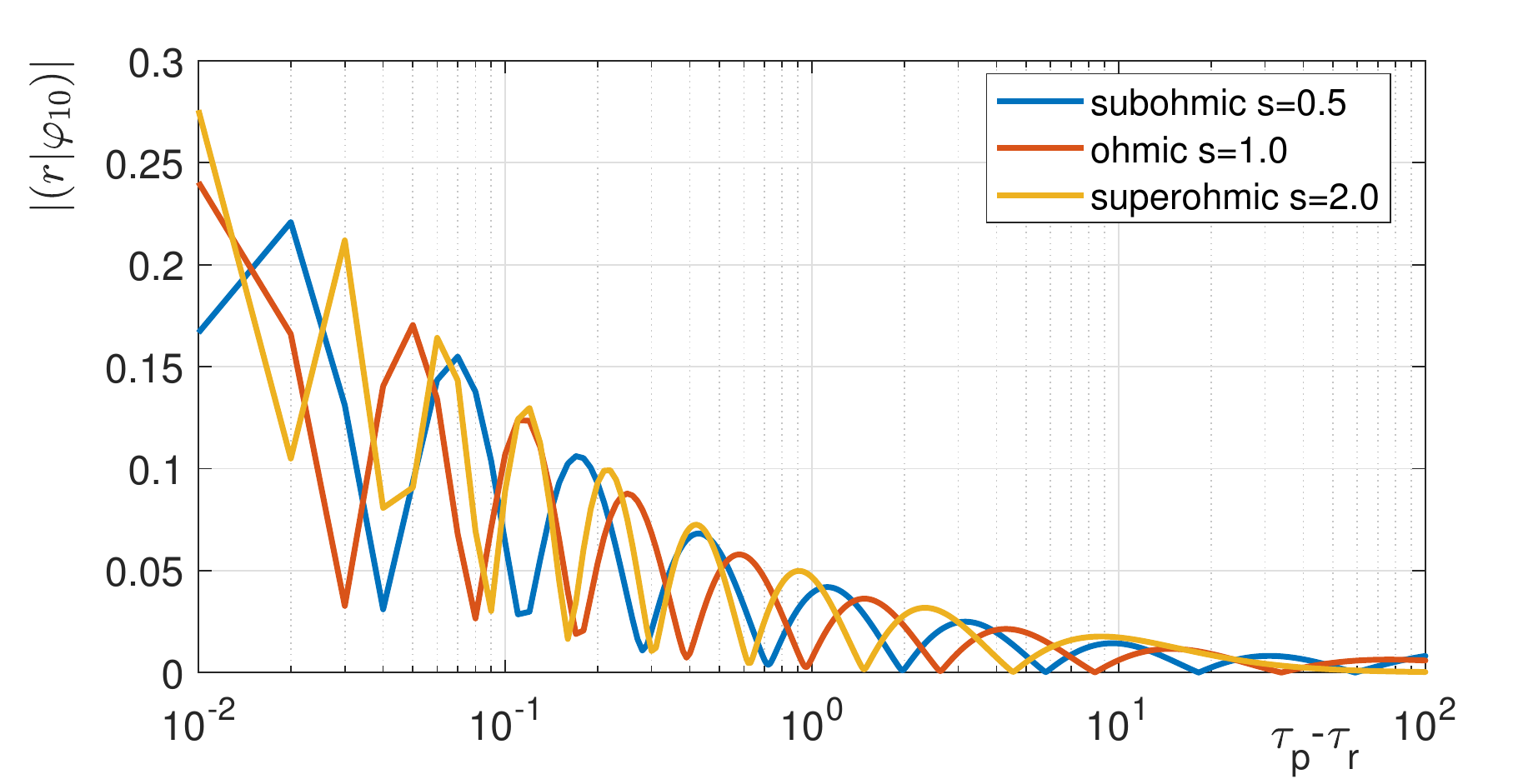}

\caption{\label{fig:mode_10_example-1}Plot of the fastest decoupling basis
function $\left|\left(r\left|\phi_{10}\right.\right)\right|$ vs $\tau_{p}-\tau_{r}$
for $t_{p}=pdt=100$, for the cases of subohmic ($s=0.5$), ohmic
($s=1.0$), and superhomic ($s=2.0$) environments. Observe that we
plot with respect to the ``time of delay'' $\tau_{p}-\tau_{r}$
between the present $\tau_{p}$ and the past $\tau_{r}$. The scales
of their oscillations are approximately constant on the logarithmic
scales of delay times. }
\end{figure}

If the intuition of Fig \ref{fig:The-spectral-content} is valid,
then the oscillations of the eigenvectors $\phi_{k}$ should rapidly
slow down so that the spectral content of $\phi_{k}$ remains in the
progressively small vicinity of the only singular frequency $\omega=0$.
In Fig. \ref{fig:mode_10_example-1} we present the plots of $\phi_{10}$
for the cases of subohmic ($s=0.5$), ohmic ($s=1.0$) and superohmic
($s=2.0$) environments. It is seen that the scale of their oscillations
is almost constant on the logarithmic scale of times $\tau_{r}$,
which supports our intuition.

\begin{figure}
\includegraphics[scale=0.5]{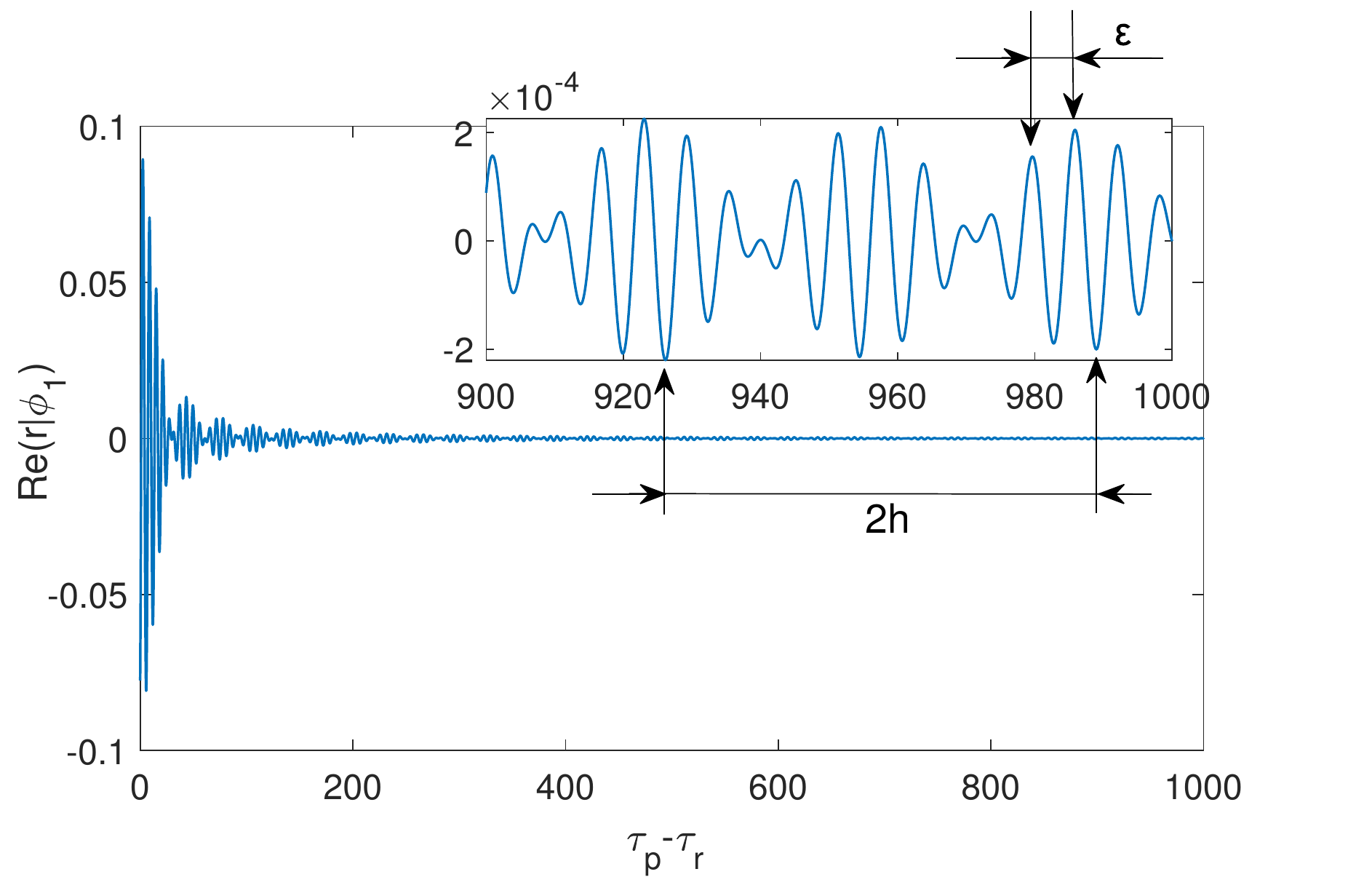}

\caption{\label{fig:Plot-of-the}Plot of the most importat mode $\phi_{1}$
for the waveguide memory function eq. (\ref{eq:waveguide_memory_function})
with $\varepsilon=1$ and $h=0.05$. It is seen on the inset that
the tail $\phi_{1}$ behaves like a superspostion of oscillations
with the frequencies $\varepsilon-2h$ and $\varepsilon-2h$. This
supports the intuition of Fig. \ref{fig:The-spectral-content} that
the coupled spectral content of the quanta emitted in the past shrinks
to the singular frequencies of the spectral function. }

\end{figure}

A more stringent test of our intuition is to compute the most important
mode $\phi_{1}$ for the waveguide memory function (\ref{eq:waveguide_memory_function}).
If the Fig. \ref{fig:The-spectral-content} is valid, and if the first
eigenvalues of $K\left(p\right)$ indeed sample the salmon-colored
spectral area, than the tails of $\phi_{1}$ should behave as a superposition
of $e^{+i\left(\varepsilon-2h\right)t}$ and $e^{+i\left(\varepsilon+2h\right)t}$.
In Fig. \ref{fig:Plot-of-the} we present the plot of $\phi_{1}$
for the waveguide with $\varepsilon=1$ and $h=0.05$. 

Taking into account the exponentially fast decoupling of the modes
$\phi_{k}$, we can keep the first $m$ modes $\phi_{1},\ldots\phi_{m}$
as the relevant modes, and consider the rest modes $\phi_{m+1},\phi_{m+2},\ldots$
as the outgoing modes. The latter are \textit{irreversibly} decoupled,
which is supported by the following property of the average intensity:
\begin{equation}
I_{q}\left[\phi;\tau_{p^{\prime}}\right]\leq I_{q}\left[\phi;\tau_{p}\right]\,\textrm{for}\,\,\tau_{p^{\prime}}\geq\tau_{p},
\end{equation}
which follows from its definition (\ref{eq:mean_intensity}).

\subsection{\label{subsec:Implications-for-the}Implications for the renormalization
group: the choice of relevant space}

\begin{figure*}
\includegraphics[scale=1.2]{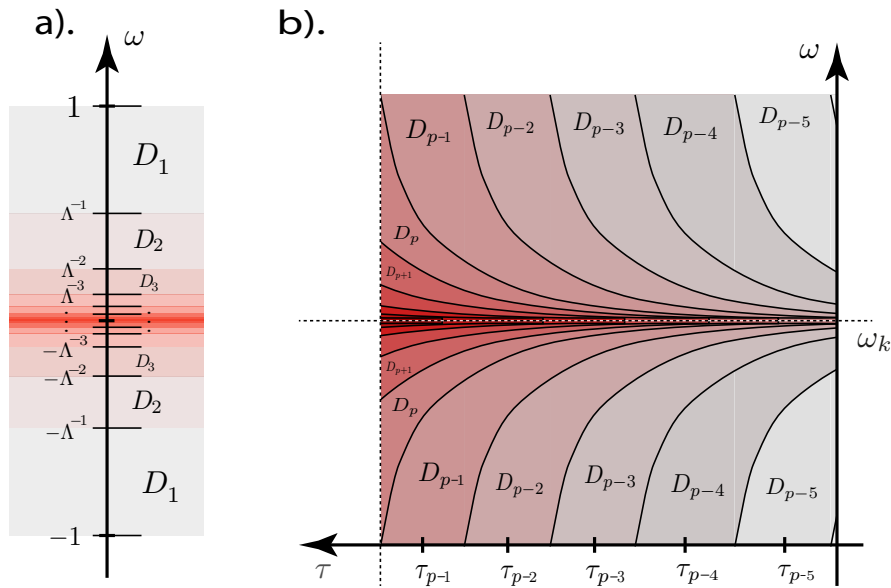}

\caption{\label{fig:ordinary_vs_realtime_rg}Comparision between the traditional
``equilibruim-inspired'' and the real-time renormalization groups
(RG). a). In the traditional RG, we have initial energy interval $\left[-1,1\right]$
(in some arbirtary units). Then one makes assumption that only the
low energy behaviour is relevant for the problem at hand. The logarithmic
discretization of the energy interval is introduced with $\Lambda>1$.
We obtain an infinite sequence of energy shells $D_{1}=\left[-1,-\Lambda^{-1}\right]\cup\left[\Lambda^{-1},1\right]$,
$\ldots,D_{k}=\left[-\Lambda^{-k+1},-\Lambda^{-k}\right]\cup\left[\Lambda^{-k},\Lambda^{-k+1}\right],\ldots$
. The renormalization group flow is obtained by sequentially tracing
out the degrees of freedom in the shells $D_{1},D_{2},\ldots D_{k},\ldots$.
b). In the real-time renormalization group, the relevant degrees of
freedom are contained in the progressively small vicinity of each
singular frequency $\omega_{k}$ where the spectral density $J\left(\omega\right)$
has sharp features (e.g. band edges). Thus we obtain the sequence
of \textit{phase-space} shells $D_{p-1},D_{p},D_{p+1},\ldots$. The
shell $D_{k}$ contains the spectral content which decouples from
future evolution after the time moment $\tau_{k}$, see Fig. \ref{fig:When-the-real-time}.
Each shell $D_{k}$ carries a degree of freedom $\phi_{\textrm{out}}\left(k\right)$,
which are constructed in sec. \ref{subsec:The-emergence-of}. These
$\phi_{\textrm{out}}\left(k\right)$ are traced out in sec. \ref{sec:TRACING-OUT-THE}
and collapsed to classical noise in sec. \ref{sec:STOCHASTIC-RENORMALIZATION-GROUP}.}
\end{figure*}

Looking at the fastest decoupling basis modes in Fig \ref{fig:mode_10_example-1},
we notice the emergent logarithmic scale when the only singular frequency
is $\omega=0$. This logarithmic scale is a characteristic feature
of the renormalization group (RG) methods \citep{Atland2010}. The
interesting difference is that the traditional (equilibrium-inspired)
RG methods perform the logarithmic discretization of the energy band,
whereas in our case it is the time behaviour which scales logarithmically.
However the core principle is the same: (i) there is a flow, (ii)
as we proceed along the flow, only the low-energy behaviour remains
to be relevant, see Fig. \ref{fig:ordinary_vs_realtime_rg} In the
equilibrium-inspired RG methods the flow is the decreasing characteristic
energy cut-off. In our case the flow is the ``aging'' (moving to
the past in time) of the emitted field, and the relevant low-energy
behaviour is the progressively small vicinity of the singular frequencies
in spectral function $J\left(\omega\right)$, as in Fig. \ref{fig:The-spectral-content}.

\subsection{The emergence of outgoing modes in real time\label{subsec:The-emergence-of}}

In this section we implement the intuition of Fig \ref{fig:When-the-real-time}
and Fig. \ref{fig:ordinary_vs_realtime_rg}: we construct the stream
of outgoing modes. That is, for each infinitesimal motion $\tau_{p}\to\tau_{p+1}$,
we identify the new mode $\phi_{\textrm{out}}\left(p\right)$ of the
environment that has just become irreversibly decoupled (which belongs
to the gray area in Fig \ref{fig:When-the-real-time} and to the shell
$D_{p}$ in Fig. \ref{fig:ordinary_vs_realtime_rg}). 

We construct the stream of outgoing modes by induction. Suppose we
have chosen to keep no more than $m$ relevant modes. This is how
we demarkate the boundary of the salmon-colored area in Fig. \ref{fig:The-spectral-content}.
This means that for $\tau_{p}$ with $p\leq m-1$ all the modes are
relevant since the matrix $K\left(p\right)$ has no more than $m$
eigenvalues. This is the basis for our inductive construction. 

\begin{figure}
\includegraphics{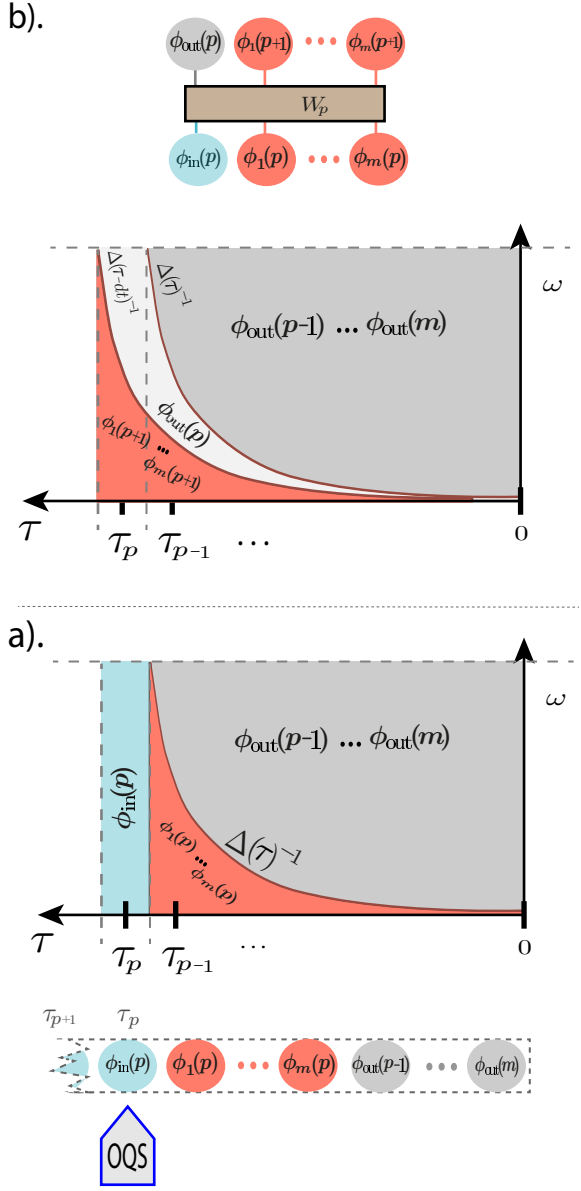}\caption{\label{fig:emergence_of_outgoing_mode}a). At a time moment $\tau_{p}$,
the open system gets coupled to the new incoming mode $\phi_{\textrm{in}}\left(p\right)$.
Since this mode is localized (in $\tau$), it has a broadband spectrum
(the vertical light turquoise spectral area). The past cells are represented
by the relevant modes $\phi_{1}\left(p\right),\ldots,\phi_{m}\left(p\right)$,
which carry the coupled part of the emitted quantum field (salmon-colored
spectral area, under the decreasing scale $\Delta\left(\tau\right)^{-1}$).
There are also outgoing modes $\phi_{\textrm{out}}\left(p-1\right),\phi_{\textrm{out}}\left(p-2\right),\ldots,\phi_{\textrm{out}}\left(m\right)$,
which represent the irreversibly decoupled part of the emitted quantum
field (the gray spectral area). b). After the propagation to the next
time moment $\tau_{p}\to\tau_{p+1}$, the incoming mode $\phi_{\textrm{in}}\left(p\right)$
becomes one of the past cells. Then the scale $\Delta\left(\tau\right)^{-1}$
shifts to the left, $\Delta\left(\tau\right)^{-1}\to\Delta\left(\tau-dt\right)^{-1}$,
as described in Fig. \ref{fig:When-the-real-time}. This leads to
the emergence of a new decoupled spectral content (light gray area).
This spectral content is represented by a new outgoing mode $\phi_{\textrm{out}}\left(p\right)$.
The latter is found via a unitary transform $W_{p}$, eq. (\ref{eq:next_step_of_stream_construction}),
to the frame of the new relevant modes $\phi_{1}\left(p+1\right),\ldots,\phi_{m}\left(p+1\right)$.}
\end{figure}

Now suppose that at a time moment $\tau_{p}$, $p\geq m-1$, we know
what are the relevant modes $\phi_{1}\left(p\right),\ldots,\phi_{m}\left(p\right)$,
and know what are (zero or more) outgoing modes $\phi_{\textrm{out}}\left(p-1\right),\phi_{\textrm{out}}\left(p-2\right),\ldots,\phi_{\textrm{out}}\left(m\right)$,
see Fig. \ref{fig:emergence_of_outgoing_mode}, a).
 At the time moment $\tau_{p}$ the open system (the write/erase head
in our tape model) points to the cell $\tau_{p}$. This cell represents
one incoming mode $\phi_{\textrm{in}}\left(p\right)$ which is localized
on this cell: $\left(r\left|\phi_{\textrm{in}}\left(p\right)\right.\right)=\delta_{rp}$,
with $\delta$ being the Kronecker delta.

The inductive step in our construction is to consider what happens
after the propagation to the next time moment $\tau_{p}\to\tau_{p+1}$.
The incoming mode $\phi_{\textrm{in}}\left(p\right)$ becomes one
of the past cells (the write/erase head moves one position to the
left). As a result, by the time moment $\tau_{p+1}$ we have the following
$m+1$ modes: $\phi_{1}\left(p\right),\ldots,\phi_{m}\left(p\right)$
and $\phi_{\textrm{in}}\left(p\right)$. Our task to is to transform
these modes into the new $m$ relevant modes $\phi_{1}\left(p+1\right),\ldots,\phi_{m}\left(p+1\right),$
and one new outgoing mode $\phi_{\textrm{out}}\left(p\right)$,
\begin{equation}
\left[\begin{array}{c}
\phi_{1}\left(p+1\right)\\
\vdots\\
\phi_{m}\left(p+1\right)\\
\phi_{\textrm{out}}\left(p\right)
\end{array}\right]=W_{p}\left[\begin{array}{c}
\phi_{1}\left(p\right)\\
\vdots\\
\phi_{m}\left(p\right)\\
\phi_{\textrm{in}}\left(p\right)
\end{array}\right],\label{eq:next_step_of_stream_construction}
\end{equation}
via some unitary transform $W_{p}$. Here $W_{p}$ acts as ordinary
matrix on the column entries e.g. $\phi_{k}\left(p+1\right)=\sum_{l}\left(W_{p}\right)_{kl}\phi_{l}\left(p\right)+\left(W_{p}\right)_{k,m+1}\phi_{\textrm{in}}\left(p\right)$.
In order to determine $W_{p}$, we choose the fastest decoupling basis.
As the fastest decoupling basis we take the eigenvectors of $K\left(p+1\right)$
in the space spanned by $\phi_{1}\left(p\right),\ldots,\phi_{m}\left(p\right)$
and $\phi_{\textrm{in}}\left(p\right)$. The matrix elements of $K\left(p+1\right)$
in this subspace are: 
\begin{equation}
\widetilde{K}\left(p+1\right)=\left[\begin{array}{cc}
K_{\textrm{rr}} & K_{\textrm{ri}}\\
K_{\textrm{ri}}^{\dagger} & K_{\textrm{ii}}
\end{array}\right],
\end{equation}
where the blocks of the matrix are: the scalar $K_{\textrm{ii}}=\left(\phi_{\textrm{in}}\left(p\right)\left|K\left(p+1\right)\right|\phi_{\textrm{in}}\left(p\right)\right)=K\left(p+1\right)_{pp}=\sum_{p^{\prime}=1}^{\infty}\left|M\left(p^{\prime}dt\right)\right|^{2}$;
the vector $\left(K_{\textrm{ri}}\right)_{q}=\left(\phi_{\textrm{in}}\left(p\right)\left|K\left(p+1\right)\right|\phi_{q}\left(p\right)\right)$;
the matrix $\left(K_{\textrm{rr}}\right)_{qq^{\prime}}=\left(\phi_{q}\left(p\right)\left|K\left(p+1\right)\right|\phi_{q^{\prime}}\left(p\right)\right)$.
This matrix has dimension $\left(m+1\right)\times\left(m+1\right)$.
Therefore, if we diagonalize it and sort the $m+1$ eigenvectors in
the descending order of their eigenvalues, 
\begin{equation}
\widetilde{K}\left(p+1\right)=U\left[\begin{array}{ccc}
\lambda_{1} &  & 0\\
 & \ddots\\
0 &  & \lambda_{m+1}
\end{array}\right]U^{\dagger},
\end{equation}
we obtain $W_{p}=U^{T}$ for eq. (\ref{eq:next_step_of_stream_construction}). 

\begin{figure}
\includegraphics[scale=1.2]{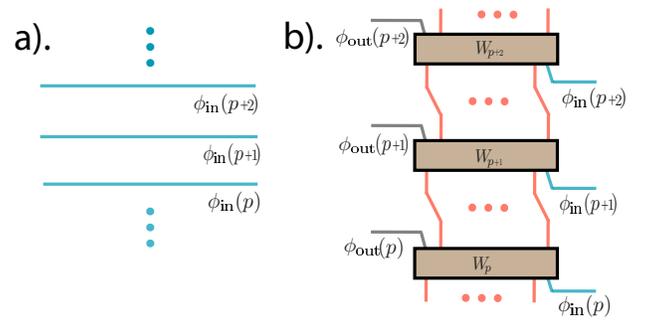}

\caption{\label{fig:construction_of_streams}a). The tape model of real-time
evolution assigns an incoming mode $\phi_{\textrm{in}}\left(p\right)$
to each time moment $\tau_{p}$. b). The iterative construction of
the $m$ most relevant modes changes the structure of modes: now we
have the two streams of incoming $\phi_{\textrm{in}}\left(p\right)$
(turquoise lines) and outgoing modes $\phi_{\textrm{out}}\left(p\right)$
(gray lines), and also the relevant modes $\phi_{1}\left(p\right)\ldots\phi_{m}\left(p\right)$
(salmon-colored lines), which represent the corresponding spectral
areas of Figs. \ref{fig:When-the-real-time} and \ref{fig:emergence_of_outgoing_mode}.}
\end{figure}

Iterating this inductive step, we construct the entire stream of outgoing
modes $\phi_{\textrm{out}}\left(p\right)$, see Fig. \ref{fig:construction_of_streams}

\section{\label{sec:EQUATIONS-OF-MOTION}EQUATIONS OF MOTION IN TERMS OF INCOMING,
OUTGOING, AND RELEVANT MODES}

In this section we give a precise meaning to the Fig. \ref{fig:entanglement_structure}.
In the previous section the treatment was on the level of superpositios
of single-quantum states of the environment. Here we derive the complete
quantum equations of motions, which fully take into account the many-body
correlations between the open quantum system and the environment. 

We remind you that our goal is to find a minimum number of degrees
of freedom which contain everything what is significant for the motion
at future time moments. We want to stay in the small relevant subspace
(spanned by these degrees of freedom) all the time. However during
each infinitesimal time interval the quantum amplitudes leak outside
this space (due to the environment dispersion). In other words, at
least one new degree of freedom gets entangled. In order to keep the
minimal number of relevant modes, we need to trace out some degree
of freedom. This degree of freedom should be independent and decoupled
from everything. Otherwise, the real-time motion will become corrupted
on large times. We conclude that we need to attach an independent
degree of freedom to each infinitesimal time moment. Otherwise there
will be nothing to trace out, and the number of coupled modes will
grow with time. Therefore, in this section we develop a formalism
which treats the infinitesimal time cells as independent degrees of
freedom. We then map the model of open quantum system (\ref{eq:density_matrix_project_away})
to the representation of independent time cells (``tape model'')
in an exact and rigorous way.

\subsection{Fock space for the tape cells\label{subsec:Fock-space-for}}

Each cell $\tau_{p}$ of the time tape (the light turquoise circles
in Figs. \ref{fig:time_as_a_tape}, \ref{fig:If-we-find}, and \ref{fig:In-the-basis})
is considered as independent bosonic degree of freedom with its own
creation operator $\widehat{\psi}_{p}^{\dagger}$, $\left[\widehat{\psi}_{q},\widehat{\psi}_{p}^{\dagger}\right]=\delta_{qp}$.
The operator $\widehat{\psi}_{p}^{\dagger}$ creates a quantum in
the mode $\phi_{\textrm{in}}\left(p\right)$. Recalling Fig. \ref{fig:time_as_a_tape},
this implies that we map the creation operator $\widehat{b}^{\dagger}\left(\tau_{p}\right)$
to $\widehat{\psi}_{p}^{\dagger}$, 
\begin{equation}
\widehat{b}^{\dagger}\left(\tau_{p}\right)\to\widehat{\psi}_{p}^{\dagger}.\label{eq:mapping_for_creation}
\end{equation}
As a result, the tape has its own vacuum $\left|0\right\rangle _{\textrm{t}}$
and the tape Fock space $\mathcal{F}_{\textrm{t}}$ is spanned by
applying $\widehat{\psi}_{p}^{\dagger}$'s to $\left|0\right\rangle _{\textrm{t}}$.
This Fock space is formally different from that $\mathcal{F}_{\textrm{b}}$
defined by the original model in eq. (\ref{eq:model_Hamiltonian}).
However the quantum mechanics permits many different formal representations
of a physical system, provided that the observable properties are
invariant. For the latter it is enough to conserve the commutation
relation (\ref{eq:temporal_commutation_relation}), which is ensured
by the mapping 
\begin{equation}
\widehat{b}\left(\tau_{l}\right)\to\sum_{r=0}^{l}M_{lr}\widehat{\psi}_{r},\label{eq:mapping_for_annihilation}
\end{equation}
were we again employ the matrix $M$ with elements $M_{lr}=M\left(\left(l-r\right)dt\right)$.
Now the tape model and the original model eq. (\ref{eq:model_Hamiltonian})
are in one-to-one correspondence. For example, the states are mapped
between these models as:
\begin{multline}
\widehat{\psi}_{0}^{\dagger n_{0}}\ldots\widehat{\psi}_{p-1}^{\dagger n_{p-1}}\left|\phi\left(n_{0},\ldots n_{p-1}\right)\right\rangle _{\textrm{s}}\otimes\left|0\right\rangle _{\textrm{t}}\\
\longleftrightarrow\widehat{b}^{\dagger n_{0}}\left(\tau_{0}\right)\ldots\widehat{b}^{\dagger n_{p-1}}\left(\tau_{p-1}\right)\left|\phi\left(n_{0},\ldots n_{p-1}\right)\right\rangle _{\textrm{s}}\otimes\left|0\right\rangle _{\textrm{b}}.
\end{multline}
Here $\left|\phi\left(n_{0},\ldots n_{p-1}\right)\right\rangle _{\textrm{s}}$
is a wavefunction of the open system which is not affected by the
mapping. The vacuum initial condition for the tape model becomes $\left|\Phi\left(0\right)\right\rangle _{\textrm{st}}=\left|\phi_{0}\right\rangle _{\textrm{s}}\otimes\left|0\right\rangle _{\textrm{t}}$.

\subsection{\label{subsec:The-Hamiltonian-for}The Hamiltonian for the tape model}

The Hamiltonian for the tape model is obtained by applying the mapping
rules (\ref{eq:mapping_for_creation})- (\ref{eq:mapping_for_annihilation})
to the original Hamitonian (\ref{eq:model_Hamiltonian}): 
\begin{equation}
\widehat{H}_{\textrm{sb}}\left(\tau_{p}\right)\to\widehat{H}_{\textrm{st}}\left(\tau_{p}\right)=\widehat{H}_{\textrm{s}}+\widehat{s}\widehat{\psi}_{p}^{\dagger}+\widehat{s}^{\dagger}\sum_{r=0}^{p}M_{pr}\widehat{\psi}_{r}.\label{eq:discrete_time_hamiltonian}
\end{equation}
Observe that in the tape model the real-time evolution happens in
discrete time steps, via jumps from one cell $\tau_{p}$ to the neighboring
one $\tau_{p+1}$. Of course, we want this discrete evolution to approximate
the continuous one from the original model (\ref{eq:model_Hamiltonian}).
It is desirable that the global propagation error is stable and vanishing
as $dt\to0$. The latter can be achieved by employing the implicit
midpoit rule as a propagator,
\begin{multline}
\left|\Phi\left(t_{p+1}\right)\right\rangle _{\textrm{st}}=\left|\Phi\left(t_{p}\right)\right\rangle _{\textrm{st}}\\
-idt\widehat{H}_{\textrm{st}}\left(\tau_{p}\right)\frac{1}{2}\left\{ \left|\Phi\left(t_{p+1}\right)\right\rangle _{\textrm{st}}+\left|\Phi\left(t_{p}\right)\right\rangle _{\textrm{st}}\right\} \\
+O\left(dt^{3}\right),\label{eq:discrete_propagation}
\end{multline}
where now we specify that the tape cells are located at the midpoint
times: $\tau_{p}=\left(p+\frac{1}{2}\right)dt$. This equation is
solved for $\left|\Phi\left(t_{p+1}\right)\right\rangle _{\textrm{st}}$
by iterations with the initial guess $\left|\Phi\left(t_{p+1}\right)\right\rangle _{\textrm{st}}=\left|\Phi\left(t_{p}\right)\right\rangle _{\textrm{st}}$.
Here the Hamiltonian $\widehat{H}_{\textrm{sb}}\left(\tau_{p}\right)$
corresponds to the head (open system) being coupled to the cell $\tau_{p}$.
Then the midpoint rule enforces us to think that the Hamiltonian $\widehat{H}_{\textrm{sb}}\left(\tau_{p}\right)$
generates the evolution from the wavefunction $\left|\Phi\left(t_{p}\right)\right\rangle _{\textrm{st}}$
at the time moment $t_{p}=pdt$ to the wavefunction $\left|\Phi\left(t_{p+1}\right)\right\rangle _{\textrm{st}}$
at the next time moment $t_{p+1}=\left(p+1\right)dt$. This ensures
the global error of $O\left(dt^{2}\right)$. The time moment $t_{p}$
corresponds to the notation of Fig. \ref{fig:entanglement_structure}.

\begin{figure}
\includegraphics[scale=1.5]{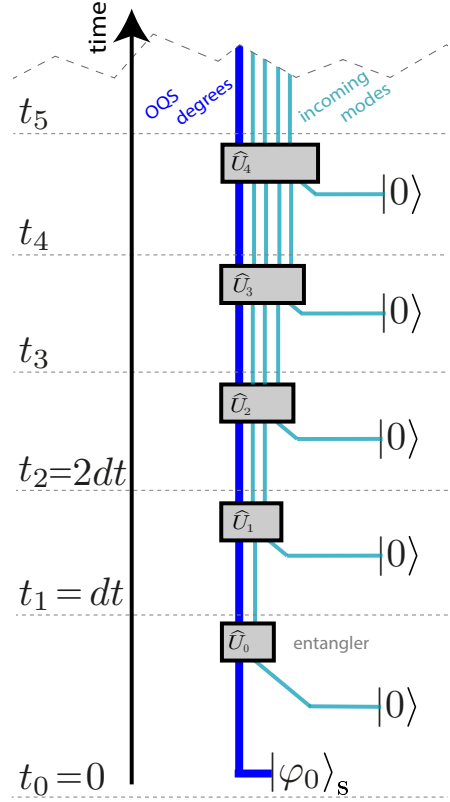}

\caption{\label{fig:discrete_time_bare_evolution}In the tape model of open
system the evolution happes between the discrete time moments $t_{0}=0$,
$t_{1}=dt,\ldots$$t_{p}=pdt,\ldots$. After propagation from $t_{p}$
to $t_{p+1}$, one new incoming mode $\phi_{\textrm{in}}\left(p\right)$
(turquoise line) corresponding to the cell $\tau_{p}$ becomes entangled
to the open system (blue line). The evolutions $\widehat{U}_{p}$
happen under the midpoint rule (\ref{eq:discrete_propagation}).}
\end{figure}

The wavefunction $\left|\Phi\left(t_{p}\right)\right\rangle _{\textrm{st}}$
depends on the modes $\widehat{\psi}_{0}^{\dagger},\ldots,\widehat{\psi}_{p-1}^{\dagger}$.
During each propagation step of eq. (\ref{eq:discrete_propagation})
it entangles one additional mode $\widehat{\psi}_{p}^{\dagger}$ (which
corresponds to $\phi_{\textrm{in}}\left(p\right)$), see Fig \ref{fig:discrete_time_bare_evolution}.
This is the usual growth of complexity which we overcome in this work.

\subsection{Partial trace\label{subsec:Partial-trace}}

To completely define the tape model, we need to discuss how to compute
the trace over the cell degrees of freedom. Suppose we compute the
reduced density matrix for open system via the partial trace over
the environment:
\begin{equation}
\widehat{\rho}_{\textrm{s}}\left(t_{p}\right)=\textrm{Tr}_{\textrm{b}}\left\{ \left|\Phi\left(t_{p}\right)\right\rangle _{\textrm{sb}}\ps{}{\textrm{sb}}{\left\langle \Phi\left(t_{p}\right)\right|}\right\} ,\label{eq:original_trace_usual}
\end{equation}
where $\left|\Phi\left(t_{p}\right)\right\rangle _{\textrm{sb}}$
is the wavefunction of the original model (\ref{eq:model_Hamiltonian}).
The effect of $\textrm{Tr}_{\textrm{b}}$ is that each $\widehat{a}^{\dagger}\left(\omega\right)$
from the ket $\left|\Phi\left(t_{p}\right)\right\rangle _{\textrm{sb}}$
becomes paired with $\widehat{a}\left(\omega\right)$'s from the bra
$\ps{}{\textrm{sb}}{\left\langle \Phi\left(t_{p}\right)\right|}$
at the same frequency. This follows from the commutation relations
$\left[\widehat{a}\left(\omega\right),\widehat{a}^{\dagger}\left(\omega^{\prime}\right)\right]=\delta\left(\omega-\omega^{\prime}\right)$.
We can rewrite the partial trace (\ref{eq:original_trace_usual})
in such a form that all the pairings become explicit:
\begin{multline}
\widehat{\rho}_{\textrm{s}}\left(t_{p}\right)=\ps{}{\textrm{b}}{\left\langle 0\right|\left\{ e^{\intop_{0}^{+\infty}d\omega\widehat{a}^{\dagger}\left(\omega\right)\widehat{a}\left(\omega\right)}\right.}\\
\left.\times:\left|\Phi\left(t_{p}\right)\right\rangle _{\textrm{sb}}\ps{}{\textrm{sb}}{\left\langle \Phi\left(t_{p}\right)\right|}:\right\} _{A}\left|0\right\rangle _{\textrm{b}}.\label{eq:original_trace_unusual}
\end{multline}
Here the notation $\left\{ \,\,\,\,:\,\,\,\,:\right\} _{A}$ means
that the creation/annihilation operators coming from the Taylor expansion
of the \textit{pairing function} $\exp\intop_{0}^{+\infty}d\omega\widehat{a}^{\dagger}\left(\omega\right)\widehat{a}\left(\omega\right)$
are antinormally ordered around the term $\left|\Phi\left(t_{p}\right)\right\rangle _{\textrm{sb}}\ps{}{\textrm{sb}}{\left\langle \Phi\left(t_{p}\right)\right|}$.
The ordering of the latter term is not affected (which is indicated
by placing it between $:\,\,\,:$ ). Observe that when we order the
Taylor expansion of the pairing function, we regard its constituent
creation/annihilation operators as commuting. The expressions (\ref{eq:original_trace_usual})
and (\ref{eq:original_trace_unusual}) are equal because they describe
the same set of pairings between the $\widehat{a}^{\dagger}\left(\omega\right)$'s
from the ket $\left|\Phi\left(t_{p}\right)\right\rangle _{\textrm{sb}}$,
and the $\widehat{a}\left(\omega\right)$'s from the bra $\ps{}{\textrm{sb}}{\left\langle \Phi\left(t_{p}\right)\right|}$.
The operation $\ps{}{\textrm{b}}{\left\langle 0\right|}\cdot\left|0\right\rangle _{\textrm{b}}$
denotes the projection of density matrix to the subspace of zero quanta
in the environment. This operation does not affect the open system. 

When defining the trace for the wavefunction $\left|\Phi\left(t_{p}\right)\right\rangle _{\textrm{st}}$
of the tape model, we should take into account that each pairing between
$\widehat{\psi}_{r}$ and $\widehat{\psi}_{s}^{\dagger}$ is possible
due to the commutator $\left[\widehat{b}\left(\tau_{r}\right),\widehat{b}^{\dagger}\left(\tau_{s}\right)\right]=M_{rs}$:
\begin{multline}
\widehat{\rho}_{\textrm{s}}\left(t_{p}\right)=\ps{}{\textrm{t}}{\left\langle 0\right|\left\{ e^{\sum_{rs=0}^{p-1}\widehat{\psi}_{r}^{\dagger}M_{rs}\widehat{\psi}_{s}}\right.}\\
\left.\times:\left|\Phi\left(t_{p}\right)\right\rangle _{\textrm{st}}\ps{}{\textrm{st}}{\left\langle \Phi\left(t_{p}\right)\right|}:\right\} _{A}\left|0\right\rangle _{\textrm{t}}.\label{eq:discrete_time_trace}
\end{multline}
This is how the partial trace is performed in the tape model. The
conceptual meaning of the pairing function is that the time-cell degrees
of freedom yield non-local contribution to the observables in quantum
mechanics, see Fig. \ref{fig:time_nonlocality} 
\begin{figure}
\includegraphics[scale=2]{temporal_nonlocality_2.ai}

\caption{\label{fig:time_nonlocality}a). It is very desirable to consider
the wavefunction as depending on quantum numbers of each infinitesimal
time moment (turquoise lines). This appoach is natural for analyzing
how the entanglement gradually develops in real time. Especially it
is interesting to track how is it happening \textit{continuously}:
what gets entangled and disentangled each infinitesimal time moment.
We succeed in deriving equations of motion which treat the infinitesimal
time cells as independent degrees of freedom: see eqs. (\ref{eq:discrete_time_hamiltonian})
and (\ref{eq:discrete_propagation}). b). Nevertheless, there are
no time-local states in quantum mechanincs. Due to the commutation
relation (\ref{eq:temporal_commutation_relation}), each time cell
$\tau_{k}$ contributes to the observables at another time moment
$t$ like a wavepacket $M_{\tau_{k}t}^{1/2}$ with long-range tails.
Here $M_{\tau_{k}t}^{1/2}=\left(2\pi\right)^{-\frac{1}{2}}\intop_{0}^{+\infty}d\omega e^{-i\omega\left(\tau_{k}-t\right)}\sqrt{J\left(\omega\right)}$.
c). We reconcile the dynamical locality and the observable non-locality
by introducing the pairing function. This effectively smears the time
cells in time during the partial trace operation. The resulting density
matrix $\widehat{\rho}_{\textrm{s}}\left(t_{p}\right)$ depends only
on quantum numbers of the open system (blue lines). d). The elementary
pairing $M_{rs}$ corresponds to the overlap (green area) between
the wavepackets of the two quanta at $\tau_{r}$ and $\tau_{s}$:
$M_{rs}=\intop_{-\infty}^{+\infty}M_{\tau_{r}t}^{1/2}M_{\tau_{s}t}^{1/2*}dt$.}
\end{figure}

\subsection{Propagation of wavefunction in terms of incoming, outgoing, and relevant
modes}

The initial condition for the propagation is $\left|\Phi\left(0\right)\right\rangle _{\textrm{st}}=\left|\phi_{0}\right\rangle _{\textrm{s}}\otimes\left|0\right\rangle _{\textrm{t}}$. 

There are two regimes of propagation: $t_{p}<t_{m}$ and $t_{p}\geq t_{m}$.

As was discussed in sec. \ref{subsec:The-emergence-of}, while $t_{p}<t_{m}$,
all the modes $\widehat{\psi}_{0}^{\dagger}\ldots\widehat{\psi}_{p-1}^{\dagger}$
are relevant: $\phi_{\textrm{in}}\left(1\right)\equiv\phi_{1}\left(p-1\right),\ldots,\phi_{\textrm{in}}\left(p-1\right)\equiv\phi_{p-1}\left(p-1\right)$.
Therefore, the propagation from $\left|\Phi\left(t_{p}\right)\right\rangle _{\textrm{st}}$
to $\left|\Phi\left(t_{p+1}\right)\right\rangle _{\textrm{st}}$ is
done via the application of the midpoint rule (\ref{eq:discrete_propagation}).
As a result, before the time moment $t_{m}$ the entanglement develops
according to the Fig. \ref{fig:discrete_time_bare_evolution}. 

The situation changes for $t_{p}\geq t_{m}$. The wavefunction $\left|\Phi\left(t_{p}\right)\right\rangle _{\textrm{st}}$
depends on the relevant modes $\phi_{1}\left(p-1\right)\ldots\phi_{m}\left(p-1\right)$
via their creation operators $\widehat{\phi}_{1}^{\dagger}\left(p-1\right)\ldots\widehat{\phi}_{m}^{\dagger}\left(p-1\right)$.
Also $\left|\Phi\left(t_{p}\right)\right\rangle _{\textrm{st}}$ depends
on zero or more outgoing modes $\phi_{\textrm{out}}\left(p-1\right)\ldots\phi_{\textrm{out}}\left(m\right)$
via their creation operators $\widehat{\phi}_{\textrm{out}}^{\dagger}\left(p-1\right)\ldots\widehat{\phi}_{\textrm{out}}^{\dagger}\left(m\right)$,
see Fig. \ref{fig:evolution_stages}, a). 
\begin{figure*}
\includegraphics[scale=2.2]{evolution_stages_v3_1-01.ai}

\caption{\label{fig:evolution_stages}The circuit model of how the wavefunction
is propagated during one time step in three stages: a). At a time
moment $t_{p}$ the wavefunction $\left|\Phi\left(t_{p}\right)\right\rangle _{\textrm{st}}$
depends on the occupations of outgoing modes $\phi_{\textrm{out}}\left(k\right)\equiv\phi_{\textrm{out}}\left(t_{k}\right),k=m\ldots p-1$
(gray lines) and of relevant modes $\phi_{k}\equiv\phi_{k}\left(p\right),k=1\ldots m$
(salmon lines). Also it depends on the quantum numbers of the open
system (blue line). Observe the matrix-product-like structure of the
wavefunction. This follows from the facts that (i) only $m$ modes
are significantly coupled to the future and (ii) each time step one
outgoing mode is produced. b). The discrete-time Hamiltonian (\ref{eq:midpoint_Hamiltonian-1-1-1})
is applied via the midpoint rule (\ref{eq:discrete_propagation}),
which approximates the evolution $\widehat{U}_{p}$. As a result,
one incoming mode (turquoise line) gets entangled to the state. c).
The frame is changed actively via $\widehat{W}_{p}$ to produce one
decoupled mode (gray line). Observe that $\widehat{U}_{p}$ and $\widehat{W}_{p}$
constitute the newly formed additional block of the matrix product
structure of $\left|\Phi\left(t_{p+1}\right)\right\rangle _{\textrm{st}}$,
see Fig. \ref{fig:MPO_block}.}
\end{figure*}
Here each line represents the occupation number of the corresponding
mode. The salmon-colored lines $\phi_{k}\equiv\phi_{k}\left(p-1\right),\,k=1\ldots m$,
correspond to the relevant modes which are populated via the creation
operators $\widehat{\phi}_{k}^{\dagger}\left(p-1\right)$. The blue
line denotes the quantum numbers of the open system. The gray lines
are the outgoing modes. 

The propagation begins with \textit{entangling step}: when the discrete-time
Hamiltonian $\widehat{H}_{\textrm{st}}\left(\tau_{p}\right)$ (\ref{eq:discrete_time_hamiltonian})
is applied, the stage b). in Fig. \ref{fig:evolution_stages}. We
express the Hamiltonian in terms of the incoming mode $\widehat{\psi}_{p}^{\dagger}$,
the relevant modes $\widehat{\phi}_{k}^{\dagger}\left(p\right)$,
and the irrelevant modes $\widehat{\phi}_{\textrm{out}}^{\dagger}\left(p\right)$:
\begin{multline}
\widehat{H}_{\textrm{st}}\left(\tau_{p}\right)=\widehat{H}_{\textrm{s}}+\widehat{s}\widehat{\psi}_{p}^{\dagger}+\widehat{s}^{\dagger}M\left(0\right)\widehat{\psi}_{p}\\
+\widehat{s}^{\dagger}\sum_{i=1}^{\textrm{min}\left(p,m\right)}M_{i}\left(p\right)\widehat{\phi}_{i}\left(p\right)\\
+\widehat{s}^{\dagger}\sum_{i=m+1}^{p-1}M_{\textrm{out}i}\left(p\right)\widehat{\phi}_{\textrm{out}}\left(i\right),\label{eq:midpoint_Hamiltonian-1-1-1}
\end{multline}
where the coupling to the relevant modes is 
\begin{equation}
M_{i}\left(p\right)=\sum_{r=0}^{p-1}M_{pr}\phi_{i}\left(p\right)_{r},
\end{equation}
and the coupling to the irrelevant modes is $M_{\textrm{out}i}\left(p\right)=\sum_{r=0}^{p-1}M_{pr}\phi_{\textrm{out}}\left(t_{i}\right)_{r}$.
Up to now the treatement was \textit{exact }(in the limit $dt\to0$).
Here we\textbf{ make an approximation for the first time}: the coupling
between the incoming mode and the irrelevant modes (the last line
of eq. (\ref{eq:midpoint_Hamiltonian-1-1-1})) is discarded, so that
we have:
\begin{multline}
\widehat{H}_{\textrm{st}}\left(\tau_{p}\right)=\widehat{H}_{\textrm{s}}+\widehat{s}\widehat{\psi}_{p}^{\dagger}+M\left(0\right)\widehat{\psi}_{p}\\
+\widehat{s}^{\dagger}\sum_{i=1}^{\textrm{min}\left(p,m\right)}M_{i}\left(p\right)\widehat{\phi}_{i}\left(p\right).\label{eq:midpoint_Hamiltonian-1-1-1-1}
\end{multline}
Below we will use this form of the Hamiltonian.

After the propagation $\widehat{U}_{p}$ via the rule (\ref{eq:discrete_propagation}),
the resulting wavefunction $\left|\widetilde{\Phi}\left(t_{p+1}\right)\right\rangle _{\textrm{st}}$
still depends on the ``old'' relevant modes $\widehat{\phi}_{k}^{\dagger}\left(p\right)$.
The wavefunction $\left|\widetilde{\Phi}\left(t_{p+1}\right)\right\rangle _{\textrm{st}}$
also entangles one additional incoming mode $\widehat{\psi}_{p}^{\dagger}$
(the turquoise line going up from the stage b). in Fig. \ref{fig:evolution_stages}):
hence the name ``entangling step''.

\begin{figure}
\includegraphics[scale=2]{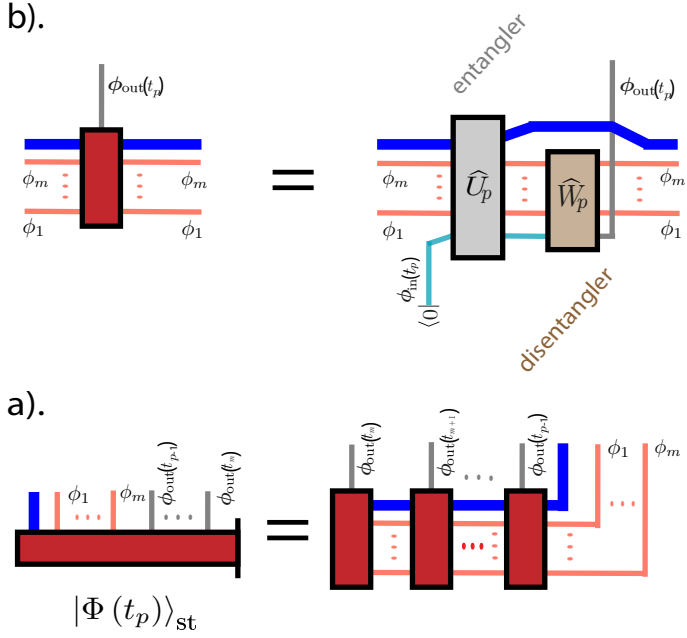}

\caption{\label{fig:MPO_block}a). In the frame of relevant/outgoing modes,
the real-time wavefunction $\left|\Phi\left(t_{p}\right)\right\rangle _{\textrm{st}}$
has the stucture of a matrix-product-state. b). Each block (tensor)
is a product of the entangling and disentangling transforms.}
\end{figure}

The propagation is finished by the \textit{disentangling step}: we
change to the basis of new relevant modes $\widehat{\phi}_{k}^{\dagger}\left(p+1\right)$,
the stage c). in Fig. \ref{fig:evolution_stages}:
\begin{equation}
\left|\Phi\left(t_{p+1}\right)\right\rangle _{\textrm{st}}=\widehat{W}_{p}\left|\widetilde{\Phi}\left(t_{p+1}\right)\right\rangle _{\textrm{st}},\label{eq:disentanglement-1}
\end{equation}
where the\textit{ disentangler} $\widehat{W}_{p}$ is given by
\begin{equation}
\widehat{W}_{p}=\exp\left(i\left[\widehat{\phi}_{1}^{\dagger}\ldots\widehat{\phi}_{m}^{\dagger}\widehat{\psi}_{p}^{\dagger}\right]h\left(p\right)\left[\widehat{\phi}_{1}\ldots\widehat{\phi}_{m}\widehat{\psi}_{p}\right]^{T}\right),\label{eq:many_particle_disentangler-1-1}
\end{equation}
and the Hermitean $\left(m+1\right)\times\left(m+1\right)$ matrix
$h\left(p\right)=i\ln W_{p}$, with the unitary matrix $W_{p}$ defined
in eq. (\ref{eq:next_step_of_stream_construction}). Observe that
in Eq. (\ref{eq:disentanglement-1}) we apply the active form of the
frame-change transformation (\ref{eq:many_particle_disentangler-1-1}).
As a result, the relevant modes loose their time dependence, $\widehat{\phi}_{k}^{\dagger}\left(p\right)\equiv\widehat{\phi}_{k}^{\dagger},\,k=1\ldots m$.
The resulting wavefunction $\left|\Phi\left(t_{p+1}\right)\right\rangle _{\textrm{st}}$
has the same structure as $\left|\Phi\left(t_{p}\right)\right\rangle _{\textrm{st}}$.
Namely, it has the same matrix-product-state structure as in Fig.
\ref{fig:evolution_stages}, a). Each block in this structure is formed
by a pair of entangler-disentangler transforms, Fig. \ref{fig:MPO_block}. 

Observe that as the result of $\widehat{W}_{p}$, one additional outgoing
mode $\phi_{\textrm{out}}\left(p\right)$ is produced in the wavefunction
$\left|\Phi\left(t_{p+1}\right)\right\rangle _{\textrm{st}}$. The
coupling of $\phi_{\textrm{out}}\left(p\right)$ to the future is
negligible by construction, so that $\phi_{\textrm{out}}\left(p\right)$
will not become entangled to the future incoming modes: hence the
name ``disentangling step''.

The iterative application of these propagation steps leads to the
circuit model presented in Fig. \ref{fig:QSA_production}. 
\begin{figure}
\includegraphics[scale=1.5]{quantum_shredder_4.ai}

\caption{\label{fig:QSA_production}The circuit model of real-time evolution
of open quantum system. Each time step $t_{p}$, one new incoming
mode $\phi_{\textrm{in}}\left(p\right)\equiv\phi_{\textrm{in}}\left(t_{p}\right)$
get entangled to the open system via the Hamiltonian evolution $\widehat{U}_{p}$.
Afterwards one new irreversibly decouled mode $\phi_{\textrm{out}}\left(p\right)\equiv\phi_{\textrm{out}}\left(t_{p}\right)$
gets \textit{disentangled }via the active change of frame $\widehat{W}_{p}$.
Observe that the product $\widehat{W}_{p}\widehat{U}_{p}$ is equal
to $\widehat{U}_{p}^{\prime}$ from Fig. \ref{fig:entanglement_structure}.}
\end{figure}

\subsection{Decoupling of the pairing function\label{subsec:Contribution-of-the}}

\begin{figure*}
\includegraphics[scale=2]{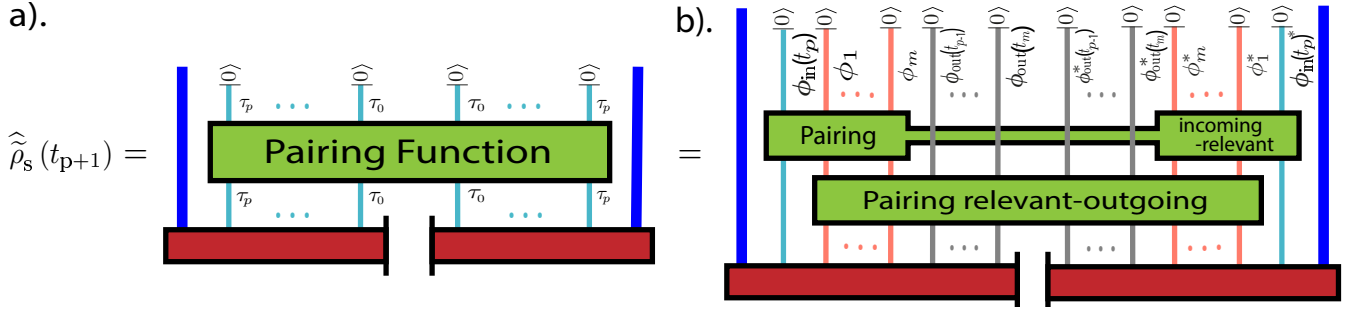}

\caption{\label{fig:decoupling_of_pairing_function}a). When evaluating observables
or taking the partial trace, the pairing function introduces additional
non-local coupling between the time cells. This coupling has the same
convolutional form as the annihilation processes considered in sec.
\ref{sec:HOW-THE-QUANTUM}. b). Therefore, in the frame of incoming/relevant/outgoing
modes, the couplings (i.e. pairings) between the past outgoing modes
and the future incoming modes are still negligible. This is the foundation
of our approach. The pairing function decouples into the two separate
parts: pairings involving relevant-outoing modes and the pairings
involving incoming-relevant modes. }

\end{figure*}

Let us recall that when computing the observables, we need to apply
the pairing function eq (\ref{eq:discrete_time_trace}). It introduces
an additional non-local coupling between the future and the past cell
modes. One might wonder if this would spoil the entanglement structure
described in Figs. \ref{fig:evolution_stages} - \ref{fig:QSA_production}.
The answer is no. The coupling in the pairing function has the same
form of the convolutional annihilation process that was thoroughly
analyzed in section \ref{sec:HOW-THE-QUANTUM}. Therefore, in the
frame of the incoming/outgoing/relevant modes, we can neglect the
pairings between the outgoing and the future incoming modes, Fig.
\ref{fig:decoupling_of_pairing_function}, 
 analogously to transition from eq. (\ref{eq:midpoint_Hamiltonian-1-1-1})
to eq. (\ref{eq:midpoint_Hamiltonian-1-1-1-1}). Below we ensure this
by explicit calculation.

\subsubsection{Change of frame in the pairing function\label{subsec:Change-of-frame}}

The pairing function in eq. (\ref{eq:discrete_time_trace}) is written
in terms of the original incoming modes $\psi_{k}^{\dagger}$. Let
us change to the frame of incoming/outgoing/relevant modes in the
pairing function. Suppose that these frames are related as
\begin{equation}
\left[\begin{array}{c}
\phi_{1}\left(p\right)\\
\vdots\\
\phi_{m}\left(p\right)\\
\phi_{\textrm{out}}\left(m\right)\\
\vdots\\
\phi_{\textrm{out}}\left(p-1\right)
\end{array}\right]=U\left[\begin{array}{c}
\phi_{\textrm{in}}\left(0\right)\\
\vdots\\
\phi_{\textrm{in}}\left(p-1\right)
\end{array}\right],
\end{equation}
where the unitary matrix $U$ is the cumulative effect of the disentanglers:
$U=W_{p}W_{p-1}\ldots W_{m}$. The creation operators are related
in the same way:
\begin{equation}
\widehat{\boldsymbol{\phi}}^{\dagger}\equiv\left[\begin{array}{c}
\widehat{\phi}_{1}^{\dagger}\\
\vdots\\
\widehat{\phi}_{m}^{\dagger}\\
\widehat{\phi}_{\textrm{out}}^{\dagger}\left(m\right)\\
\vdots\\
\widehat{\phi}_{\textrm{out}}^{\dagger}\left(p-1\right)
\end{array}\right]=U\left[\begin{array}{c}
\widehat{\psi}_{0}^{\dagger}\\
\vdots\\
\widehat{\psi}_{p-1}^{\dagger}
\end{array}\right],
\end{equation}
where $\widehat{\boldsymbol{\phi}}^{\dagger}$ is a shorthand for
the column vector of creation operators. Its components $\widehat{\boldsymbol{\phi}}_{1}^{\dagger}\ldots\widehat{\boldsymbol{\phi}}_{m}^{\dagger}$
refer to $\widehat{\phi}_{1}^{\dagger}\ldots\widehat{\phi}_{m}^{\dagger}$
correspondignly, and $\widehat{\boldsymbol{\phi}}_{m+1}^{\dagger}\ldots\widehat{\boldsymbol{\phi}}_{m+p}^{\dagger}$
refer to $\widehat{\phi}_{\textrm{out}}^{\dagger}\left(m\right)\ldots\widehat{\phi}_{\textrm{out}}^{\dagger}\left(p-1\right)$
correspondingly. The inverse transforms are given by $U^{\dagger}$.
We have for the pairing function: 
\begin{multline}
\sum_{rs=0}^{p-1}\widehat{\psi}_{r}^{\dagger}M_{rs}\widehat{\psi}_{s}=\sum_{rs=0}^{p-1}U_{ru}^{\dagger}\widehat{\boldsymbol{\phi}}_{u}^{\dagger}M_{rs}U_{sv}^{T}\widehat{\boldsymbol{\phi}}_{v}\\
=\sum_{uv}^{p-1}\widehat{\boldsymbol{\phi}}_{u}^{\dagger}\widetilde{M}_{uv}\left(p\right)\widehat{\boldsymbol{\phi}}_{v},
\end{multline}
where the matrix $\widetilde{M}$ has the block structure which reflects
the pairings between the relevant and ougoing modes:
\begin{equation}
\widetilde{M}\left(p\right)=\left[\begin{array}{cc}
M_{\textrm{rr}} & M_{\textrm{ro}}\\
M_{\textrm{ro}}^{\dagger} & M_{\textrm{oo}}
\end{array}\right].
\end{equation}
Here $M_{\textrm{rr}}$ generate the pairings between the relevant
modes: $\left(M_{\textrm{rr}}\right)_{rs}=\left(\phi_{r}\left(p\right)\left|M\right|\phi_{s}\left(p\right)\right)$;
$\left(M_{\textrm{oo}}\right)_{rs}=\left(\phi_{\textrm{out}}\left(r\right)\left|M\right|\phi_{\textrm{out}}\left(s\right)\right)$
generates the pairings between the outgoing modes; $\left(M_{\textrm{ro}}\right)_{rs}=\left(\phi_{r}\left(p\right)\left|M\right|\phi_{\textrm{out}}\left(s\right)\right)$
defining the pairings between the relevant-outgoing modes. 

\subsubsection{Negligible pairings between the past outgoing and the future incoming
modes}

Now suppose we have a new incoming mode $\phi_{\textrm{in}}\left(p\right)$
with the creation operator $\widehat{\psi}_{p}^{\dagger}$. It will
yield additional terms in the pairing function: 
\begin{multline}
\widehat{\psi}_{p}^{\dagger}M\left(0\right)\widehat{\psi}_{p}+\sum_{s=0}^{p-1}\left\{ \widehat{\psi}_{p}^{\dagger}M_{ps}\widehat{\psi}_{s}+\widehat{\psi}_{s}^{\dagger}M_{sp}\widehat{\psi}_{p}\right\} \\
=\widehat{\psi}_{p}^{\dagger}M\left(0\right)\widehat{\psi}_{p}\\
+\widehat{\psi}_{p}^{\dagger}\left\{ \sum_{i}M_{i}\left(p\right)\widehat{\phi}_{i}+\sum_{i}M_{\textrm{out}i}\left(p\right)\widehat{\phi}_{\textrm{out}}\left(i\right)\right\} +\textrm{c.c.}.
\end{multline}
The last term in the second line, $M_{\textrm{out}i}\left(p\right)\widehat{\phi}_{\textrm{out}}\left(i\right)$,
can be discarded by construction of the streams of outgoing modes.
Therefore, when taking the partial trace over the time cells, we can
assume that there are no pairings between the past outgoing and the
future incoming modes, and the described above structure of entanglement,
Fig. \ref{fig:QSA_production}, is preserved. 

\section{\label{sec:TRACING-OUT-THE}TRACING OUT THE OUTGOING MODES: RENOMALIZATION
GROUP FOR DENSITY MATRICES}

In this section we implement the idea that the ``idle'' bricks in
Fig. \ref{fig:entanglement_structure} can be removed as soon as they
emerge. In the conventional RG approach, we trace out the irrelevant
degrees of freedom as soon as they appear. Therefore, in this section
we introduce the joint density matrix $\widehat{\rho}_{\textrm{rel}}\left(t_{p}\right)$
of the open system and of the relevant degrees of freedom of the environment
as a partial trace over the irrelevant degrees. After that an iterative
propagation procedure is derived for $\widehat{\rho}_{\textrm{rel}}\left(t_{p}\right)$.

\subsection{\label{subsec:The-convention-of}The convention of growing Fock space}

We consider the case of the vacuum initial condition for the environment.
In this case the incoming modes are always in vacuum. Then the notation
will be more concise if we assume that the wavefunction $\left|\Phi\left(t_{p}\right)\right\rangle _{\textrm{st}}$
is defined on a growing Hilbert space $\mathcal{H}_{\textrm{s}}\otimes\mathcal{F}_{\textrm{t}}\left(p\right)$
of the open system, the present and the past tape cells. Namely, at
$t_{0}=0$ there are no cells at all, so that $\mathcal{F}_{\textrm{t}}\left(0\right)=\oslash$,
and the initial condition is just $\left|\Phi\left(0\right)\right\rangle _{\textrm{st}}=\left|\phi_{0}\right\rangle _{\textrm{s}}$.
After each propagation $t_{p}\to t_{p+1}$, one additional incoming
mode $\psi_{p}^{\dagger}$ gets coupled, and the Fock space is enlarged:
$\mathcal{F}_{\textrm{t}}\left(p+1\right)=\mathcal{F}_{\textrm{t}}\left(p\right)\otimes\left\{ \textrm{space spanned by \ensuremath{\psi_{p}^{\dagger}} }\right\} $.
Below we follow this convention. 

\subsection{\label{subsec:Density-matrix-in-the-time-domain}Density matrix in
the time domain}

We recall the formula eq. (\ref{eq:discrete_time_trace}) for the
partial trace over the time cells. We can rewrite it as 
\begin{equation}
\widehat{\rho}_{\textrm{s}}\left(t_{p}\right)=\ps{}{\textrm{t}}{\left\langle 0\right|\widehat{\rho}_{\textrm{st}}\left(t_{p}\right)\left|0\right\rangle _{\textrm{t}}},\label{eq:tracing_out_the_bath}
\end{equation}
where we introduce the joint density matrix for the open system and
the time cells,
\begin{figure*}
\includegraphics[scale=2]{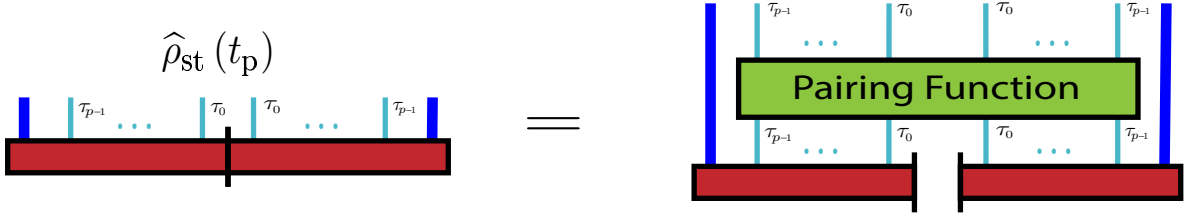}

\caption{\label{fig:definition_of_time_domain_density_matrix}The joint density
matrix $\widehat{\rho}_{\textrm{st}}\left(t_{p}\right)$ for the open
system (blue lines) and the time cells (turquoise lines) is defined
by applying the pairing function to $\left|\Phi\left(t_{p}\right)\right\rangle _{\textrm{st}}\protect\ps{}{\textrm{st}}{\left\langle \Phi\left(t_{p}\right)\right|}$.}
\end{figure*}
\begin{multline}
\widehat{\rho}_{\textrm{st}}\left(t_{p}\right)\\
=\left\{ e^{\sum_{rs=0}^{p-1}\widehat{\psi}_{r}^{\dagger}M_{rs}\widehat{\psi}_{s}}:\left|\Phi\left(t_{p}\right)\right\rangle _{\textrm{st}}\ps{}{\textrm{st}}{\left\langle \Phi\left(t_{p}\right)\right|}:\right\} _{A},\label{eq:density_matrix_in_time_domain}
\end{multline}
see also Fig \ref{fig:definition_of_time_domain_density_matrix}.
This density matrix is Hermitean and positive semidefinite. The latter
follows from the fact that $\left|\Phi\left(t_{p}\right)\right\rangle _{\textrm{st}}\ps{}{\textrm{st}}{\left\langle \Phi\left(t_{p}\right)\right|}$
is positive semidefinite, and the action of the pairing function is
a completely positive map since it has the ``sandwitch'' form required
by the Stinespring factorization theorem. Indeed, each elementary
pairing has the form
\begin{equation}
\sum_{rs}M_{rs}\widehat{\psi}_{s}:\ldots:\widehat{\psi}_{r}^{\dagger}=\intop_{-\infty}^{+\infty}d\tau\widehat{K}\left(\tau\right):\ldots:\widehat{K}^{\dagger}\left(\tau\right),\label{eq:sandwitch_form_of_pairing_update}
\end{equation}
where $\widehat{K}\left(\tau\right)=\sum_{s}M_{\tau\tau_{S}}^{1/2}\widehat{\psi}_{s}$,
and $M_{\tau\tau^{\prime}}^{1/2}=\left(2\pi\right)^{-\frac{1}{2}}\intop_{0}^{+\infty}d\omega e^{-i\omega\left(\tau-\tau^{\prime}\right)}\sqrt{J\left(\omega\right)}$
is the wavepacket of a single emission event from Fig. \ref{fig:time_nonlocality}. 

However the density matrix $\widehat{\rho}_{\textrm{st}}\left(t_{p}\right)$
need not be normalized in the conventional sense, $\textrm{Tr}\widehat{\rho}_{\textrm{st}}\left(t_{p}\right)\neq1$.
Instead, its vaccum part is always normalized: 
\begin{equation}
\textrm{Tr}\left\{ \ps{}{\textrm{t}}{\left\langle 0\right|\widehat{\rho}_{\textrm{st}}\left(t_{p}\right)\left|0\right\rangle _{\textrm{t}}}\right\} =1.\label{eq:normalization_in_the_time_domain}
\end{equation}
This follows from the fact that density matrix of open system is normalized,
$\textrm{Tr}\widehat{\rho}_{\textrm{s}}\left(t_{p}\right)=1$, and
that the partial trace over the environment involves the vacuum projection
in the tape model, eq. (\ref{eq:discrete_time_trace}). Let us remind
that we have derived the trace formula eq. (\ref{eq:discrete_time_trace})
from the formal requirement that all the commutators are conserved,
see sections \ref{subsec:Fock-space-for} and \ref{subsec:Partial-trace}.
However later, in sec. \ref{subsec:Pairing-function-as}, we find
a physical interpretation of this relation: partial trace eq. (\ref{eq:discrete_time_trace})
and the normalization (\ref{eq:normalization_in_the_time_domain})
can be interpreted as a result of vacuum fluctuations of the classical
background of the environment.

\subsection{Definition of the relevant density matrix}

\begin{figure*}
\includegraphics[scale=2]{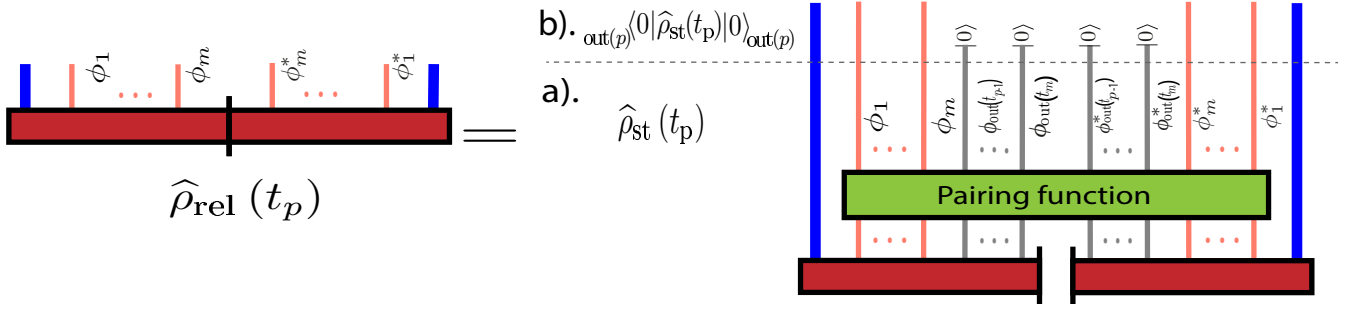}

\caption{\label{fig:relevant_density_matrix}Renormalization group method amounts
to tracing out the the irrelevant degrees of freedom as they emerge
during the flow. This results in the density matrix $\widehat{\rho}_{\textrm{rel}}$
which is the joint reduced state of relevant degrees of freedom and
of open system. This matrix is defined in two steps: a). we take the
joint density matrix of open system and time cells $\widehat{\rho}_{\textrm{st}}\left(t_{p}\right)$
in the frame of relevant/outgoing modes; b). then we project to the
vacuum all the outgoing modes which have emerged by the time moment
$t_{p}$. }

\end{figure*}

Here we define the relevant density matrix $\widehat{\rho}_{\textrm{rel}}\left(t_{p}\right)$,
and in the next section we discuss how it can be computed. In order
to introduce the relevant density matrix $\widehat{\rho}_{\textrm{rel}}\left(t_{p}\right)$,
we first switch to the frame of incoming/relevant/outgoing modes in
the definition (\ref{eq:density_matrix_in_time_domain}) of $\widehat{\rho}_{\textrm{st}}\left(t_{p}\right)$,
Fig. \ref{fig:relevant_density_matrix} a). 
This change of frame was described in sec. \ref{subsec:Change-of-frame}.
After that all the outgoing modes which have emerged by the time moment
$t_{p}$ are projected to vacuum, Fig. \ref{fig:relevant_density_matrix}
b) : 

\begin{multline}
\widehat{\rho}_{\textrm{rel}}\left(t_{p}\right)=\ps{}{\textrm{out}\left(p\right)}{\left\langle 0\right|}\left\{ e^{\sum_{rs=0}^{p-1}\widehat{\psi}_{r}^{\dagger}M_{rs}\widehat{\psi}_{s}}\right.\\
\left.\times:\left|\Phi\left(t_{p}\right)\right\rangle _{\textrm{st}}\ps{}{\textrm{st}}{\left\langle \Phi\left(t_{p}\right)\right|}:\right\} _{A}\left|0\right\rangle _{\textrm{out}\left(p\right)},\label{eq:relevant_density_matrix-1}
\end{multline}
where 
\begin{equation}
\left|0\right\rangle _{\textrm{out}\left(p\right)}=\left|0\right\rangle _{\phi_{\textrm{out}}\left(p-1\right)}\otimes\ldots\otimes\left|0\right\rangle _{\phi_{\textrm{out}}\left(m\right)}
\end{equation}
is the joint vacuum of the outgoing modes $\phi_{\textrm{out}}\left(m\right),\ldots,\phi_{\textrm{out}}\left(p-1\right)$
which have emerged by the time moment $t_{p}$. The operation $\ps{}{\textrm{out}\left(p\right)}{\left\langle 0\right|}\cdot\left|0\right\rangle _{\textrm{out}\left(p\right)}$
in eq. (\ref{eq:relevant_density_matrix-1}) denotes the projection
of density matrix to the subspace of zero quanta in the outgoing modes
$\phi_{\textrm{out}}\left(m\right),\ldots,\phi_{\textrm{out}}\left(p-1\right)$.
This operation does not affect the other degrees of freedom. Actually
this projection restricts the growing Fock space $\mathcal{F}_{\textrm{t}}\left(p\right)$
to the subspace $\mathcal{F}_{\textrm{rel}}$ spanned by the relevant
modes $\phi_{1}\ldots\phi_{m}$. The reduced density matrix for the
open system is obtained via the projection
\begin{equation}
\widehat{\rho}_{\textrm{s}}\left(t_{p}\right)=\ps{}{\textrm{rel}}{\left\langle 0\right|}\widehat{\rho}_{\textrm{rel}}\left(t_{p}\right)\left|0\right\rangle _{\textrm{rel}},
\end{equation}
where here and below under $\left|0\right\rangle _{\textrm{rel}}$
we designate the joint vacuum for the relevant modes $\widehat{\phi}_{1}^{\dagger},\ldots,\widehat{\phi}_{m}^{\dagger}$. 

\subsection{\label{subsec:Propagation-of-density}Propagation of density matrix}

Given an environment with a spectral density $J\left(\omega\right)$,
one computes its memory function $M\left(t-t^{\prime}\right)$. Then
the incoming/outgoing/relevant modes, and disentanglers $W_{p}$ are
found according to the algorithm of sec. \ref{subsec:The-emergence-of}.
This needs to be done once for a given $J\left(\omega\right)$.

The initial condition for the propagation is $\widehat{\rho}_{\textrm{rel}}\left(0\right)=\left|\phi_{0}\right\rangle _{\textrm{s}}\ps{}{\textrm{s}}{\left\langle \phi_{\textrm{0}}\right|}$,
since there is no relevant modes at $t_{0}=0$. 

\begin{figure*}
\includegraphics[scale=2.5]{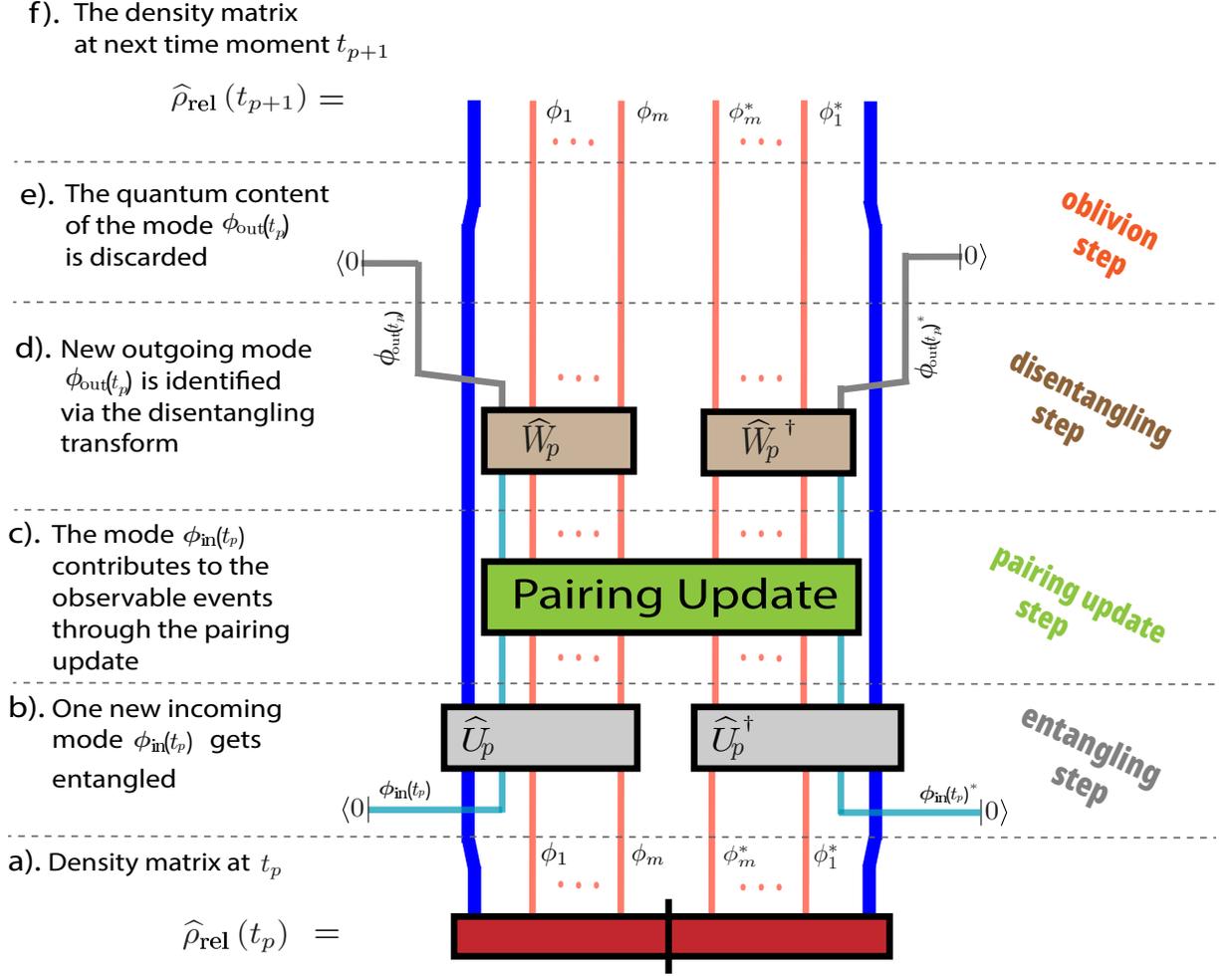}

\caption{\label{fig:density_matrix_rg}The lifecycle of entanglement in course
of real time evolution. It directly maps to the numerical recipe of
the renormalization group for density matrices. }
\end{figure*}

Now suppose we have $\widehat{\rho}_{\textrm{rel}}\left(t_{p}\right)$
which is an operator acting in $\mathcal{H}_{\textrm{s}}\otimes\mathcal{F}_{\textrm{rel}}$,
Fig. \ref{fig:density_matrix_rg}, a). The first step of propagation
is the\textit{ entangling step}, Fig. \ref{fig:density_matrix_rg},
b). 
Here the Hamiltonian $\widehat{H}_{\textrm{st}}\left(\tau_{p}\right)$
of eq. (\ref{eq:midpoint_Hamiltonian-1-1-1-1}) is applied. We use
the following three facts. First, since $\widehat{H}_{\textrm{st}}\left(\tau_{p}\right)$
is not Hermitean, the bra state $\ps{}{\textrm{st}}{\left\langle \Phi\left(t_{p}\right)\right|}$
evolves under $\widehat{H}_{\textrm{st}}^{\dagger}\left(\tau_{p}\right)$,
whereas the ket state $\left|\Phi\left(t_{p}\right)\right\rangle _{\textrm{st}}$
evolves under $\widehat{H}_{\textrm{st}}\left(\tau_{p}\right)$. Second,
the Hamiltonian $\widehat{H}_{\textrm{st}}\left(\tau_{p}\right)$
can be commuted to the left of the pairing function in eq. (\ref{eq:relevant_density_matrix-1}).
Analogously, the Hamiltonian $\widehat{H}_{\textrm{st}}^{\dagger}\left(\tau_{p}\right)$
can be commuted to the right of the pairing function. Third, the Hamiltonian
$\widehat{H}_{\textrm{st}}\left(\tau_{p}\right)$ entangles one additional
incoming mode $\widehat{\psi}_{p}^{\dagger}$. Therefore, the Fock
space is enlarged from $\mathcal{F}_{\textrm{rel}}$ to $\mathcal{F}_{\textrm{rel+i}}$
due to the states generated by $\widehat{\psi}_{p}^{\dagger}$. The
the density matrix $\widehat{\rho}_{\textrm{rel}}\left(t_{p}\right)$
is embedded into the space $\mathcal{H}_{s}\otimes\mathcal{F}_{\textrm{rel+i}}$
as $\widehat{\rho}_{\textrm{rel+i}}\left(t_{p}\right)=\widehat{\rho}_{\textrm{rel}}\left(t_{p}\right)\otimes\left|0\right\rangle _{\varphi_{\textrm{in}}\left(p\right)}\ps{}{\varphi_{\textrm{in}}\left(p\right)}{\left\langle 0\right|}$.
It propagates as 
\begin{multline}
\widehat{\rho}_{\textrm{rel+i}}^{\left(1\right)}\left(t_{p+1}\right)=\widehat{\rho}_{\textrm{rel+i}}\left(t_{p}\right)\\
-idt\left\{ \widehat{H}_{\textrm{st}}\left(\tau_{p}\right)\widehat{\rho}_{\textrm{rel+i}}^{\left(1/2\right)}\left(t_{p}\right)-\widehat{\rho}_{\textrm{rel+i}}^{\left(1/2\right)}\left(t_{p}\right)\widehat{H}_{\textrm{st}}^{\dagger}\left(\tau_{p}\right)\right\} ,\label{eq:hamiltonian_propagation_of_density_matrix}
\end{multline}
where $\widehat{\rho}_{\textrm{rel+i}}^{\left(1/2\right)}\left(t_{p}\right)=\frac{1}{2}\left(\widehat{\rho}_{\textrm{rel+i}}^{\left(1\right)}\left(t_{p+1}\right)+\widehat{\rho}_{\textrm{rel+i}}\left(t_{p}\right)\right)$.
This propagation corresponds to the operators $\widehat{U}_{p}$ and
$\widehat{U}_{p}^{\dagger}$ on Fig. \ref{fig:density_matrix_rg}. 

Observe that $\widehat{\rho}_{\textrm{rel+i}}^{\left(1\right)}\left(t_{p+1}\right)$
entangles one additional incoming mode $\widehat{\psi}_{p}^{\dagger}$,
and the pairing function in eq. (\ref{eq:relevant_density_matrix-1})
does not yet contain the pairings for this mode. As follows from Fig.
\ref{fig:decoupling_of_pairing_function} and section \ref{subsec:Contribution-of-the},
only the pairings of incoming mode with itself and with the relevent
modes should be included. Therefore, the second step is the \textit{pairing
update}, Fig. \ref{fig:density_matrix_rg} c)., 
\begin{multline}
\widehat{\rho}_{\textrm{rel+i}}^{\left(2\right)}\left(t_{p+1}\right)=\left\{ e^{\widehat{\psi}_{p}^{\dagger}M\left(0\right)\widehat{\psi}_{p}+\left(\widehat{\psi}_{p}^{\dagger}\sum_{i}M_{i}\left(p\right)\widehat{\phi}_{i}+\textrm{c.c}.\right)}\right.\\
\left.:\widehat{\rho}_{\textrm{rel+i}}^{\left(1\right)}\left(t_{p+1}\right):\right\} _{A}.\label{eq:nonMarkovain_jump}
\end{multline}

During the propagation the pairing update is the only operation which
can make the density matrix impure. This is the non-Markovian analog
of the jump $L\widehat{\rho}L^{\dagger}$-terms in the Markovian Lindblad
dissipator \citep{Plenio1998,Breuer2011}. 

In the third step of propagation, the \textit{disentangling step},
Fig. \ref{fig:density_matrix_rg} d)., we apply the disentangler $\widehat{W}_{p}$
to convert the incoming mode $\widehat{\psi}_{p}$ into the new outgoing
mode $\widehat{\phi}_{\textrm{out}}\left(p\right)$: 
\begin{equation}
\widehat{\rho}_{\textrm{rel+o}}^{\left(3\right)}\left(t_{p+1}\right)=\widehat{W}_{p}\widehat{\rho}_{\textrm{rel+i}}^{\left(2\right)}\left(t_{p+1}\right)\widehat{W}_{p}^{-1}.\label{eq:density_matrix_disentangler}
\end{equation}
Here we change the subscript from 'rel+i' to 'rel+o' in order to make
explicit the fact that (i) we start with the Fock space spanned by
the relevant modes $\widehat{\phi}_{1}^{\dagger}\ldots\widehat{\phi}_{m}^{\dagger}$
and one incoming mode $\widehat{\psi}_{p}^{\dagger}$, and (ii) after
the disentangler transform we end with the Fock space spanned by the
relevant modes $\widehat{\phi}_{1}^{\dagger}\ldots\widehat{\phi}_{m}^{\dagger}$
and one outgoing mode $\widehat{\phi}_{\textrm{out}}^{\dagger}\left(p\right)$.

The last step, \textit{the oblivion}, \ref{fig:density_matrix_rg}
e)., is to discard the quantum content of this irrelevant mode:
\begin{equation}
\widehat{\rho}_{\textrm{rel}}\left(t_{p+1}\right)=\ps{}{\phi_{\textrm{out}}\left(t_{p}\right)}{\left\langle 0\right|}\widehat{\rho}_{\textrm{rel+o}}^{\left(3\right)}\left(t_{p+1}\right)\left|0\right\rangle _{\phi_{\textrm{out}}\left(t_{p}\right)}.\label{eq:density_matrix_project_away}
\end{equation}
As a result, we return to the smaller Fock space $\mathcal{F}_{\textrm{rel}}$
. Then the propagation procedure is repeated for the next time moment
$t_{p+1}$.

These four steps, eqs. (\ref{eq:hamiltonian_propagation_of_density_matrix}),
(\ref{eq:nonMarkovain_jump}), (\ref{eq:density_matrix_disentangler}),
(\ref{eq:density_matrix_project_away}), constute one complete iteration
of the real-time RG flow. They can be implemented numerically in a
Fock space of the tape cells (see Fig \ref{fig:The-renormalization-group_Fock_space}),
see appendix \ref{sec:SOLVING-EQUATIONS-IN}.
\begin{figure}
\includegraphics[scale=1.2]{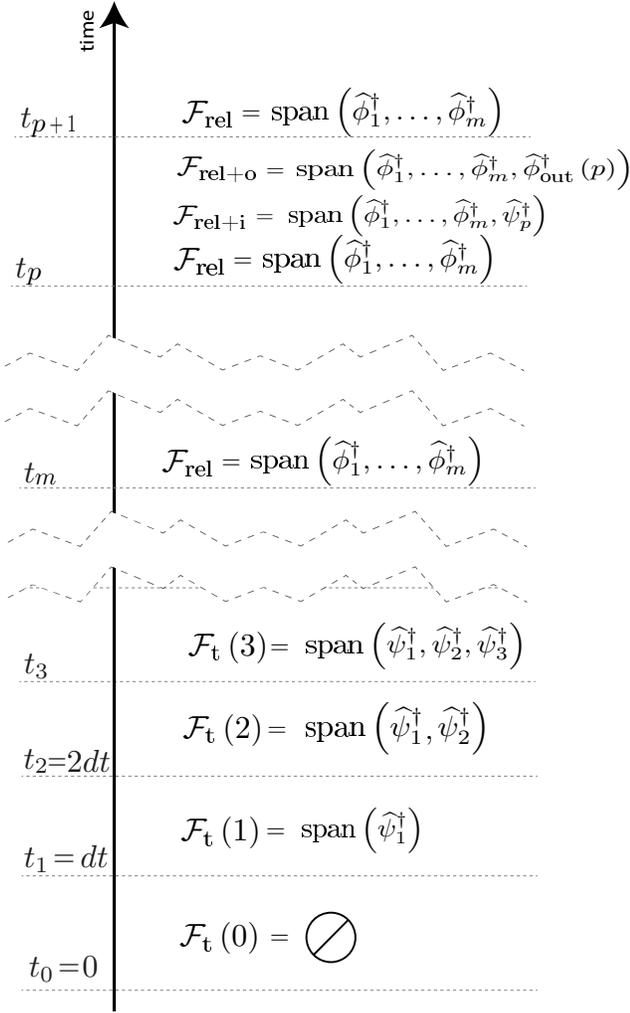}

\caption{\label{fig:The-renormalization-group_Fock_space}The renormalization
group for density matrices has two regimes. 1). Before the time moment
$t_{m}$ all the time cells are relevant, and the tape Fock space
continuously grows. 2). After the time moment $t_{m}$ the Fock space
stops growing and starts to ``oscillate'' between $\mathcal{F}_{\textrm{rel}}$,
$\mathcal{F}_{\textrm{rel+i}}$, $\mathcal{F}_{\textrm{rel+o}}$.
Here $\textrm{span}\left(\ldots\right)$ designates the Fock space
generated by applying creation operators inside the brackets to the
vacuum state. In a numerical simulation the Fock spaces are truncated
in the maximal total occupation of the modes. Such a truncation is
justified by the conjectured balance of complexity: see sec. \ref{subsec:Balance-of-complexity}.
In a numerical simulation the Fock spaces $\mathcal{F}_{\textrm{rel+i}}$
and $\mathcal{F}_{\textrm{rel+o}}$ are the same: it is our interpretation
that the additional mode alternatively plays the role of the incoming
and outgoing mode.}
\end{figure}
which is truncated in the maximal total occupation of the modes. The
pairing update (\ref{eq:nonMarkovain_jump}) is inexpensive since
we can truncate it to second order, see appendix \ref{sec:COMPUTING-THE-PAIRING}.
The disentangler (\ref{eq:density_matrix_project_away}) can be efficiently
implemeted using Lanczos-like algorithms: see appendix \ref{sec:LANCZOS-INSPIRED-ALGORITHM-FOR}. 

\subsection{Example calculation}

\begin{figure}
\includegraphics[scale=0.45]{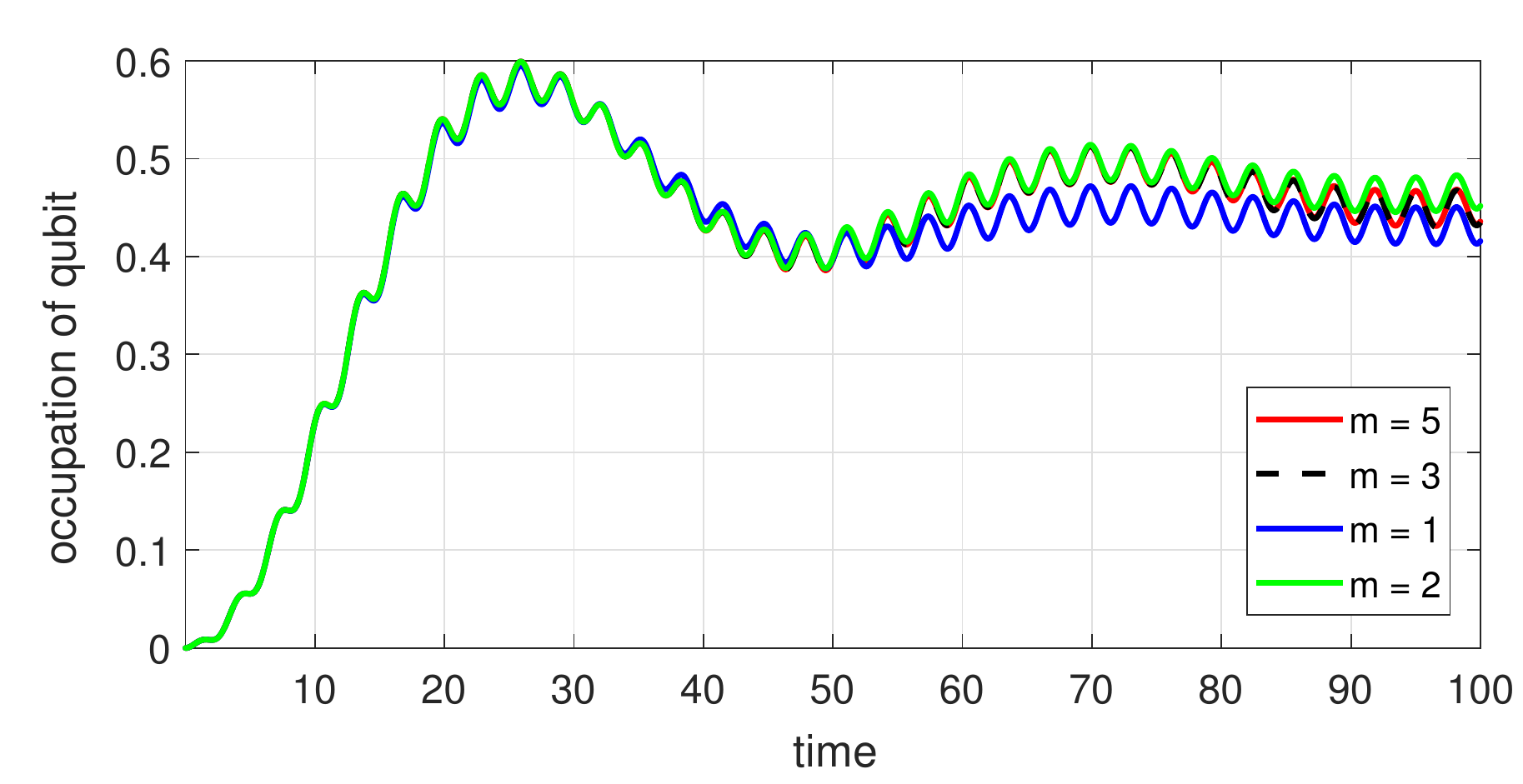}

\caption{\label{fig:For-the-system}For the system of a driven qubit in a highly
non-Markovian waveguide, only 3 relevant modes at a maximal occupation
of 2 quanta is enough to achieve the convergence of observable properties.
The calculation is done by RG for density matrices.}
\end{figure}

\begin{figure}
\includegraphics[scale=0.45]{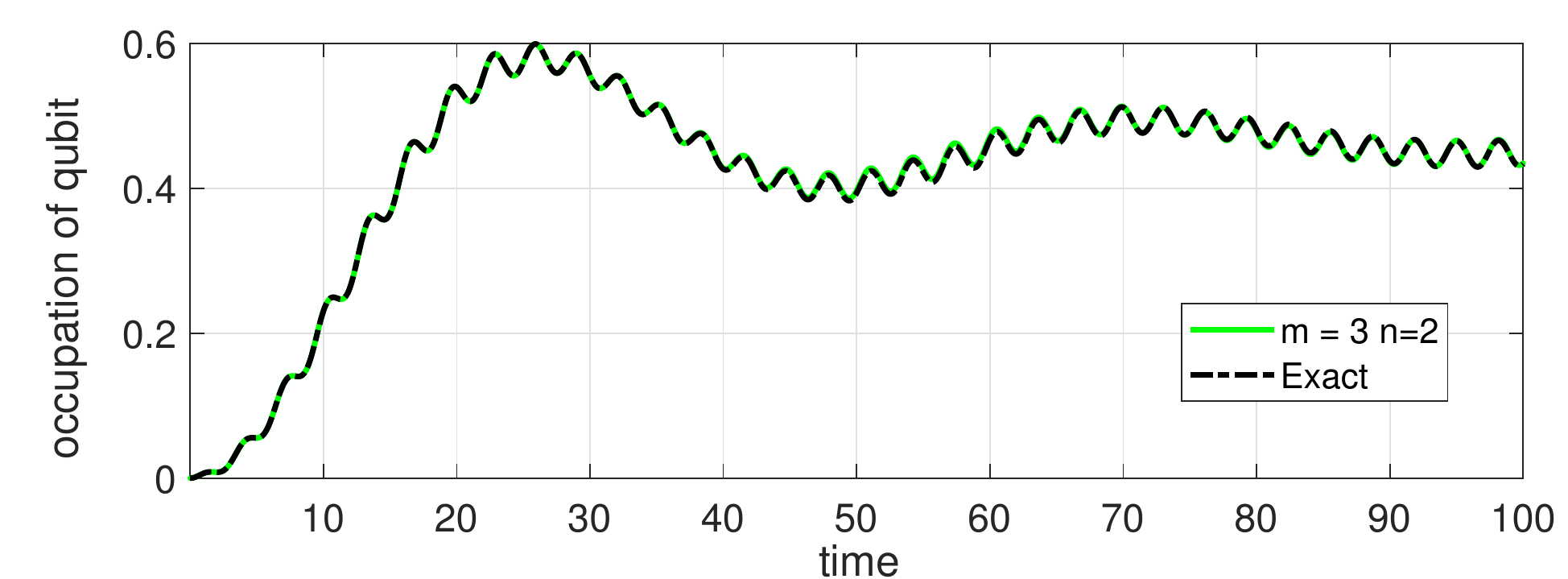}

\caption{\label{fig:For-the-same}For the same situation as in previous Fig.
\ref{fig:For-the-system}, we compare the converged RG for density
matrices with the numerically exact solution of Schrodinger equation.
The latter required us to keep $14$ lattice sites in eq. (\ref{eq:waveguide_system})
and all states of maximum total occupation of $7$ quanta. }

\end{figure}

In Fig. \ref{fig:For-the-system} we present the results of calculation
of the RG for density matrices for the system of a driven qubit coupled
to the waveguide eq. (\ref{eq:waveguide_system}) with $\varepsilon=1$
and $h=0.05$. We have: $\widehat{H}_{\textrm{s}}=\widehat{\sigma}_{+}\widehat{\sigma}_{-}+0.1\cos t$,
and $\widehat{s}=\widehat{\sigma}_{-}$.
 We conduct a series of simulations with one ($m=1$) , two ($m=2$),
three ($m=3$), and five ($m=5$) relevant modes. The convergence
is achieved when only the states of no more than two quanta were kept
in the tape Fock spaces $\mathcal{F}_{\textrm{t}}\left(p\right)$,
$\mathcal{F}_{\textrm{rel}}$, $\mathcal{F}_{\textrm{rel+i}}$, $\mathcal{F}_{\textrm{rel+o}}$.
It is seen that the simulation converges already for the three ($m=3$)
relevant modes. In Fig. \ref{fig:For-the-same} we compare the converged
result with the numerically-exact solution of the Schrodinger equation.
The Schrodinger equation was solved in the truncated Hilbert space
of the model (\ref{eq:waveguide_system}). The truncation was done
by keeping the first $14$ lattice sites $\widehat{a}_{1}^{\dagger}\ldots\widehat{a}_{14}^{\dagger}$
and by keeping all the Fock states $\widehat{a}_{1}^{\dagger n_{1}}\ldots\widehat{a}_{14}^{\dagger n_{14}}\left|0\right\rangle _{\textrm{b}}$
with maximal total occupation $n_{1}+\ldots+n_{14}\leq7$. Our RG
procedure yields a benefit in comparison with the ``bare'' solution
of the Schrodinger equation: $3$ modes vs $14$ and $2$ quanta vs
$7$. The physical explanation is that among the $14$ modes only
$3$ are non-neglibigly coupled to the future motion, and their occupation
is rather small, so that $2$ quanta are enough.  

\subsection{\label{subsec:Balance-of-complexity}Balance of complexity}

As we discussed in the Introduction, the quantum complexity is expected
to saturate if the rates of occurences of incoming and outgoing modes
are equal, and the incoing and outgoing fluxes are also equal. Here
we discuss there questions on the example of the driven qubit in the
waveguide from the previous section.

\subsubsection{Rates of occurences of incoming and ougoing modes}

As it is seen from the Fig. \ref{fig:The-renormalization-group_Fock_space},
the rates of ougoing modes are not always equal to the rates of incoming
modes. On the time interval $\left[0,t_{m}\right]$ every incoming
mode becomes relevant: no outgoing modes are produced at all. During
this initial period the dimension of the relevant space grows linearly
with time. However, due to the fastest decoupling property, Fig. \ref{fig:decay_of_eigenvals-1},
when the dimension becomes equal to $m$, the contribution of new
relevant modes becomes exponentially small. The relevant space saturates.
We can keep the dimension fixed to $m$ and start to produce the outgoing
modes with the rate equal to the rate of incoming modes. 

\subsubsection{Balance of fluxes}

\begin{figure}
\includegraphics[scale=0.5]{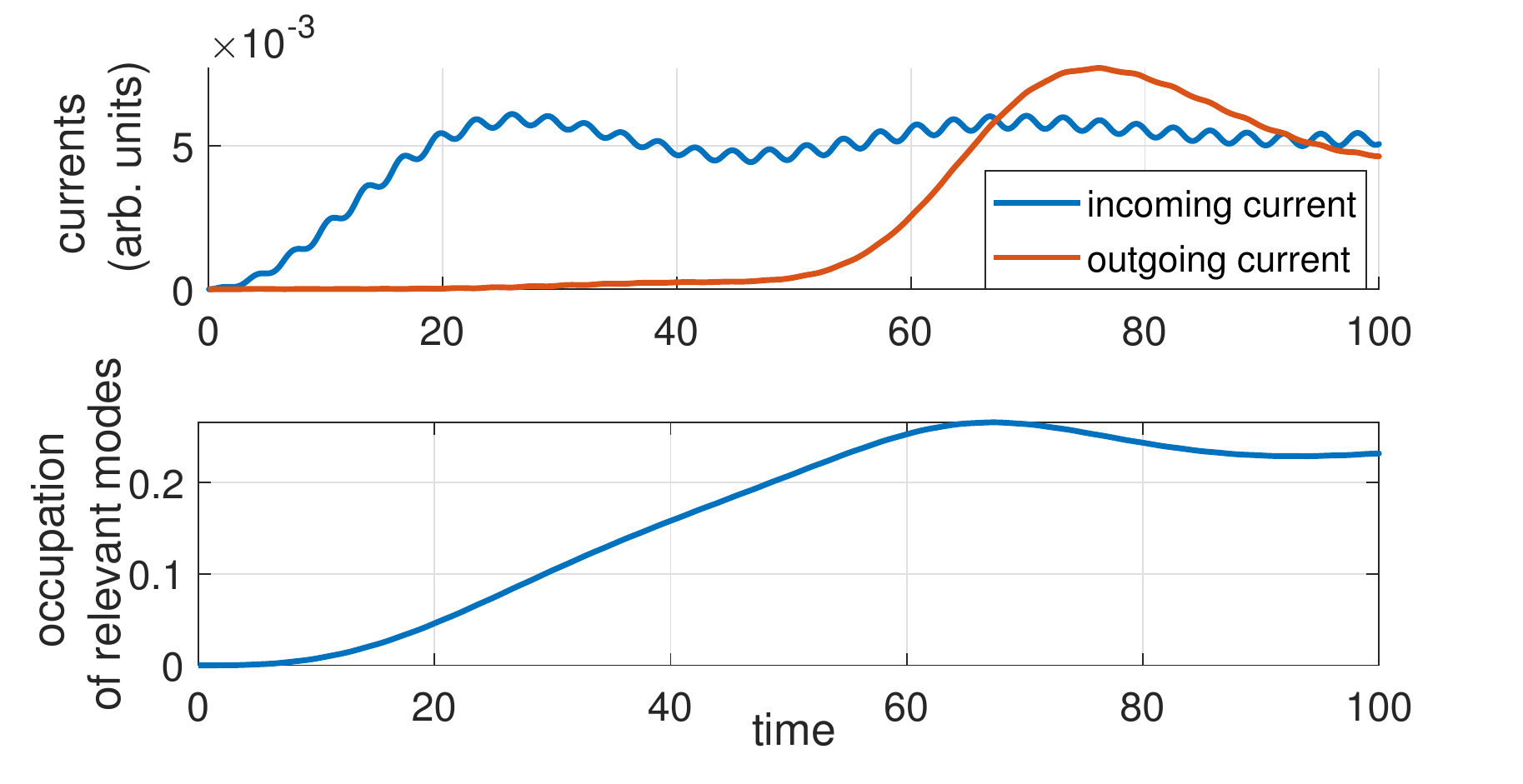}

\caption{\label{fig:The-outgoing-current}The system of a driven qubit in a
waveguide. The outgoing current of quanta which are discarded in the
outgoing modes balances the incoming flux of quanta which are emitted
into the incoming modes. The units for the currents are common but
otherwise arbitrary. As a consequence, the total occupation of the
relevant modes saturates. }

\end{figure}

For the same system of a driven qubit in the waveguide, we check the
proposed conjecture that the flux of the quanta emitted into the incoming
mode (the \textit{incoming current}) and the flux of the quanta discarded
in the ougoing mode (the \textit{outgoing current}) should balance
each other. The results of the Fig. \ref{fig:The-outgoing-current}
support our conjecture. The incoming current $j_{\textrm{in}}\left(t_{p}\right)$
is computed as the population of the incoming mode $\varphi_{\textrm{in}}\left(p\right)$
at the end of stage b) in Fig. \ref{fig:density_matrix_rg} (after
propagation under $\widehat{U}_{p}\ldots\widehat{U}_{p}^{\dagger}$,
but before the pairing update):
\begin{equation}
j_{\textrm{in}}\left(t_{p}\right)=\frac{1}{dt}\textrm{Tr}\left\{ \widehat{\rho}_{\textrm{rel+i}}^{\left(1\right)}\left(t_{p+1}\right)\widehat{\psi}_{p}^{\dagger}\widehat{\psi}_{p}\right\} .
\end{equation}
The outgoing current $j_{\textrm{out}}\left(t_{p}\right)$ is computed
as the population of the outgoing mode $\phi_{\textrm{out}}\left(p\right)$
at the end of stage d) on Fig. \ref{fig:density_matrix_rg} (after
disentangling under $\widehat{W}_{p}\ldots\widehat{W}_{p}^{\dagger}$
but before the oblivion):
\begin{equation}
j_{\textrm{out}}\left(t_{p}\right)=\frac{1}{dt}\textrm{Tr}\left\{ \widehat{\rho}_{\textrm{rel+o}}^{\left(3\right)}\left(t_{p+1}\right)\widehat{\phi}_{\textrm{out}}^{\dagger}\left(p\right)\widehat{\phi}_{\textrm{out}}\left(p\right)\right\} .
\end{equation}
Finally, the total occupation of relevant modes is computed as 
\begin{equation}
n_{\textrm{tot}}\left(t_{p}\right)=\textrm{Tr}\left\{ \widehat{\rho}_{\textrm{rel+i}}^{\left(1\right)}\left(t_{p+1}\right)\left(\widehat{\psi}_{p}^{\dagger}\widehat{\psi}_{p}+\sum_{k=1}^{m}\widehat{\phi}_{k}^{\dagger}\widehat{\phi}_{k}\right)\right\} .
\end{equation}

\begin{figure*}
\includegraphics[scale=2]{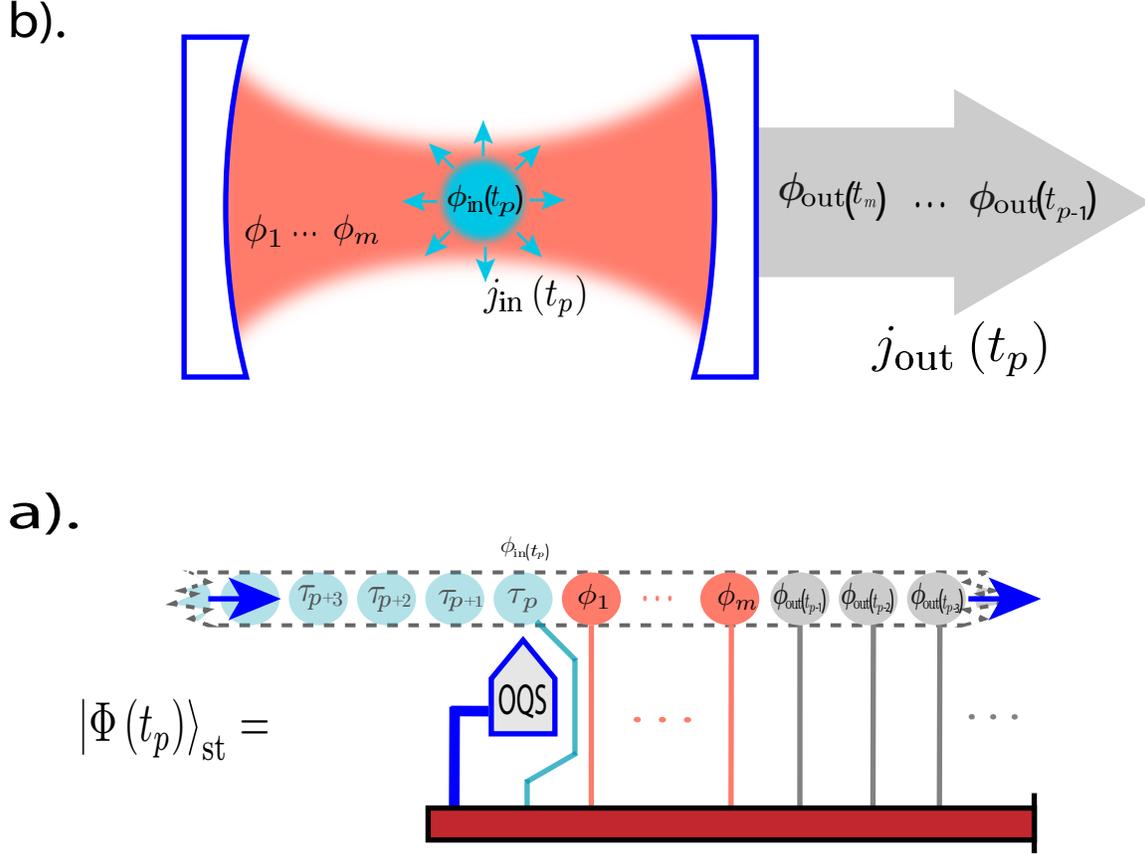}

\caption{\label{fig:complexity_balance_interpretation}We conjecture that the
quantum complexity of the real-time motion of open quantum systems
is asymptotically bounded. The interpretation is the following a).
Suppose we are given a joint quantum state $\left|\Phi\left(t_{p}\right)\right\rangle _{\textrm{st}}$
for the open system and the time cells, in the frame of incoming/relevant/outgoing
modes. b). It can be interpreted that the relevant modes $\phi_{1}\ldots\phi_{m}$
play the role of the ``cavity'' modes. The ``mirrors'' of the
cavity are the coupling to the future evolution. New quanta are injected
into the ``cavity'' with a flux $j_{\textrm{in}}\left(t_{p}\right)$
via the incoming mode $\varphi_{\textrm{in}}\left(p\right)$. However
the ``mirrors'' are imperfect: the outgoing modes are continuously
leaking out and carry away with them a flux $j_{\textrm{out}}\left(t_{p}\right)$
of quanta. As a result, the total occupation $n_{\textrm{tot}}\left(t_{p}\right)$
of the ``cavity'' modes is expected to be bounded due to the balance
of the fluxes. }

\end{figure*}

This balance of compelxity has the following physical interpretation,
Fig. \ref{fig:complexity_balance_interpretation}. The incoming modes
inject new quanta with the flux $j_{\textrm{in}}\left(t_{p}\right)$
into the ``cavity'' which is formed by the relevant modes. The ``mirrors''
of the cavity are formed by the coupling to the future motion. These
``mirrors'' are imperfect: the outgoing modes are continuously leaking
out and carry away with them a flux $j_{\textrm{out}}\left(t_{p}\right)$
of quanta. 

\subsection{Analogies with the time-symmetric formulation of quantum mechanics }

\begin{figure*}
\includegraphics[scale=1.2]{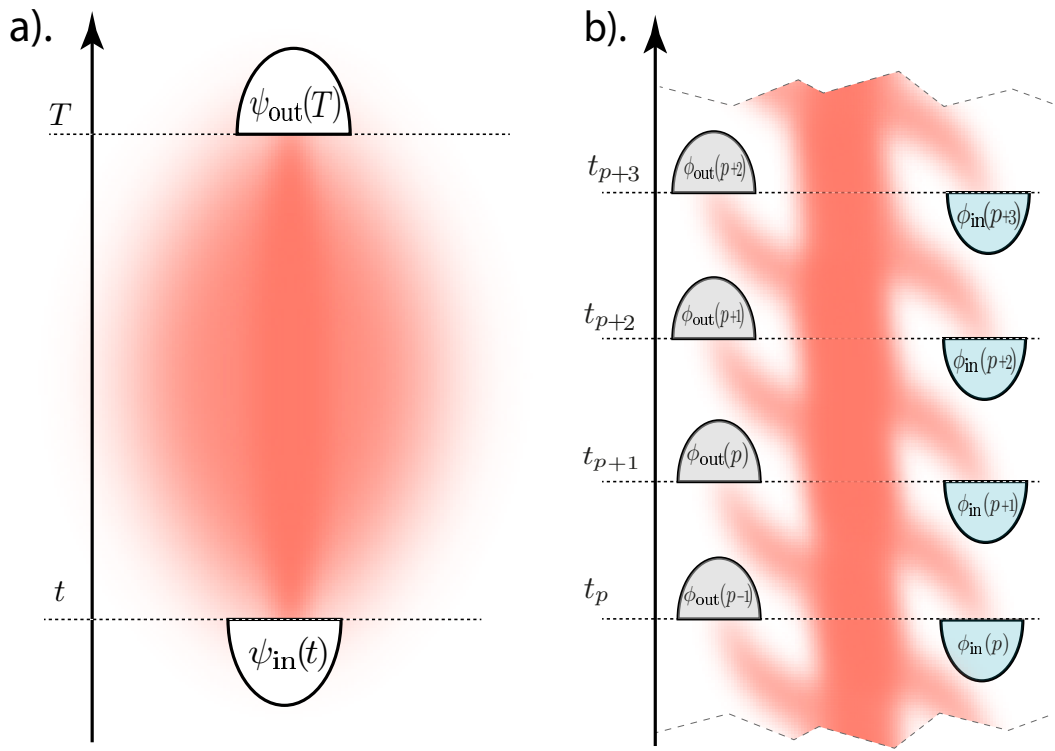}

\caption{\label{fig:time_symmetric_analogy}a). In the time-symmetric formulation
of quantum mechanics (also called the two-state vector formalism,
TSVF) of Yakir Aharonov et al. \citep{Aharonov2008}, the quantum
evolution is considered which starts from some initial state $\psi_{\textrm{in}}\left(t\right)$.
Then one assumes that there exists a final boundary condition in the
future $\psi_{\textrm{out}}\left(T\right)$. The conjecture is that
this final boundary condition is ``fine-tuned'' to yield an effective
collapse to definite outcomes of all the quantum measurements happening
inside the interval $\left[0,T\right]$ \citep{Aharonov2017}. b).
However the real-time structure of entanglement suggests an interesting
alternative: instead of the speculated global future boundary condition,
there are \textit{streams }of initial $\varphi_{\textrm{in}}\left(p\right)$
and final $\varphi_{\textrm{out}}\left(p\right)$ states, so that
there is a continuous competition between the birth and the death
of the quantum reality during each infinitesimal time interval.}
\end{figure*}

Observe that the general structure of the RG flow described above
is time-symmetric: if we apply the steps in Fig. \ref{fig:density_matrix_rg}
forward in time, then i) one incoming mode in the vacuum gets entangled;
ii) one outgoing mode gets disentangled and projected to the vacuum.
If we reverse this procedure in time, we get: i') one outgoing mode
in the vacuum gets entangled; ii') one incoming mode gets disentangled
and projected to the vacuum. This has interesting analogies to the
two-state vector formulation (TSVF) of quantum mechanics by Yakir
Aharonov et al. \citep{Aharonov2008}. In TSVF one considers the two
measurements. The outcome of the first measurement at time $t$ produces
the state $\psi_{\textrm{in}}\left(t\right)$. The outcome of the
second measurement at $T>t$ produces the state $\psi_{\textrm{out}}\left(T\right)$.
Then TSVF considers the quantum evolution inside the interval $\left[t,T\right]$
with the fixed initial and final boundary conditions $\psi_{\textrm{in}}\left(t\right)$
and $\psi_{\textrm{out}}\left(T\right)$ correspondingly, Fig. \ref{fig:time_symmetric_analogy}
This formalism is applied when analysing the experiments with the
postselection of measurement results \citep{Aharonov2014}. This formalism
also yields an interpretation of quantum mechanics: one may assume
that there is a final boundary condition for the Universe which singles
out a definite outcome for all the measurements \citep{Aharonov2017}.
However the real-time entanglement structure suggests an interesting
alternative: instead of global initial and final states, one may consider
the \textit{streams }of initial and final states, so that there is
a continuous competition between the birth and the death of the quantum
reality in real time, see Fig. \ref{fig:time_symmetric_analogy}

\section{\label{sec:MARKOVIAN-LIMIT}MARKOVIAN LIMIT}

In this section we relate our renormalization group procedure Fig.
\ref{fig:density_matrix_rg} to the Lindblad master equation which
is applied in the Markovian regime \citep{Plenio1998,Gardiner2004}.
This will help us to develop intuition about the steps of our RG procedure. 

There are many ways to perform the Markovian limit. A good way should
be based solely on the spectral decoupling mechanism represented in
sec. \ref{sec:HOW-THE-QUANTUM}. A rough way is to apply the conventional
formal arguments which lead to the Lindblad master equation. Here
we choose the latter in order to connect the RG procedure to the conventional
terms. Moreover, later it will help us to assess the additional insights
provided by the spectral decoupling mechanism. 

We choose the derivation of Markovian limit via the quantum stochastic
differential equations \citep{Plenio1998,Gardiner2004}. Essentially
this approach assumes that we are dealing with a small-coupling, near-resonant,
and broadband setup. Then the time cells can be considered as local
independent degrees of freedom, $\left[\widehat{b}\left(t\right),\widehat{b}^{\dagger}\left(t^{\prime}\right)\right]=\Gamma\delta\left(t-t^{\prime}\right)$.
In terms of the stream of discrete time moments this reads
\begin{equation}
\left[\widehat{b}\left(\tau_{p}\right),\widehat{b}^{\dagger}\left(\tau_{q}\right)\right]=dt^{-1}\Gamma\delta_{pq}.\label{eq:broadband_memory}
\end{equation}
This means that we have $m=0$ relevant modes: only incoming mode
appears in Fig. \ref{fig:density_matrix_rg}. The disentangler in
Fig. \ref{fig:density_matrix_rg}, d) is the identity operator: $\phi_{\textrm{out}}\left(p\right)=\phi_{\textrm{in}}\left(p\right)$.
The small-coupling assumption means that we can consider at most one
quantum in the incoming mode. 

The stage a) in Fig. \ref{fig:density_matrix_rg} considers the relevant
density matrix $\widehat{\rho}_{\textrm{rel}}\left(t_{p}\right)$
at a time moment $t_{p}$, which is simply 
\begin{equation}
\widehat{\rho}_{\textrm{rel}}\left(t_{p}\right)=\widehat{\rho}_{\textrm{s}}\left(t_{p}\right),
\end{equation}
where $\widehat{\rho}_{\textrm{s}}\left(t_{p}\right)$ is a reduced
density matrix of open system. 

Let us assume for a moment that $\widehat{\rho}_{\textrm{rel}}\left(t_{p}\right)$
is a pure state $\left|\phi\right\rangle _{\textrm{rel}}$. After
the entangling step, Fig. \ref{fig:density_matrix_rg} b)., this state
becomes 
\begin{multline}
\left|\phi^{\left(1\right)}\right\rangle _{\textrm{rel+i}}=\left(1-i\widehat{H}_{\textrm{s}}dt\right)\left|\phi\right\rangle _{\textrm{rel}}\otimes\left|0\right\rangle _{\varphi_{\textrm{in}}\left(p\right)}\\
-i\widehat{s}\left|\phi\right\rangle _{\textrm{rel}}\otimes\left|1\right\rangle _{\varphi_{\textrm{in}}\left(p\right)}dt\\
-\frac{1}{2}\Gamma\widehat{s}^{\dagger}\widehat{s}\left|\phi\right\rangle _{\textrm{rel}}\otimes\left|0\right\rangle _{\varphi_{\textrm{in}}\left(p\right)}dt.\label{eq:markovian_entanglement_step}
\end{multline}
Here we keep only $O\left(dt\right)$ terms in the midpoint propagation
under (\ref{eq:midpoint_Hamiltonian-1-1-1-1}). In the last line we
see the second-order virtual process because there is $dt^{-1}$ in
the memory function commutator (\ref{eq:broadband_memory}). We observe
that the first line yields the unitary free motion of open system
$-i\left[\widehat{H}_{\textrm{s}},\widehat{\rho}_{\textrm{rel}}\right]$
in the Lindblad equation, and the last line yields the standard term
$-\frac{1}{2}\Gamma\left\{ \widehat{s}^{\dagger}\widehat{s},\widehat{\rho}_{\textrm{rel}}\right\} $
which describes the decay of probability of no-emission event. 

After applying the pairing update, Fig. \ref{fig:density_matrix_rg}
c)., to the state $\left|\phi^{\left(1\right)}\right\rangle _{\textrm{rel+i}}$,
we obtain the quantum jump term from the second line of eq. (\ref{eq:markovian_entanglement_step}):
\begin{multline}
\widehat{\rho}_{\textrm{rel+i}}^{\left(2\right)}\left(t_{p}\right)=\left(\widehat{\rho}_{\textrm{rel}}-i\left[\widehat{H}_{\textrm{s}},\widehat{\rho}_{\textrm{rel}}\right]dt\right)\otimes\left|0\right\rangle _{\varphi_{\textrm{in}}\left(p\right)}\ps{}{\varphi_{\textrm{in}}\left(p\right)}{\left\langle 0\right|}\\
-\frac{1}{2}\Gamma\left\{ \widehat{s}^{\dagger}\widehat{s},\widehat{\rho}_{\textrm{rel}}\right\} dt\otimes\left|0\right\rangle _{\varphi_{\textrm{in}}\left(p\right)}\ps{}{\varphi_{\textrm{in}}\left(p\right)}{\left\langle 0\right|}\\
+\Gamma\widehat{s}\widehat{\rho}_{\textrm{rel}}\widehat{s}^{\dagger}dt\otimes\left|0\right\rangle _{\varphi_{\textrm{in}}\left(p\right)}\ps{}{\varphi_{\textrm{in}}\left(p\right)}{\left\langle 0\right|}\\
+\left\{ \textrm{terms containing either}\,\,\left|1\right\rangle _{\varphi_{\textrm{in}}\left(p\right)}\,\,\textrm{or}\,\,\ps{}{\varphi_{\textrm{in}}\left(p\right)}{\left\langle 1\right|}\right\} .\label{eq:pairing_update_leads_to_quantum_jump}
\end{multline}
Finally, the oblivion step, Fig. \ref{fig:density_matrix_rg} e).,
removes the emitted quanta and the incoming mode. We obtain the conventional
Lindblad equation:
\begin{multline}
\widehat{\rho}_{\textrm{s}}\left(t_{p+1}\right)=\widehat{\rho}_{\textrm{s}}\left(t_{p}\right)\\
-i\left[\widehat{H}_{\textrm{s}},\widehat{\rho}_{\textrm{s}}\right]dt-\frac{1}{2}\Gamma\left\{ \widehat{s}^{\dagger}\widehat{s},\widehat{\rho}_{\textrm{s}}\right\} dt+\Gamma\widehat{s}\widehat{\rho}_{\textrm{s}}\widehat{s}^{\dagger}dt.
\end{multline}

Therefore, we obtain the following ``rough'' picture of the Markovian
mode. Each infinitesimal time interval one incoming mode $\varphi_{\textrm{in}}\left(p\right)$
is coupled to the open system; a quantum is emitted into $\varphi_{\textrm{in}}\left(p\right)$
during the entangling step; the quantum jump is taken into account
via the pairing update. After that, the incoming mode (which is equal
to the outgoing mode in this case) is irreversibly decoupled and the
quantum is discarded. 

In summary, our RG procedure can be considered as an \textit{entanglement-assisted}
non-Markovian generalization of the Lindblad master equation \citep{Breuer2011}.

\section{\label{sec:STOCHASTIC-RENORMALIZATION-GROUP}STOCHASTIC UNRAVELLING:
RENORMALIZATION GROUP ALONG A QUANTUM TRAJECTORY}

Here we implement our intuition that the outgoing modes in Fig. \ref{fig:entanglement_structure}
can be replaced by a classical stochastic signal. In other words,
we explore the second alternative for the irrelevant degrees of freedom:
instead of tracing them out, we collapse them to a classical noise
as soon as they emerge. This results in a stochastic variant of renormalization
group which we call the renormalization group along a quantum trajectory.
This way we completely implement the intuitive picture of Introduction
that the ultimate fate of the emitted field is not only to decouple
but also to become observed. 

\subsection{Ensembles vs quantum trajectories}

Let us recall that in the Markovian regime the motion of open quantum
system can be described in the two alternative but equivalent ways
\citep{Plenio1998}. The first way is the ensemble description. It
is given in terms of a reduced density matrix and its evolution is
governed by the Lindblad master equation. The second way is the description
by the quantum trajectories. Here one takes into account the fact
that every open system will exercise a random motion under the influence
of the environment. The environment stores a classical record of the
history of such a motion. The quantum trajectory is the quantum evolution
which is conditioned on a particular realization of the record. This
results in a stochastic pure-state evolution which is known under
various names in literature: quantum jumps, quantum state diffusion,
stochastic wavefunction etc. Then the description in terms of the
reduced density matrix is recovered as average over the ensemle of
all the possible records (hence the name ``ensemble description'').
One also says about this in other words: that the quantum trajectories
provide a stochastic unravelling of the dissipative master equations.

The merit of quantum trajectories is two-fold. First, they yield efficient
Monte Carlo simulation methods for the dissipative dynamics. This
is because the size of density matrix scales as $N^{2}$ with the
dimension $N$ of the Hilbert space \citep{Plenio2007,Breuer2011}.
While the size of a pure state of the quantum trajectory scales as
$N$. Second, they provide the interpretation for the experiments
on observing and manipulating a single quantum system, under the Markovian
conditions \citep{Plenio1998}.

Our renormalization group for density matrices can also be considered
as an ensemble description. Then the question arises what is the corresponding
quantum trajectory description. The merit of this is again two-fold:
we obtain a). efficient Monte-Carlo simulations for pure states and
b). new insights into how the classical records are stored in the
non-Markovian environment.

\subsection{Pairing function as average over the vacuum fluctuations of environment\label{subsec:Pairing-function-as}}

\begin{figure*}
\includegraphics[scale=1.6]{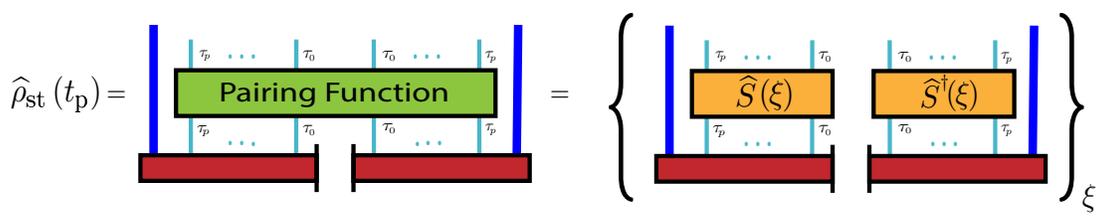}

\caption{\label{fig:statistical_interpretation_pairing}The pairing function
can be stochastically decoupled into the two separate displacement
operators $\widehat{S}\left(\xi\right)$ and $\widehat{S}^{\dagger}\left(\xi\right)$
which act on the ket and the bra states correspondingly. The averaging
is over the displacement $\xi$ which is a classical noise of vacuum
fluctuations at the coupling site $\widehat{b}^{\dagger}\left(t\right)$. }

\end{figure*}

In the Markovian regime the ``sandwitch'' term $\Gamma\widehat{s}\widehat{\rho}_{\textrm{s}}\widehat{s}^{\dagger}$
arises as average over the ensemble of quantum-jump histories \citep{Plenio1998,Breuer2011,Gardiner2004}.
In our renormalization group the pairing function plays the role of
the non-Markovian ``sandwitch'' term, see eq. (\ref{eq:sandwitch_form_of_pairing_update})
and (\ref{eq:pairing_update_leads_to_quantum_jump}). Therefore, we
want to represent it as an ensemble of some observable events. Here
we show that as such events one can choose the vacuum fluctuations
of the classical field in the environment, see Fig. \ref{fig:statistical_interpretation_pairing}.

The classical field $z$ is carried by the coherent state
\begin{equation}
\left|z\right\rangle _{\textrm{b}}=\mathcal{Q_{\textrm{vac}}}\left(z\right)^{\frac{1}{2}}\exp\left(\int d\omega z^{*}\left(\omega\right)\widehat{a}^{\dagger}\left(\omega\right)\right)\left|0\right\rangle _{\textrm{b}},
\end{equation}
where the normalization factor $\mathcal{Q_{\textrm{vac}}}\left(z\right)$
is the probability to observe a vacuum fluctuation z of the field,
\begin{equation}
\mathcal{Q_{\textrm{vac}}}\left(z\right)=\left|\ps{}{\textrm{b}}{\left\langle z\left|0\right.\right\rangle _{\textrm{b}}}\right|^{2}\propto\exp\left(-\int d\omega\left|z\left(\omega\right)\right|^{2}\right).
\end{equation}

At the coupling site $\widehat{b}^{\dagger}\left(t\right)$ the vaccum
fluctuation $z$ appears as a colored (non-Markovian) noise $\xi^{*}\left(t\right)$:
\begin{equation}
\xi^{*}\left(t\right)=\int d\omega c^{*}\left(\omega\right)z^{*}\left(\omega\right)e^{i\omega t}.
\end{equation}
This noise has the Gaussian statistics
\begin{equation}
\overline{\xi_{r}}=\overline{\xi_{r}^{*}}=0,\,\,\overline{\xi_{r}\xi_{s}^{*}}=M_{rs},
\end{equation}
where we consider the noise $\xi$ at the midpoint times: $\xi_{r}=\xi\left(\tau_{r}\right)$.
Each realization of $z$ leads to a single realization of the entire
time-trajectory of $\xi^{*}\left(t\right)$. 

\begin{figure*}
\includegraphics[scale=1.7]{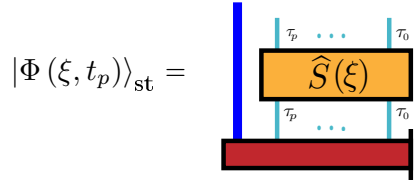}

\caption{\label{fig:displaced_pure_state}The density matrix $\widehat{\rho}_{\textrm{st}}\left(t_{p}\right)$
can be represented as the ensemble of displaced pure states $\left|\Phi\left(\xi,t_{p}\right)\right\rangle _{\textrm{st}}$.
This state is displaced by $\widehat{S}\left(\xi\right)$. The result
is that the emitted quanta are ``centered'' on a classical signal
$\xi$, see eqs. (\ref{eq:displacement_operator}), (\ref{eq:displacement_of_annihilation}),
(\ref{eq:displacement_of_creation})). }

\end{figure*}

It turns out that the pairing function can be represented as an average
over these vacuum fluctuations, Fig. . \ref{fig:statistical_interpretation_pairing}:
\begin{equation}
\left\{ e^{\sum_{rs=0}^{\infty}\widehat{\psi}_{r}^{\dagger}M_{rs}\widehat{\psi}_{s}}\right\} _{A}=\overline{\left[\left\{ e^{\sum_{r=0}^{\infty}\xi_{r}\widehat{\psi}_{r}^{\dagger}}e^{\sum_{s=0}^{\infty}\xi_{s}^{*}\widehat{\psi}_{s}}\right\} _{A}\right]}_{\xi},\label{eq:stochastic_unraveling_of_pairing_function}
\end{equation}
where $\left\{ \cdot\right\} _{A}$ is the antinormal averaging. Observe
that in the pairing function we extend the summation range from $\left[0,p-1\right]$
(as in eq. (\ref{eq:density_matrix_in_time_domain})) to $\left[0,\infty\right)$.
This is possible since the incoming modes $\phi_{\textrm{in}}\left(p\right),\phi_{\textrm{in}}\left(p+1\right),\ldots$
are empty at a time moment $t_{p}$. Then the density matrix eq. (\ref{eq:density_matrix_in_time_domain})
is represented as average over the ensemble of pure states, Fig. \ref{fig:displaced_pure_state},
\begin{equation}
\widehat{\rho}_{\textrm{st}}\left(t_{p}\right)=\overline{\left[\left|\Phi\left(\xi,t_{p}\right)\right\rangle _{\textrm{st}}\ps{}{\textrm{st}}{\left\langle \Phi\left(\xi,t_{p}\right)\right|}\right]}_{\xi}.
\end{equation}
Here the joint pure state of the time cells and of the open system
is 
\begin{equation}
\left|\Phi\left(\xi,t_{p}\right)\right\rangle _{\textrm{st}}=e^{\sum_{s=0}^{\infty}\xi_{s}^{*}\widehat{\psi}_{s}}\left|\Phi\left(t_{p}\right)\right\rangle _{\textrm{st}}.\label{eq:dressed_state-1}
\end{equation}
Observe the interpretation of the state $\left|\Phi\left(\xi,t_{p}\right)\right\rangle _{\textrm{st}}$.
The action of operator 
\begin{equation}
\widehat{S}\left(\xi\right)=e^{\sum_{s=0}^{\infty}\xi_{s}^{*}\widehat{\psi}_{s}}\label{eq:displacement_operator}
\end{equation}
induces a non-unitary Bogoliubov transform of the creation/annihilation
operators in $\left|\Phi\left(t_{p}\right)\right\rangle _{\textrm{st}}$:
\begin{equation}
\widehat{S}\left(\xi\right)\widehat{\psi}_{r}\widehat{S}^{-1}\left(\xi\right)=\widehat{\psi}_{r},\label{eq:displacement_of_annihilation}
\end{equation}
\begin{equation}
\widehat{S}\left(\xi\right)\widehat{\psi}_{r}^{\dagger}\widehat{S}^{-1}\left(\xi\right)=\widehat{\psi}_{r}^{\dagger}+\xi_{r}^{*}.\label{eq:displacement_of_creation}
\end{equation}
In other words, the operation $\widehat{S}\left(\xi\right)$ introduces
a classical background $\xi$ in the state $\left|\Phi\left(t_{p}\right)\right\rangle _{\textrm{st}}$,
see Fig. \ref{fig:displaced_pure_state}. As a result, each emitted
quantum $\widehat{\psi}_{r}^{\dagger}$ in $\left|\Phi\left(\xi,t_{p}\right)\right\rangle _{\textrm{st}}$
becomes a superposition of the classical background $\xi_{r}^{*}$
and of a purely quantum part:
\begin{equation}
\widehat{\psi}_{r}^{\dagger}\to\xi_{r}^{*}+\widehat{\psi}_{r}^{\dagger}.
\end{equation}
We call $\left|\Phi\left(\xi,t_{p}\right)\right\rangle _{\textrm{st}}$
the quantum state over a classical background $\xi$. When we perform
the partial trace (\ref{eq:tracing_out_the_bath}), 
\begin{equation}
\widehat{\rho}_{\textrm{s}}\left(t_{p}\right)=\overline{\left[\ps{}{\textrm{t}}{\left\langle 0\right.}\left|\Phi\left(\xi,t_{p}\right)\right\rangle _{\textrm{st}}\ps{}{\textrm{st}}{\left\langle \Phi\left(\xi,t_{p}\right)\right|\left.0\right\rangle _{\textrm{t}}}\right]}_{\xi},\label{eq:partial_trace_as_stochastic_average}
\end{equation}
the purely quantum part gets discarded due to projection to the vacuum.
In other words, all the quanta \textit{collapse} to the classical
background,
\begin{equation}
\widehat{\psi}_{r}^{\dagger}\to\xi_{r}^{*},
\end{equation}
and then we average over all the possible vacuum fluctuations.

\subsection{Renormalization group along a quantum trajectory}

Now we derive the evolution equations in which the irrelevant degrees
are collapsed to a classical noise as soon as they emerge. 

Below we follow the convention of the growing Fock space of sec. \ref{subsec:The-convention-of}.
Then the wavefunction $\left|\Phi\left(t_{p}\right)\right\rangle _{\textrm{st}}$
belongs to the space $\mathcal{H}_{\textrm{s}}\otimes\mathcal{F}_{\textrm{t}}\left(p\right)$,
with the Fock space $\mathcal{F}_{\textrm{t}}\left(p\right)$ being
spanned by the time cells $\psi_{p-1}^{\dagger}\ldots\psi_{0}^{\dagger}$. 

\begin{figure*}
\includegraphics[scale=2.5]{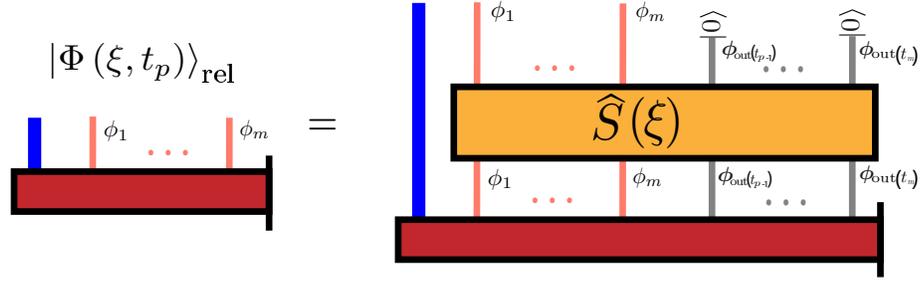}

\caption{\label{fig:relevant_wavefunction}The relevant wavefunction $\left|\Phi\left(\xi,t_{p}\right)\right\rangle _{\textrm{rel}}$
is defined as a joint pure state of open system and relevant modes.
All the outgoing modes $\phi_{\textrm{out}}\left(m\right),\ldots,\phi_{\textrm{out}}\left(p-1\right)$
which have emerged by the time moment $t_{p}$ are projected to vacuum
(their quantum content is discarded).}

\end{figure*}

Let us introduce the relevant wavefunction 
\begin{equation}
\left|\Phi\left(\xi,t_{p}\right)\right\rangle _{\textrm{rel}}=\ps{}{\textrm{out}\left(p\right)}{\left\langle 0\right|}e^{\sum_{s=0}^{p-1}\xi_{s}^{*}\widehat{\psi}_{s}}\left|\Phi\left(t_{p}\right)\right\rangle _{\textrm{st}},\label{eq:dressed_state}
\end{equation}
which means that we take the quantum state $\left|\Phi\left(\xi,t_{p}\right)\right\rangle _{\textrm{st}}$
over the classical background $\xi$ and collapse all the outgoing
modes $\phi_{\textrm{out}}\left(m\right),\ldots,\phi_{\textrm{out}}\left(p-1\right)$
which have emerged by the time moment $t_{p}$, Fig. \ref{fig:relevant_wavefunction}.
As a result, the relevant wavefunction belongs to the space $\mathcal{H}_{\textrm{s}}\otimes\mathcal{F}_{\textrm{rel}}$.

Given an environment with a spectral density $J\left(\omega\right)$,
one computes its memory funNction $M\left(t-t^{\prime}\right)$. Then
the incoming/outgoing/relevant modes, and disentanglers $W_{p}$ are
found according to the algorithm of sec. \ref{subsec:The-emergence-of}.
This needs to be done once for a given $J\left(\omega\right)$.

Suppose we have a random noise sample $\xi$. The initial condition
for evolution is $\left|\Phi\left(\xi,0\right)\right\rangle _{\textrm{rel}}=\left|\phi_{0}\right\rangle _{\textrm{s}}$,
since there are no relevant modes at $t_{0}=0$. 

\begin{figure*}
\includegraphics[scale=2.5]{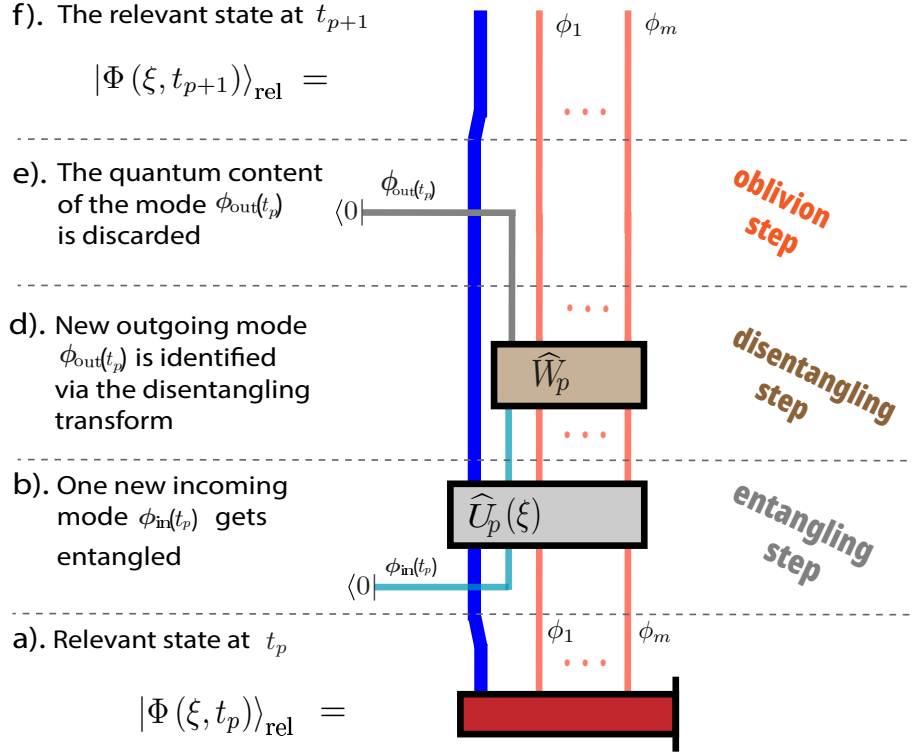}

\caption{\label{fig:rg_qtrajectory}The renormalization group along a quantum
trajectory. For each realization (trajectory) of the classical noise
$\xi$ of vacuum fluctuations of the environment, the relevant wavefunction
$\left|\Phi\left(\xi,t_{p}\right)\right\rangle _{\textrm{rel}}$ is
propagated in time. The full quantum dynamics is obtained as average
over $\xi$. The realizations of $\xi$ can be generated via Monte
Carlo sampling. On a conceptual level, this structure shows the lifecycle
of the quantum complexity during the real-time evolution. First the
complexity is generated at the step b). However eventually the quantum
complexity is destroyed in the step e). This is because the decoupled
entanglement is equivalent to the stochastic ensemble of classical
records. }
\end{figure*}

The steps of how $\left|\Phi\left(\xi,t_{p}\right)\right\rangle _{\textrm{rel}}$
is propagated in time closely mirror those of RG for density matrices
of sec. \ref{subsec:Propagation-of-density}, see Fig. \ref{fig:rg_qtrajectory}.
The first step, \textit{the entangling step}, begins by enlarging
the Hilbert space $\mathcal{H}_{\textrm{s}}\otimes\mathcal{F}_{\textrm{rel}}$
with the states of the incoming mode via the embedding $\left|\Phi\left(\xi,t_{p}\right)\right\rangle _{\textrm{rel}}\to\left|\Phi\left(\xi,t_{p}\right)\right\rangle _{\textrm{rel+i}}=\left|\Phi\left(\xi,t_{p}\right)\right\rangle _{\textrm{rel}}\otimes\left|0\right\rangle _{\varphi_{\textrm{in}}\left(p\right)}$.
It is followed by the Hamiltonian propagation: 
\begin{multline}
\left|\Phi^{\left(1\right)}\left(\xi,t_{p+1}\right)\right\rangle _{\textrm{rel+i}}=\left|\Phi\left(\xi,t_{p}\right)\right\rangle _{\textrm{rel+i}}\\
-idt\widehat{H}_{\textrm{st}}\left(\xi,\tau_{p}\right)\frac{1}{2}\left\{ \left|\Phi^{\left(1\right)}\left(\xi,t_{p+1}\right)\right\rangle _{\textrm{rel+i}}+\left|\Phi\left(\xi,t_{p}\right)\right\rangle _{\textrm{rel}}\right\} ,\label{eq:rg_pure_hamiltonian}
\end{multline}
where the Hamiltonian
\begin{multline}
\widehat{H}_{\textrm{st}}\left(\xi,\tau_{p}\right)=\widehat{S}\left(\xi\right)\widehat{H}_{\textrm{st}}\left(\tau_{p}\right)\widehat{S}^{-1}\left(\xi\right)\\
=\widehat{H}_{\textrm{s}}+\widehat{s}\left\{ \widehat{\psi}_{p}^{\dagger}+\xi_{p}^{*}\right\} +\widehat{s}^{\dagger}M\left(0\right)\widehat{\psi}_{p}\\
+\widehat{s}^{\dagger}\sum_{i=1}^{\textrm{min}\left(p,m\right)}M_{i}\left(p\right)\widehat{\phi}_{i}\label{eq:dressed_Hamiltonian}
\end{multline}
takes into account the classical noise of vacuum fluctuations. Here
the subscript 'rel+i' means that the Hilbert space is enlarged by
additional incoming mode $\phi_{\textrm{in}}\left(p\right)$. 

The second step, \textit{the disentangling step }identifies the new
outgoing mode $\phi_{\textrm{out}}\left(p\right)$ which has already
decoupled: 
\begin{equation}
\left|\Phi^{\left(2\right)}\left(\xi,t_{p+1}\right)\right\rangle _{\textrm{rel+o}}=\widehat{W}_{p}\left|\Phi^{\left(1\right)}\left(\xi,t_{p+1}\right)\right\rangle _{\textrm{rel+i}}.
\end{equation}
Please note that since we apply the active change of frame $\widehat{W}_{p}$,
the quantum content of the outgoing mode $\phi_{\textrm{out}}\left(p\right)$
just replaces the content of $\phi_{\textrm{in}}\left(p\right)$,
so that the Hilbert space 'rel+o' is actually equal to 'rel+i'.

Finally, in\textit{ the oblivion step} the quantum content of the
newly formed outgoing mode is discarded (it collapses to the classical
noise ): 
\begin{equation}
\left|\Phi\left(\xi,t_{p+1}\right)\right\rangle _{\textrm{rel}}=\ps{}{\phi_{\textrm{out}}\left(t_{p}\right)}{\left\langle 0\right.}\left|\Phi^{\left(2\right)}\left(\xi,t_{p+1}\right)\right\rangle _{\textrm{rel+o}}.\label{eq:wavefunction_disentangler}
\end{equation}
Observe that as a result we return to the Hilbert space $\mathcal{F}_{\textrm{rel}}\otimes\mathcal{H}_{\textrm{s}}$.
Then the propagation procedure is repeated for the next time moment
$t_{p+1}$.

\subsection{Computing the observables}

The observables $\widehat{o}$ for the open system are computed as
\begin{multline}
\left\langle \widehat{o}\left(t_{p}\right)\right\rangle \equiv\textrm{Tr}\left\{ \widehat{o}\widehat{\rho}_{\textrm{s}}\left(t_{p}\right)\right\} \\
=\overline{\left[\ps{}{\textrm{rel}}{\left\langle \Phi\left(\xi,t_{p}\right)\right|}\left.0\right\rangle _{\textrm{rel}}\widehat{o}\,\ps{}{\textrm{rel}}{\left\langle 0\right.}\left|\Phi\left(\xi,t_{p}\right)\right\rangle _{\textrm{rel}}\right]}_{\xi}Z^{-1}\left(t_{p}\right),\label{eq:nominator}
\end{multline}
where $\left|0\right\rangle _{\textrm{rel}}$ is the joint vacuum
for the relevant modes $\widehat{\phi}_{1}^{\dagger},\ldots,\widehat{\phi}_{m}^{\dagger}$;
the state $\,\ps{}{\textrm{rel}}{\left\langle 0\right.}\left|\Phi\left(\xi,t_{p}\right)\right\rangle _{\textrm{rel}}$
belongs to $\mathcal{H}_{\textrm{s}}$; the normalization
\begin{equation}
Z\left(t_{p}\right)=\overline{\left\Vert \ps{}{\textrm{rel}}{\left\langle 0\right.}\left|\Phi\left(\xi,t_{p}\right)\right\rangle _{\textrm{rel}}\right\Vert ^{2}}_{\xi}\label{eq:denominator}
\end{equation}
is equal to $1$ in the exact simuation, but should be included in
the approximate computation to keep the normalization of the probability.
The observables for the environment can also be computed via the mappings
(\ref{eq:mapping_for_creation}), (\ref{eq:mapping_for_annihilation}).

\subsection{\label{subsec:Monte-Carlo-sampling-of}Monte-Carlo sampling of vacuum
fluctuations}

The realizations of the noise $\xi$ can be sampled stochastically
by 
\begin{equation}
\xi_{r}=\sqrt{\Delta\omega}\sum_{k}c\left(\omega_{k}\right)e^{-i\omega_{k}\tau_{r}}z_{k},\label{eq:discretized_noise}
\end{equation}
where we have introduced a discretization of the frequency axis $\omega_{k}$.
The coefficients $c\left(\omega_{k}\right)=\sqrt{J\left(\omega_{k}\right)/\pi}$,
which follows from (\ref{eq:coupling_site}) and (\ref{eq:temporal_commutation_relation}).
The complex random numbers $z_{k}$ have the statistics 
\begin{equation}
\overline{z_{k}}=\overline{z_{k}^{*}}=0,\,\,\,\overline{z_{k}z_{l}^{*}}=\delta_{kl}.\label{eq:pseudorandom_number}
\end{equation}
We generate a sample of $M$ noise instances $\xi^{1}\ldots\xi^{M}$,
and the averages (\ref{eq:nominator}), (\ref{eq:denominator}) are
computed as the sample means. 

The states $\left|\Phi\left(\xi,t_{p+1}\right)\right\rangle _{\textrm{rel}}$
are represented in the Hilbert space $\mathcal{H}_{\textrm{s}}\otimes\mathcal{F}_{\textrm{rel}}$
which is truncated in the maximal total occupation of the relevant
modes. Such a truncation is expected to converge uniformly on wide
time scales if we make the reasonable assumption that there is a balance
between the flux of quanta $j_{\textrm{in}}\left(t_{p}\right)$ being
emitted by the open system, and the flux of quanta $j_{\textrm{out}}\left(t_{p}\right)$
which are eventually discarded in the outgoing modes, see sec. \ref{subsec:Balance-of-complexity}.

\subsection{Importance sampling}

Observe that the Hamiltonian $\widehat{H}_{\textrm{st}}\left(\xi,\tau_{p}\right)$
is not Hermitean and does not conserve the norm. As a consequence,
it may turn out that the ensemble of noise samples (\ref{eq:discretized_noise})-(\ref{eq:pseudorandom_number})
becomes non-representative: the rarest noise realizations recieve
the highest weights. This can be fixed by applying an importance sampling
technique. Any such technique will require to shift the noise sample,
$\xi\to\xi+\Delta\xi$. In this work we employ the continuous shift
of the noise with time. We refer the interested reader to the appendix
\ref{sec:IMPORTANCE-SAMPLING} for the derivation. Here we present
the resulting simulation procedure.

Given an environment with a spectral density $J\left(\omega\right)$,
one computes its memory function $M\left(t-t^{\prime}\right)$. Then
the incoming/outgoing/relevant modes, and disentanglers $W_{p}$ are
found according to the algorithm of sec. \ref{subsec:The-emergence-of}.
This needs to be done once for a given $J\left(\omega\right)$.

The initial condition for the simulation is $\left|\Phi\left(\xi,0\right)\right\rangle _{\textrm{rel}}=\left|\phi_{0}\right\rangle _{\textrm{s}}$.
A noise sample $\xi$ is generated as described in sec. \ref{subsec:Monte-Carlo-sampling-of}.
Then the relevant wavefunction $\left|\Phi\left(\xi,t_{p}\right)\right\rangle _{\textrm{rel}}$
is propagated for the noise sample $\xi$ starting from the initial
condition $\left|\Phi\left(\xi,0\right)\right\rangle _{\textrm{rel}}$. 

For a given noise sample $\xi$, the open system observable $\widehat{o}$
is averaged at a time $t_{p}$ as

\begin{equation}
\overline{o}\left(t_{p};\xi\right)=\frac{\textrm{Tr}\left\{ \widehat{o}\times\ps{}{\textrm{rel}}{\left\langle 0\right.}\left|\Phi\left(\xi,t_{p}\right)\right\rangle _{\textrm{rel}}\ps{}{\textrm{rel}}{\left\langle \Phi\left(\xi,t_{p}\right)\right|\left.0\right\rangle _{\textrm{rel}}}\right\} }{\left\Vert \ps{}{\textrm{rel}}{\left\langle 0\right.}\left|\Phi\left(\xi,t_{p}\right)\right\rangle _{\textrm{rel}}\right\Vert ^{2}}.\label{eq:conditional_average}
\end{equation}
Then the full quantum average of $\widehat{o}$ is given by averaging
over $\xi$:
\begin{equation}
\left\langle \widehat{o}\left(t\right)\right\rangle \equiv\textrm{Tr}\left\{ \widehat{o}\widehat{\rho}_{\textrm{s}}\left(t\right)\right\} =\overline{\left[\overline{o}\left(t_{p};\xi\right)\right]}_{\xi},
\end{equation}
which is implemented by generating a sample of $M$ noise instances
$\xi^{1}\ldots\xi^{M}$, and computing the sample mean of $\overline{o}\left(t_{p};\xi^{k}\right),k=1\ldots M$. 

The propagation of $\left|\Phi\left(\xi,t_{p}\right)\right\rangle _{\textrm{rel}}$
is done as follows. Suppose we know $\left|\Phi\left(\xi,t_{p}\right)\right\rangle _{\textrm{rel}}$
at a time step $t_{p}$. The propagation to the next time moment begins
with \textit{the entangling step}: the embedding $\left|\Phi\left(\xi,t_{p}\right)\right\rangle _{\textrm{rel}}\to\left|\Phi\left(\xi,t_{p}\right)\right\rangle _{\textrm{rel+i}}=\left|\Phi\left(\xi,t_{p}\right)\right\rangle _{\textrm{rel}}\otimes\left|0\right\rangle _{\varphi_{\textrm{in}}\left(p\right)}$
and the Hamiltonian evolution 
\begin{multline}
\left|\Phi^{\left(1\right)}\left(\xi,t_{p+1}\right)\right\rangle _{\textrm{rel+i}}=\left|\Phi\left(\xi,t_{p}\right)\right\rangle _{\textrm{rel+i}}\\
-idt\widehat{H}_{\textrm{st}}^{\prime}\left(\xi,\tau_{p}\right)\frac{1}{2}\left\{ \left|\Phi^{\left(1\right)}\left(\xi,t_{p+1}\right)\right\rangle _{\textrm{rel+i}}+\left|\Phi\left(\xi,t_{p}\right)\right\rangle _{\textrm{rel+i}}\right\} .\label{eq:rg_pure_hamiltonian-1}
\end{multline}
The Hamiltonian $\widehat{H}_{\textrm{st}}^{\prime}\left(\xi,\tau_{p}\right)$
takes into account the continuous noise shift $f_{p}$,
\begin{multline}
\widehat{H}_{\textrm{st}}^{\prime}\left(\xi,\tau_{p}\right)=\widehat{H}_{\textrm{s}}+\widehat{s}\left\{ \widehat{\psi}_{p}^{\dagger}+\xi_{p}^{*}+f_{p}^{*}\right\} \\
+\left(\widehat{s}^{\dagger}-\overline{s}^{*}\left(\tau_{p};\xi\right)\right)\left\{ M\left(0\right)\widehat{\psi}_{p}+\sum_{i=1}^{\textrm{min}\left(p,m\right)}M_{i}\left(p\right)\widehat{\phi}_{i}\right\} .\label{eq:shifting_noise_hamiltonian-1}
\end{multline}
The shift $f_{p}$ is computed as a convolution of averages $\overline{s}\left(\tau_{p};\xi\right)$
of $\widehat{s}$ at all the previous midpoint times $\tau_{p}$ with
the memory function: 
\begin{multline}
f_{p}=-i\frac{dt}{2}M\left(\frac{dt}{2}\right)\overline{s}\left(\tau_{p};\xi\right)\\
-idt\sum_{l=0}^{p-1}M\left(\left(p-l\right)dt\right)\overline{s}\left(\tau_{l};\xi\right).\label{eq:midpoint_shift}
\end{multline}
Here one can approximate $\overline{s}\left(\tau_{l};\xi\right)\approx\left\{ \overline{s}\left(t_{l+1};\xi\right)+\overline{s}\left(t_{l};\xi\right)\right\} /2$,
and $\overline{s}\left(t_{l};\xi\right)$ is the average of $\widehat{s}$
according to eq. (\ref{eq:conditional_average}). Observe that when
solving (\ref{eq:rg_pure_hamiltonian-1}) by iteration, the shift
$f_{p}$ should also be recomputed for each iteration since it contains
the midpoint term $\overline{s}\left(\tau_{p};\xi\right)$ which changes
from iteration to iteration.

The second step, \textit{the disentangling step }identifies the new
outgoing mode $\phi_{\textrm{out}}\left(p\right)$ which has already
decoupled: 
\begin{equation}
\left|\Phi^{\left(2\right)}\left(\xi,t_{p+1}\right)\right\rangle _{\textrm{rel+o}}=\widehat{W}_{p}\left|\Phi^{\left(1\right)}\left(\xi,t_{p+1}\right)\right\rangle _{\textrm{rel+i}}.
\end{equation}

Finally, in\textit{ the oblivion step} the quantum content of the
newly formed outgoing mode is discarded (it collapses to the classical
noise ): 
\begin{equation}
\left|\Phi\left(\xi,t_{p+1}\right)\right\rangle _{\textrm{rel}}=\ps{}{\phi_{\textrm{out}}\left(t_{p}\right)}{\left\langle 0\right.}\left|\Phi^{\left(2\right)}\left(\xi,t_{p+1}\right)\right\rangle _{\textrm{rel+o}}.\label{eq:pure_state_oblivion_step-1}
\end{equation}

Then the propagation procedure is repeated for the next time moment
$t_{p+1}$.

\subsection{Example calculations}

\subsubsection{Driven qubit}

\begin{figure}
\includegraphics[scale=0.5]{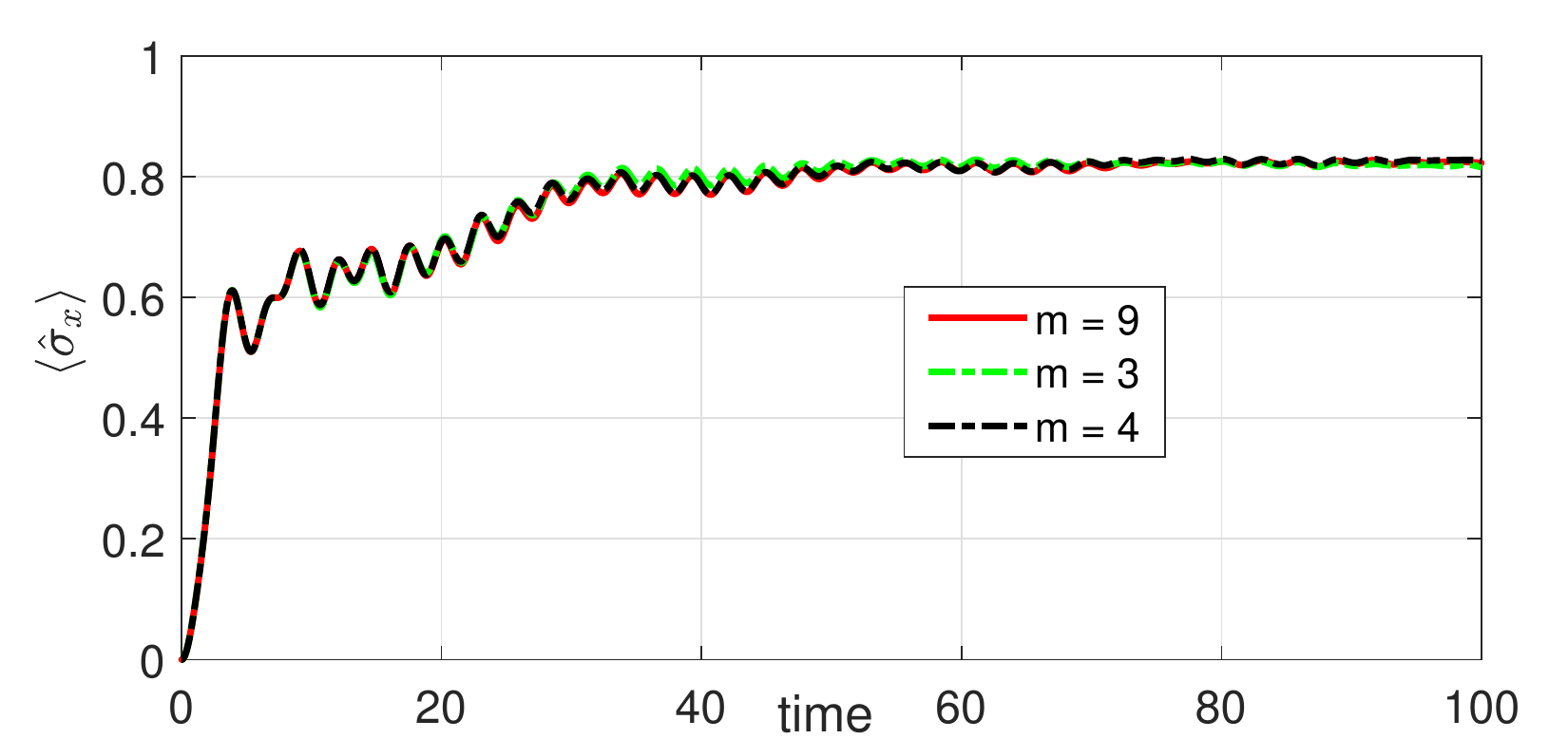}

\caption{\label{fig:For-the-system-1}For the system of a driven qubit with
a non-RWA coupling to the subohmic environment, good almost quantitative
result is provided by $m=3$ relevant modes. The convergence is achieved
for $m=4$ relevant modes. The computation is by RG along a quantum
trajectory.}

\end{figure}

\begin{figure}
\includegraphics[scale=0.52]{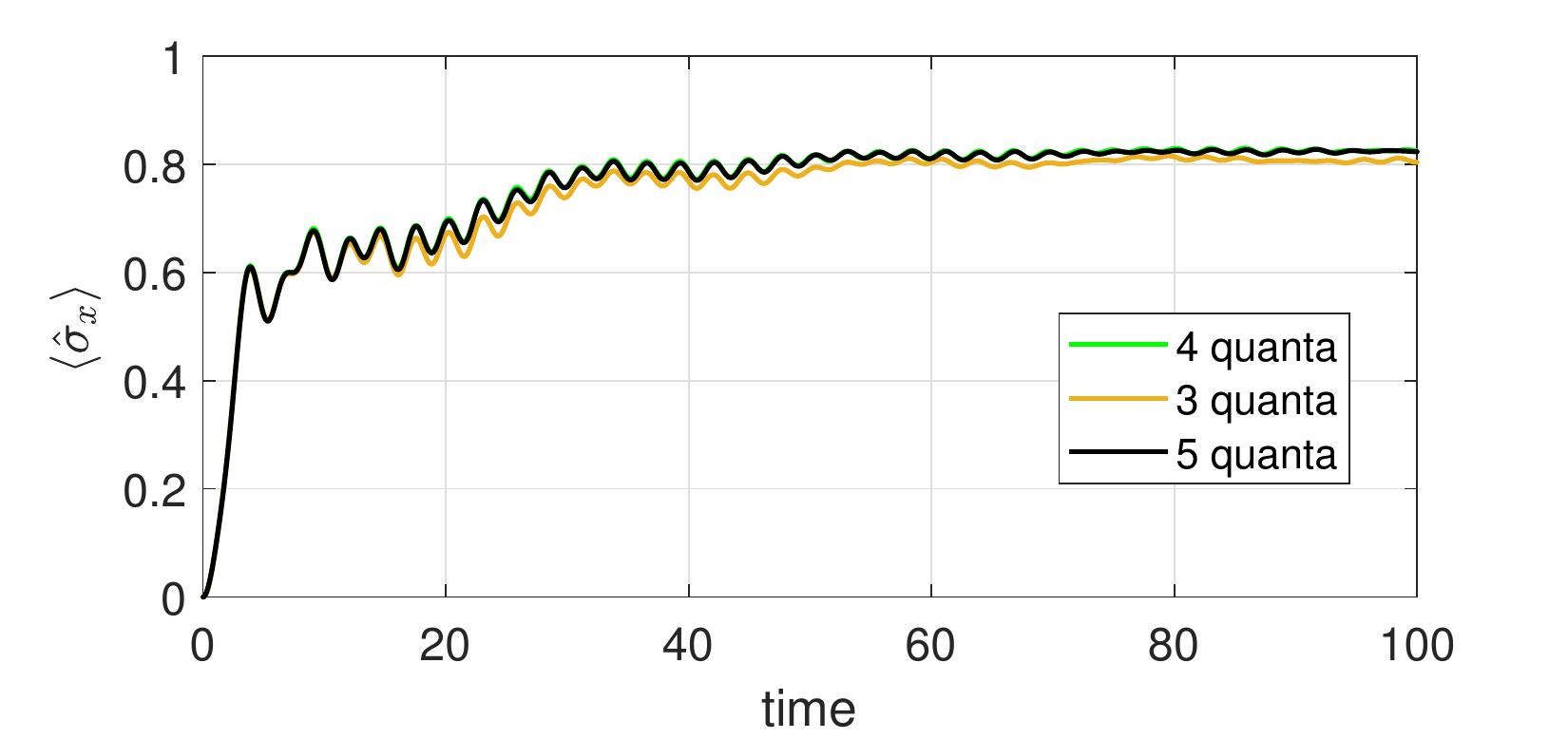}

\caption{\label{fig:RG_vs_quanta}The system of a driven qubit with a non-RWA
coupling to the subohmic environment. Uniform convergence of the RG
along a quantum trajectory with respect to the maximal number of quanta
in the relevant modes. }
\end{figure}

\begin{figure}
\includegraphics[scale=0.55]{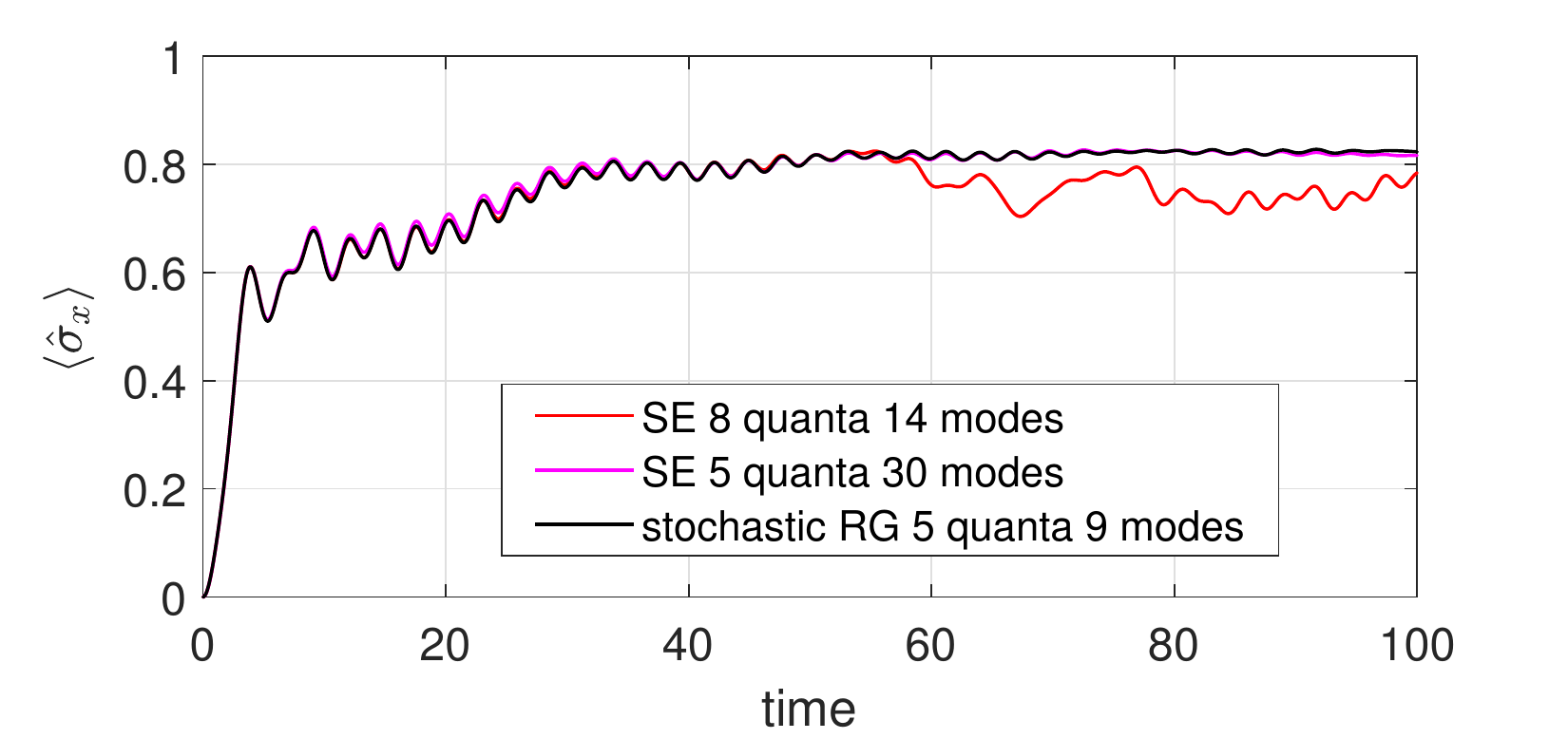}

\caption{\label{fig:RG_vs_ED}The system of a driven qubit with a non-RWA coupling
to the subohmic environment. Comparison of the RG along a quantum
trajectory with the solution of Schrodinger equation (SE). In RG the
number of relevant modes is $m=9$, the maximal occupation is truncated
at $n=5$ quanta. SE was computed for the two set of parameters: to
validate the large-time asymptotics, the Fock space was truncated
at $n=5$ quanta and $m=30$ sites of the semiinfinite chain; in order
to validate the mid-time solution, $n=8$ and $m=14$ was used. }
\end{figure}

Here we provide an example calculation for the model of the driven
open quantum system
\begin{equation}
\widehat{H}_{\textrm{sb}}\left(t\right)=-\frac{\Delta}{2}\widehat{\sigma}_{x}+\widehat{\sigma}_{z}f\cos\omega t+\widehat{\sigma}_{z}\left(\widehat{b}\left(t\right)+\widehat{b}^{\dagger}\left(t\right)\right),\label{eq:subohic_driven}
\end{equation}
with the Heaviside-regularized spetral density $J\left(\omega\right)=2\pi\alpha\omega_{\textrm{c}}^{1-s}\omega^{s}\theta\left(\omega-\omega_{\textrm{c}}\right)$
for $\alpha=0.1$, $s=0.5$, $\Delta=1$, $\omega_{\textrm{c}}=1$,
$f=0.1$, $\omega=1$. This leads to the memory function $M\left(\tau\right)=\frac{2\alpha\omega_{\textrm{c}}^{2}}{s+1}\times\exp\left(-i\tau\omega_{\textrm{c}}\right)\times\ps{}1F_{1}\left(1,s+2,i\tau\omega_{\textrm{c}}\right)$.
This situation corresponds to the non-RWA resonant driving and a highly
non-Markovian behaviour due to the subohmic spectrum at the origin
and because the transition of the open system is at the sharp edge
of the band. The is a good benchmark problem since we expect that
the resulting dynamics is strongly determined by long-range memory
effects. In Fig. \ref{fig:For-the-system-1} we demonstrate that only
$m=3$ relevant modes are required to yield almost quantitative agreement.
The complete convergence is achieved for $m=4$. In Fig. \ref{fig:RG_vs_quanta}
we demontrate the uniform convergence of RG with respect to the maximal
occupation of the relevant modes, for the same system. Again, 3 quanta
yield almost quantitative agreement, and the full convergence is provided
by $5$ quanta. In order to test our approach, we also solve the Schrodinger
equation for this model represented in the semiinfinite chain form
\citep{Chin2010}. In Fig. \ref{fig:RG_vs_ED} we compare the stochastic
RG vs the numerical solution of Schrodinger equation (SE) for $\alpha=0.1$,
$s=0.5$, $\Delta=1$, $\omega_{\textrm{c}}=1$, $f=0.1$, $\omega=1$.
It is seen that the RG simulation quantitatively reproduces the time-dependence
of observable up to steady state. 

\subsubsection{Information backflow}

When the dynamics of the open quantum system is non-Markovian, it
is accompanied by such a phenomenon as information backflow \citep{Breuer2009,Bylicka2017,Buscemi2016}.
This phenomenon refers to the special time behaviour of the distinguishability
of states of open quantum system. The distiguishability of two quantum
states $\widehat{\rho}_{1}\left(t\right)$ and $\widehat{\rho}_{2}\left(t\right)$
(reduced density matrices) of open quantum system is measured by the
trace distance \citep{Breuer2009}
\begin{equation}
T\left(\widehat{\rho}_{1},\widehat{\rho}_{2};t\right)=\frac{1}{2}\textrm{Tr}\sqrt{\left(\widehat{\rho}_{1}\left(t\right)-\widehat{\rho}_{2}\left(t\right)\right)^{2}}.\label{eq:trace_distance}
\end{equation}
In the Markovian regime $T\left(\widehat{\rho}_{1},\widehat{\rho}_{2};t\right)$
should monotonically nonincrease with time for any pair of states
\citep{Ruskai1994}. On the contrary, in the non-Markovian mode, we
can identify at least one pair of states for which $T\left(\widehat{\rho}_{1},\widehat{\rho}_{2};t\right)$
is increasing for some time interval. This is the information backflow,
and is one of the signatures of non-Markovian behaviour. 

The concept of relevant modes presented in this paper proposes a physical
mechanism behind this phenomemon. As shown in Fig. \ref{fig:complexity_balance_interpretation},
the relevant modes play the role of a finite imperfect cavity. Then
the interaction between the open quantum system and these modes leads
to the vacuum Rabi oscillations, i. e. oscillating exchange of energy
between the open system and the cavity. These damped oscillations
lead to the non-monotonous behaviour of the measures of information
backflow. Roughly speaking, when the open quantum system losses its
energy, the available phase space shrinks, the states get closer:
become less distinguishable. However when the energy flow is reversed,
the available phase space of open quantum system expands, and the
states move away from each other (become more distinguishable).

\begin{figure}
\includegraphics[scale=0.45]{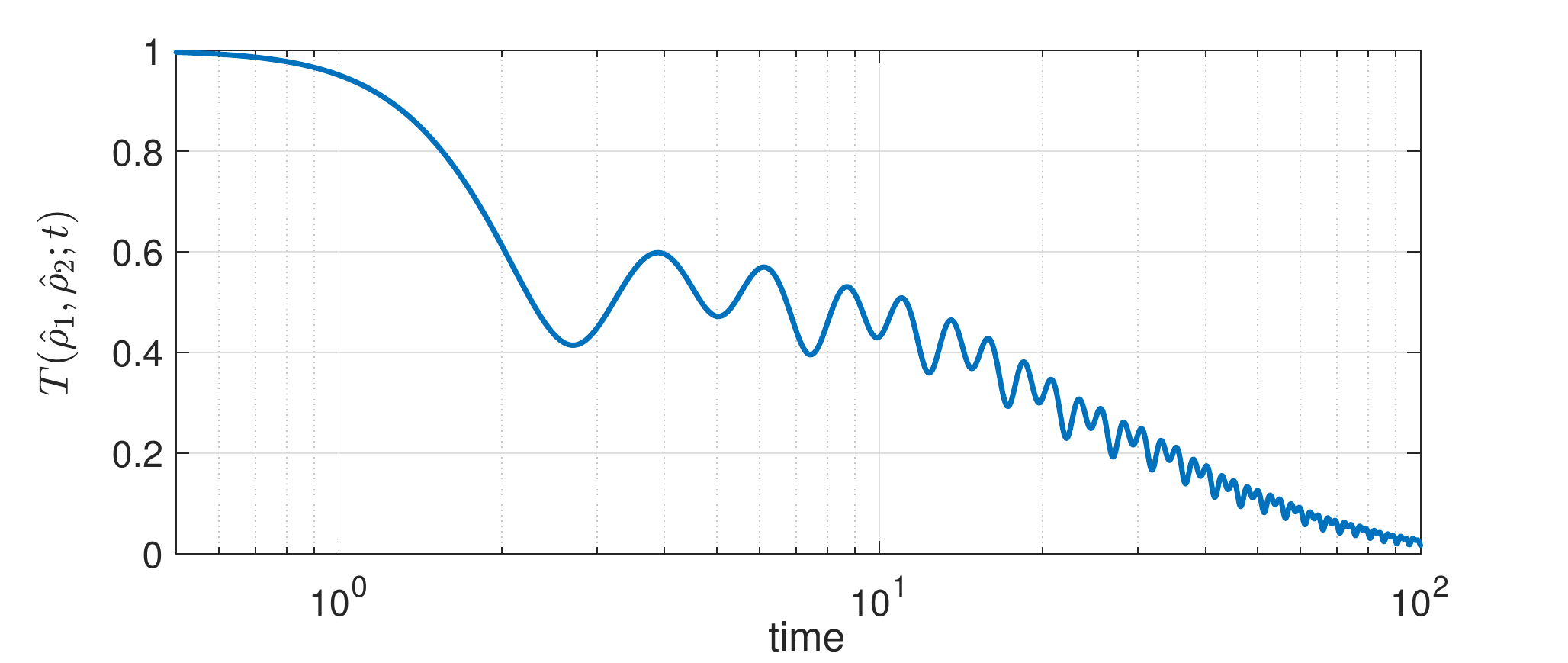}

\caption{\label{fig:trace_distance}Time dependence for the trace distance
$T\left(\widehat{\rho}_{1},\widehat{\rho}_{2};t\right)$ eq. (\ref{eq:trace_distance})
between the initial open systen states $\widehat{\rho}_{1}\left(0\right)=\left|\uparrow\right\rangle \left\langle \uparrow\right|$
and $\widehat{\rho}_{2}\left(0\right)=\left|\downarrow\right\rangle \left\langle \downarrow\right|$
for the model (\ref{eq:subohic_driven}) without driving ($f=0$).
The nonmonotonic behaviour inicates that the we are in the non-Markovian
regime.}
\end{figure}

Here we present the calculation of the time dependence of trace distance
(\ref{eq:trace_distance}) for the system (\ref{eq:subohic_driven}).
We take $\widehat{\rho}_{1}\left(0\right)=\left|\uparrow\right\rangle \left\langle \uparrow\right|$
and $\widehat{\rho}_{2}\left(0\right)=\left|\downarrow\right\rangle \left\langle \downarrow\right|$.
In Fig. \ref{fig:trace_distance} we present the plot of $T\left(\widehat{\rho}_{1},\widehat{\rho}_{2};t\right)$
for the model (\ref{eq:subohic_driven}) without driving ($f=0$)
: we observe the periodic increase of distinguishability, so we are
indeed in the non-Markovian regime. 

\subsubsection{Comparison with the tensor network methods}

\begin{figure}
\includegraphics[scale=0.45]{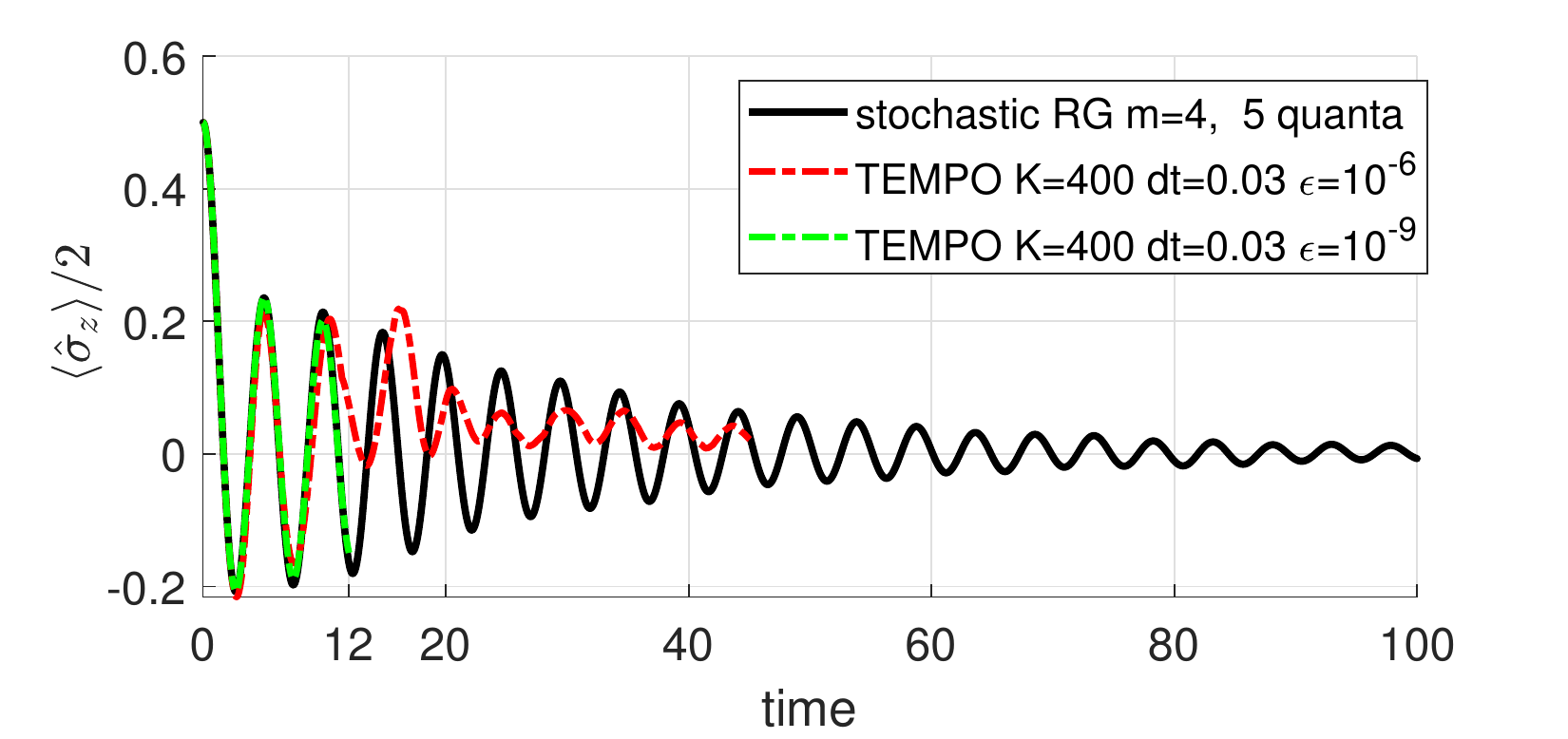}

\caption{\label{fig:TEMPO_vs_stochastic_RG}Comparison of stochastic RG along
trajectories and TEMPO. Time dependence of $\left\langle \widehat{\sigma}_{z}\right\rangle /2$
is computed for model (\ref{eq:subohic_driven}) without driving ($f=0$).
Initial state is $\widehat{\rho}_{\textrm{s}}\left(0\right)=\left|\uparrow\right\rangle \left\langle \uparrow\right|$.
Stochastic RG (black line) is for $m=4$ relevant modes and is truncated
at $n=5$ quanta. It yields the converged result on the presented
time scales. The TEMPO (dashed green and red lines) employs a matrix
product operator of length $K=400$ with bond dimensions up to 358.
However it can capture correctly the dynamics only up to the revival
time $t_{\textrm{cut}}=Kdt=12$, with $dt=0.03$. TEMPO with SVD treshold
$\epsilon=10^{-9}$ (green dashed line) agrees with the stochastic
RG results before the revival at $t_{\textrm{cut}}$. However each
time step $dt=0.03$ becomes rather time consuming ($\approx$40 min
on a 40 CPU core cluster). TEMPO with larger SVD treshold $\epsilon=10^{-6}$
(red dashed line) starts to deviate even before $t_{\textrm{cut}}$,
but can be efficiently propagated to larger times, so that we see
clearly see the revival behaviour.}
\end{figure}

\begin{figure}
\includegraphics[scale=0.45]{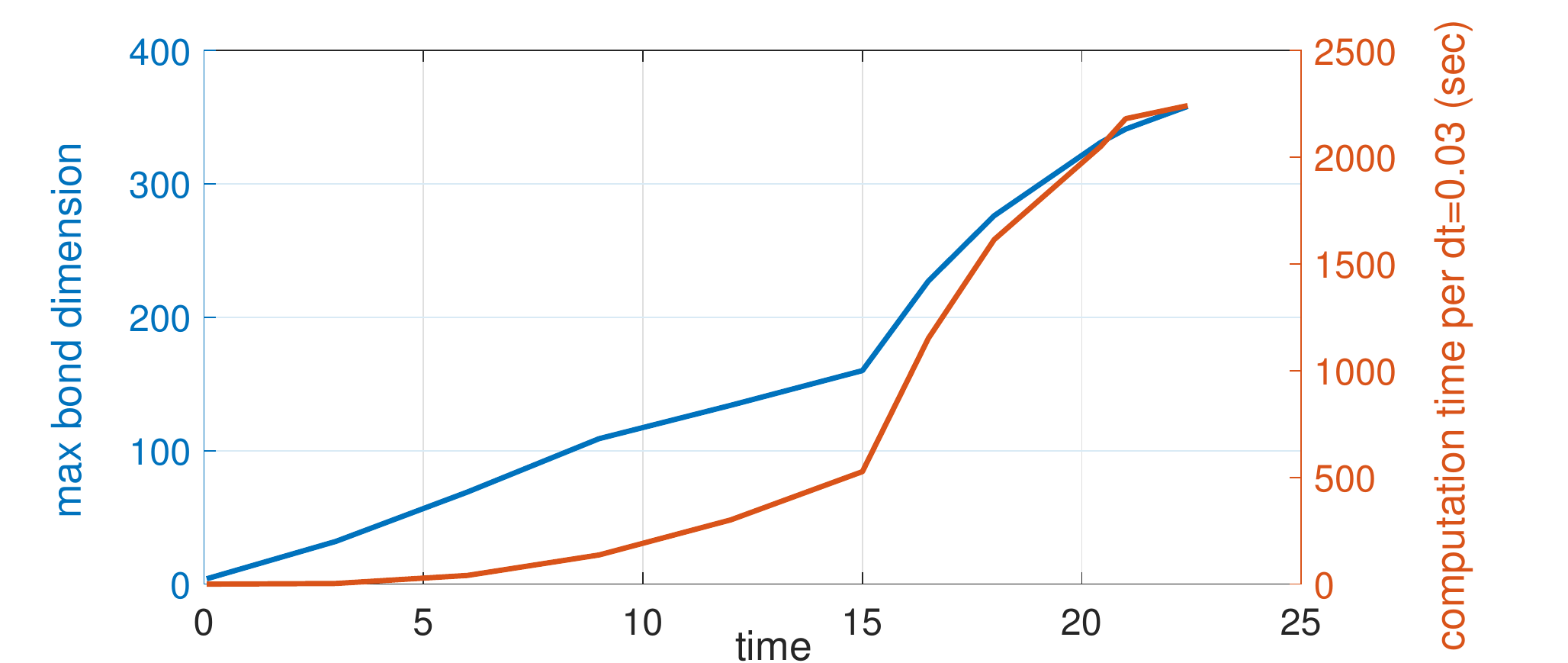}

\caption{\label{fig:TEMPO_complexity_growth}For the same situation as in previous
Fig. \ref{fig:TEMPO_vs_stochastic_RG}. The maximal bond dimension
of TEMPO ($K=400,\epsilon=10^{-9},dt=0.03$) is growing with the propagation
time. This growth is accompanied with the increase in the computation
time per one time step $dt$. }

\end{figure}

Here we provide a numerical comparison with the time-evolving matrix
product operator (TEMPO) method \citep{Strathearn2018}. TEMPO is
a state-of-the-art approach for the computation of open quantum system
dynamics. It is used in recent works e.g. to calculate the quantum
heat statistics \citep{Popovic2021}. We use the package provided
by the authors \citep{Strathearn2018b}. TEMPO represents the correlated
state of the open system and the environment as a matrix product operator
(MPO) state \citep{Strathearn2018}. It depends on parameters $dt$,
$K$, and $\epsilon$. The evolution is discretized with a time step
$dt$. The length $K$ of the MPO corresponds to cutoff time scale
$t_{\textrm{cut}}=Kdt$. All the correlation effects of the environment
at time scales larger than $t_{\textrm{cut}}$ are neglected. Finally,
$\epsilon$ defines the treshold below which the SVD components are
discarded. In Fig. \ref{fig:TEMPO_vs_stochastic_RG} we present the
time-dependence of $\left\langle \widehat{\sigma}_{z}\right\rangle /2$
computed by the RG along a quantum trajectory and compare it with
the TEMPO computation. The system under consideration is again given
by eq. (\ref{eq:subohic_driven}) without driving ($f=0$). We start
from $\widehat{\rho}_{\textrm{s}}\left(0\right)=\left|\uparrow\right\rangle \left\langle \uparrow\right|$.
We observe that only $m=4$ relevant modes at a finite occupation
(5 quanta) is enough to achieve the converged time dependence on large
time scales. At the same time, the finite length $K$ of the TEMPO
corresponds to the finite revival time $t_{\textrm{cut}}$ after which
the computed evolution is corrupted. The longer we want to simulate
the dynamics, the longer MPO should we employ. TEMPO cannot correctly
capture the large-time asyptotics with a finite number of variables.
Also, TEMPO requires rather small SVD treshold $\epsilon=10^{-9}$
in order to quantitatively reproduce the observable properties of
the open quantum system.
 This leads to a rapid growth of the bond dimension, Fig. \ref{fig:TEMPO_complexity_growth}
The computational time per one time step ($dt=0.03$) also increases.
As a result of these two effects (increasing length of MPO with the
increasing bond dimensions), the large time asymptotics of nonstationary
evolution is a computationally hard property for the state-of-the-art
tensor network methods. That is why the conjectured boundedness of
complexity (as a function of simulation time) for the proposed RG
methods is a nontrivial result. 

\section{\label{sec:ENTANGLEMENT-IN-THE}ENTANGLEMENT IN THE TIME DOMAIN}

In the previous sections we repeatedly refer to the entanglement as
a \textit{temporal} property. For example, we say that the outgoing
mode $\phi_{\textrm{out}}\left(t\right)$ is not entangled to the
future (after $t$) in Fig. \ref{fig:entanglement_structure} We also
introduce the entanglers and disentanglers in Figs. \ref{fig:evolution_stages}
and \ref{fig:QSA_production}. However the cautious reader may notice
that the entanglement is usually considered as a spatial property
and not a temporal one. The reason is that the entanglement is defined
by dividing a quantum system into parts with strictly independent
degrees of freedom. Due to the non-locality of the time degrees of
freedom, Fig. \ref{fig:time_nonlocality}, it is impossible to perform
such a strict partitioning in the time domain. Thus, the question
arises, what is the exact content of our concept of temporal entanglement.

\subsection{Absence of entanglement}

We believe there are no doubts how should one define the absence of
entanglement. Suppose we have a pure state $\left|\Phi\left(t_{p}\right)\right\rangle _{\textrm{st}}$
which depends on the time cell degrees of freedom $\widehat{\psi}_{0}^{\dagger},\ldots,\widehat{\psi}_{p-1}^{\dagger}$
and on the quantum numbers of the open system. Let us divide it into
the two subsystems. The subsystem $A$ consists of the open quantum
system and of the $q$ most recent time cells $\widehat{\psi}_{p-q}^{\dagger},\ldots,\widehat{\psi}_{p-1}^{\dagger}$.
The subsystem $B$ is chosen to contain the remaining time cells $\widehat{\psi}_{0}^{\dagger},\ldots,\widehat{\psi}_{p-q-1}^{\dagger}$.
There is no entanglement between $A$ and $B$ if after tracing out
the $B$'s degrees of freedom the reduced state of $A$ is still pure.
Recalling the trace relation (\ref{eq:discrete_time_trace}) for the
tape model, we have for the reduced density matrix of $A$:
\begin{equation}
\widehat{\rho}_{A}=\ps{}{\tau_{p-q-1}}{\left\langle 0\right|}\ldots\ps{}{\tau_{0}}{\left\langle 0\right|}\widehat{\rho}_{A|B}\left|0\right\rangle _{\tau_{0}}\ldots\left|0\right\rangle _{\tau_{p-q-1}},\label{eq:reduced_state_for_entanglement}
\end{equation}
where $\left|0\right\rangle _{\tau_{k}}$ is the vacuum of mode $\widehat{\psi}_{k}^{\dagger}$,
and 
\begin{multline}
\widehat{\rho}_{A|B}=\left\{ e^{\sum_{r=0}^{p-q-1}\sum_{s=p-q}^{p-1}\left(\widehat{\psi}_{r}^{\dagger}M_{rs}\widehat{\psi}_{s}+\textrm{c.c.}\right)}e^{\sum_{rs=0}^{p-q-1}\widehat{\psi}_{r}^{\dagger}M_{rs}\widehat{\psi}_{s}}\right.\\
\left.\times:\left|\Phi\left(t_{p}\right)\right\rangle _{\textrm{st}}\ps{}{\textrm{st}}{\left\langle \Phi\left(t_{p}\right)\right|}:\right\} _{A}\label{eq:bipartite_pairings}
\end{multline}
takes into account the pairings between the time cells of $B$ and
the cross-pairings between the time-cells of $A$ and $B$. Observe
that there is no pairings between the time cells of $A$ in $\widehat{\rho}_{A|B}$:
otherwise $\widehat{\rho}_{A}$ could be mixed irrespectively of its
entanglement to $B$. We say that there is no entanglement between
$A$ and $B$ if $\widehat{\rho}_{A}$ is pure, $\widehat{\rho}_{A}=\textrm{const}\times\left|f\right\rangle _{A}\ps{}A{\left\langle f\right|}$.
In other words, after tracing out $B$ we still can consider that
$A$ is defined by a pure state $\left|f\right\rangle _{A}$.

One example situation when $A$ and $B$ are not entangled is the
state of the form
\begin{equation}
\left|\Phi\left(t_{p}\right)\right\rangle _{\textrm{st}}=\left|f\right\rangle _{A}\otimes\left|f_{\textrm{out}}\left(t_{p-q-1}\right)\right\rangle _{B},
\end{equation}
where in the state $\left|f_{\textrm{out}}\left(t_{p-q-1}\right)\right\rangle _{B}$
only the outgoing modes $\phi_{\textrm{out}}\left(m\right),\ldots,\phi_{\textrm{out}}\left(p-q-1\right)$
are populated by quanta. Then the cross-pairings in eq. (\ref{eq:bipartite_pairings})
are negligible by construction of the incoming and outgoing streams
in sec. \ref{subsec:The-emergence-of}, and $\widehat{\rho}_{A}\propto\left|f\right\rangle _{A}\ps{}A{\left\langle f\right|}$
is pure. This is the exact meaning of saying that the outgoing modes
at time $t_{p-q-1}$ are not entangled to the future incoming modes. 

\subsection{Measure of temporal entanglement}

There are situations when we need to measure the amount of entanglement
between the subsystems. For example, one usually wants to know whether
the entanglement is short-range or long-range. If the entanglement
is long-range, then it is interesting to know its asymptotical scaling.
Therefore, here we propose a measure of temporal entanglement.

The entanglement is measured by the entropy of distribution of non-zero
eigenvalues over the ensemble of eigenvectors. If $\widehat{\rho}_{A}$
(from eq. (\ref{eq:reduced_state_for_entanglement})) is pure then
this distibution is localized on a single eigenvector: the entropy
is zero and there is no entanglement. In the opposite case of a maximally
mixed $\widehat{\rho}_{A}$, the distribution is evenly smeared over
the eigenvectors: the entropy is maximal and so is the entanglement.

Therefore, we define the entanglement entropy as 
\begin{multline}
S_{A|B}=-\log\textrm{Tr}_{A}\left\{ \left|\widehat{\rho}_{A}\right|/\textrm{Tr}_{A}\left|\widehat{\rho}_{A}\right|\right\} ^{2}\\
=2\log\textrm{Tr}_{A}\left|\widehat{\rho}_{A}\right|-\log\textrm{Tr}_{A}\left|\widehat{\rho}_{A}\right|^{2}.\label{eq:temporal_entanglement_measure}
\end{multline}
Here we base our definition on the Renyi entropy of the second order.
The absolute value $\left|\widehat{\rho}_{A}\right|$ is employed
because $\widehat{\rho}_{A}$ may have negative eigenvalues. This
is due to the restriction of the set of pairings, see sec. \ref{subsec:Density-matrix-in-the-time-domain}.
Also we normalize the density matrix by $\textrm{Tr}\left|\widehat{\rho}_{A}\right|$
because $\widehat{\rho}_{A}$ is in general not normalized, $\textrm{Tr}_{A}\widehat{\rho}_{A}\neq1$.
Here, $\textrm{Tr}_{A}$ is the trace in the conventional sense, over
some basis in the Fock space of the time cells $\mathcal{F}_{\textrm{t}}$.

Our definition of the entanglement entropy reduces to the conventional
Renyi entropy in the limit of infinite-band environment with $M\left(t-t^{\prime}\right)=\delta\left(t-t^{\prime}\right)$.
It is reasonable because in this limit the time cells are local independent
degrees of freedom, and the conventional partitioning into the independent
subsystems is possible. 

\subsection{Entanglement is long-range in time }

\begin{figure}
\includegraphics[scale=0.5]{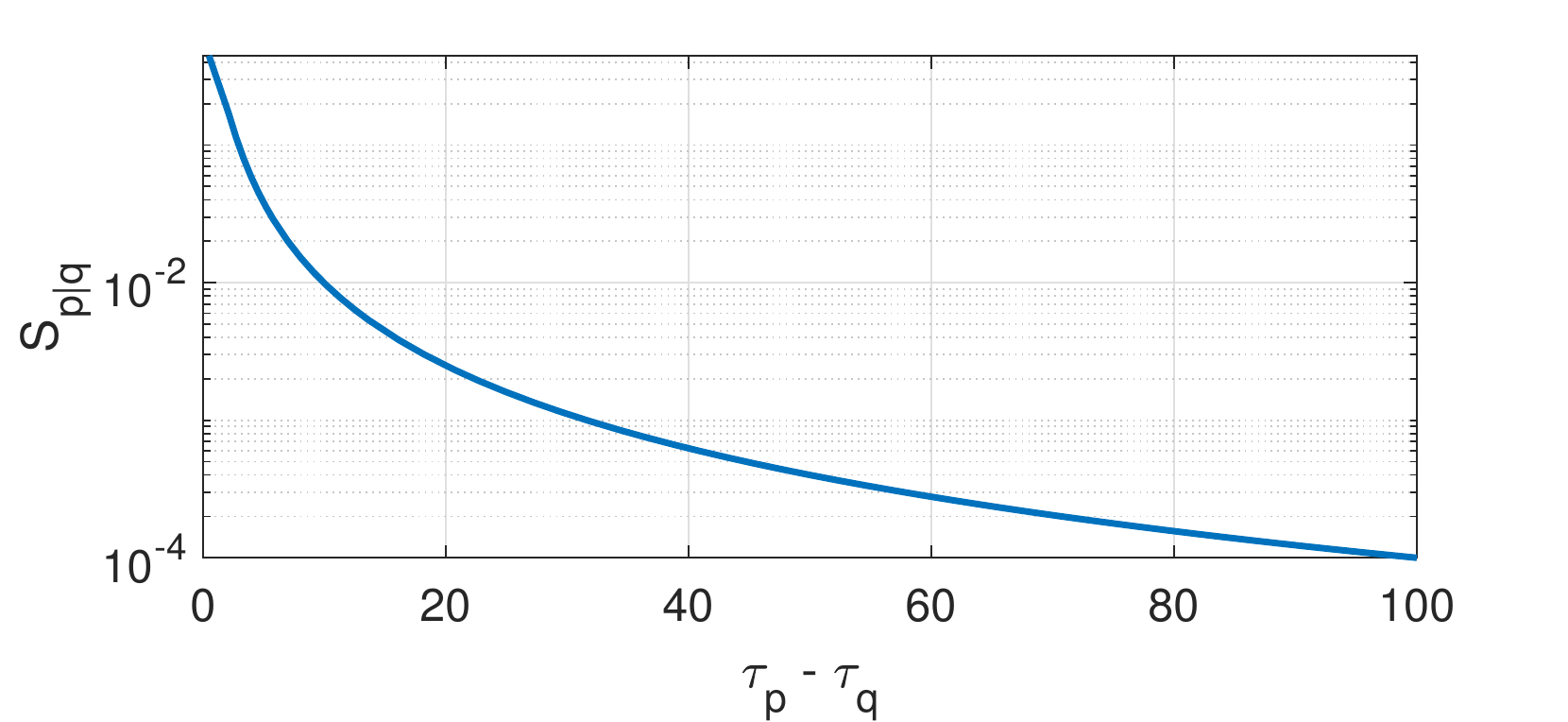}

\caption{\label{fig:The-entanglement-entropy}The entanglement entropy $S_{p|q}$
of the state $\widehat{\psi}_{p}^{\dagger}\widehat{\psi}_{q}^{\dagger}\left|0\right\rangle _{\textrm{t}}$
with two quanta emitted at $\tau_{p}$ and $\tau_{p}$, $\tau_{p}>\tau_{q}$.
It is a long-range function of the time distance $\tau_{p}-\tau_{q}$:
$S_{p|q}\propto\left|M_{pq}\right|^{2}\propto\left(\tau_{p}-\tau_{q}\right)^{-s}$.
The plot is for the case of Ohmic bath with the memory function $\frac{1}{2}\left(1+it\right)^{-2}$.
Observe that althrough the state $\widehat{\psi}_{p}^{\dagger}\widehat{\psi}_{q}^{\dagger}\left|0\right\rangle _{\textrm{t}}$
seems to be factorized, actually the states of emitted quanta are
non-local in time, see. Fig. \ref{fig:time_nonlocality}. That is
why when we trace out one of the time cells, the reduced state of
the other time cell becomes mixed: the quanta are distributed non-locally
between the both time cells.}
\end{figure}

Let us consider as a simple example the pure state created by the
two emission events: one at $\tau_{p}$ and the other at $\tau_{q}$,
with $\tau_{p}>\tau_{q}$, 
\begin{equation}
\left|\Phi\right\rangle _{\textrm{t}}=\widehat{\psi}_{p}^{\dagger}\widehat{\psi}_{q}^{\dagger}\left|0\right\rangle _{\textrm{t}}.
\end{equation}
Let us compute the entanglement entropy $S_{p|q}$ between these two
time moments. Applying (\ref{eq:reduced_state_for_entanglement}),
(\ref{eq:bipartite_pairings}), and (\ref{eq:temporal_entanglement_measure}),
we find:

\begin{equation}
S_{p|q}=2\log\left(p_{1}+p_{2}\right)-\log\left(p_{1}^{2}+p_{2}^{2}\right),\label{eq:formula_for_entanglement_S}
\end{equation}
where $p_{1}=M\left(0\right)/\left(M\left(0\right)+\left|M_{pq}\right|^{2}\right)$,
$p_{2}=\left|M_{pq}\right|^{2}/\left(M\left(0\right)+\left|M_{pq}\right|^{2}\right)$.
At a large time interval between the emission times the memory function
decays as inverse power law, $\left|M_{pq}\right|^{2}\propto\left(\tau_{p}-\tau_{q}\right)^{-s}$,
and the entanglement entropy aslo decays according to the inverse
power law $S_{p|q}\propto\left|M_{pq}\right|^{2}\propto\left(\tau_{p}-\tau_{q}\right)^{-s}$.
In Fig. \ref{fig:The-entanglement-entropy}
 we present the plot of $S_{p|q}$ vs $\tau_{p}-\tau_{q}$ for the
Ohmic memory function (\ref{eq:power_memory}) with $\alpha=1$, $\omega_{\textrm{c}}=1$,
$s=1$. 

\subsection{Entanglement is finite-range in the frame of incoming/relevant/outgoing
modes}

\begin{figure}
\includegraphics[scale=0.5]{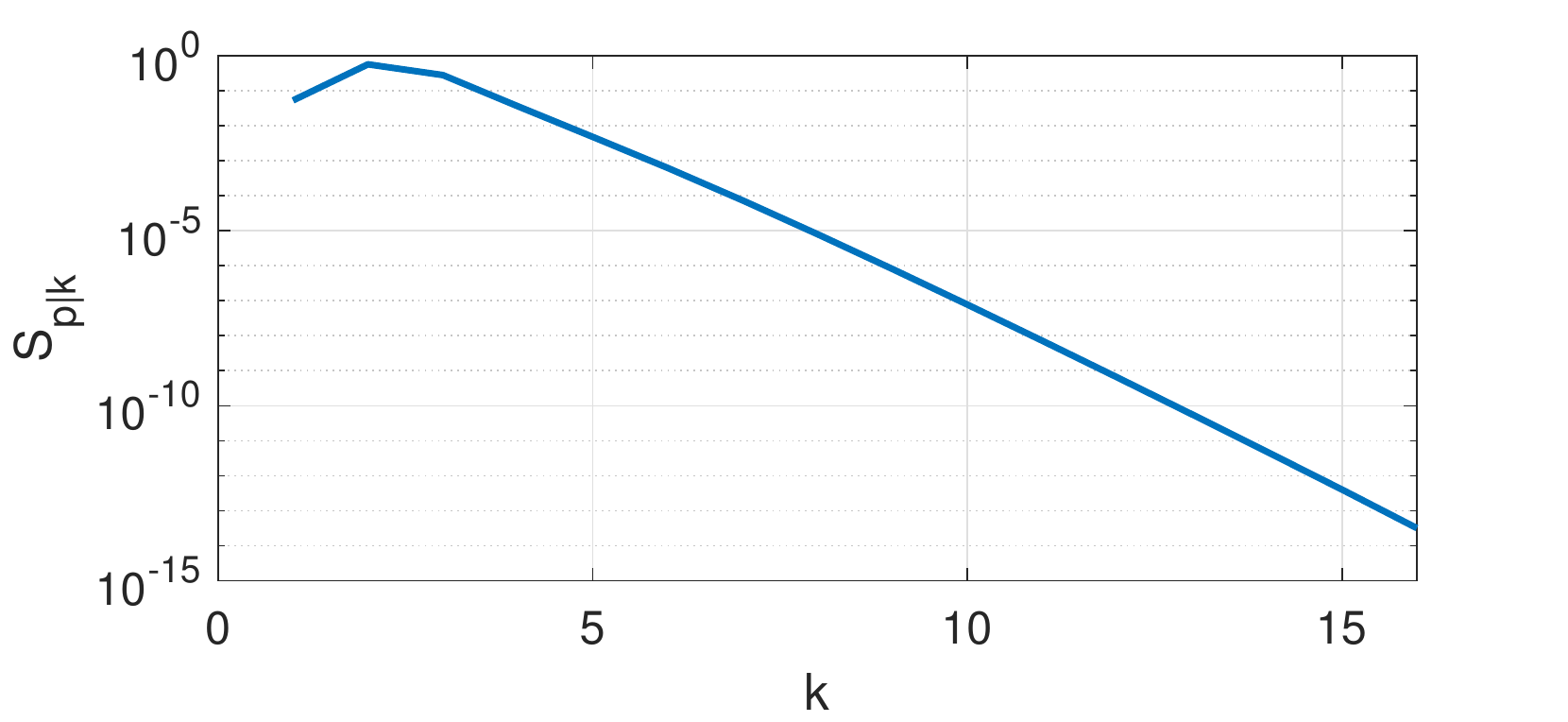}

\caption{\label{fig:The-entanglement-entropy-in_fdms}The entanglement entropy
$S_{p|k}$ of the state $\widehat{\psi}_{p}^{\dagger}\widehat{\phi}_{k}^{\dagger}\left|0\right\rangle _{\textrm{t}}$
with one quantum emitted at $\tau_{p}$, and the other quantum in
the fastest decoupling basis state $\phi_{k}$, computed for the time
moment $t_{p}$, see the section \ref{subsec:How-to-find}. . The
case of Ohmic bath with the memory function $\frac{1}{2}\left(1+it\right)^{-2}$.
The plot is for $S_{p|k}$ wrt the fastest decoupling basis element
number $k$. It is seen that the entanglement entropy decays exponentially
fast with $k$. Therefore, in the frame of incoming/relevant/outgoing
modes the entanglement becomes effectively finite-ranged. }

\end{figure}

Now let us consider the pure state $\left|\Phi\right\rangle _{\textrm{t}}=\widehat{\psi}_{p}^{\dagger}\widehat{\phi}_{k}^{\dagger}\left|0\right\rangle _{\textrm{t}}$
when one quantum is emitted at a time moment $\tau_{p}$, and the
other is in the fastest decoupling basis state $\phi_{k}$, computed
for the time moment $t_{p}$, see the section \ref{subsec:How-to-find}.
The formula (\ref{eq:formula_for_entanglement_S}) for the entanglement
entropy still holds, with the substitution $S_{p|q}\to S_{p|k}$ and
$M_{pq}\to\left(p\left|M\right|\phi_{k}\right)$. In Fig. \ref{fig:The-entanglement-entropy-in_fdms}
we plot the dependence of $S_{p|k}$ with respect to the fastest decoupling
basis element number $k$. We see that in the frame of incoming/relevant/outgoing
modes the entanglement becomes effectively finite-ranged. 

\section{\label{sec:CONCLUSION}CONCLUSION}

\textit{Structure of entanglement.} In this paper we propose a constructive
way of how to characterize the structure of entanglement. By tracking
the degrees of freedom (the ``bricks'') which couple (for the first
time) to the evolution and decouple (irreversibly) from the evolution,
we obtain a useful and \textit{intrinsic }(free from adhoc assumptions)
characterization. Then the entanglement starts to look like a Lego
built from these bricks. Here the evolution can be understood in a
generalized sense, e.g. as an RG flow parameter.

\textit{Real-time motion of open quantum systems. }In this paper we
describe the entire lifecycle of the entanglement: from its generation
through the entanglement of incoming modes to its death in the oblivion
of outgoing modes. This sheds light on a number of questions. First,
this shows that the ``true'' quantum complexity \citep{Poulin2011,Brandao2019,Luchnikov2019,Ali2019,Bhattacharyya2020}
of the real-time motion survives only in the modes of the environment
which are non-negligibly coupled to the future quantum motion. If
there is a balance between the flux of emitted quanta and the flux
of quanta which are forgotten in the outgoing modes, then this complexity
is asymptotically bounded on large times. Then the real-time motion
of open quantum system becomes efficiently computable on a classical
computer. Second, we find how the decoherence happens continuously
in the non-Markovian regime. We find how tiny parts of quantum information
decouple and disappear each infinitesimal time moment. Interesting
future direction is to study the distributed open quantum systems
when there is a number $r$ of coupling sites $\widehat{b}_{\alpha}^{\dagger}\left(t\right)$,
$\alpha=1\ldots r$, to the environment. We expect that the presented
approach should be straightforwardly exended to this case. The main
modification is that the memory function (\ref{eq:temporal_commutation_relation}),
the matrix $K\left(p\right)$ eq. (\ref{eq:K_matrix_element}), and
the relevant modes will acquire additional indices $\alpha,\alpha^{\prime}=1\ldots r$.
The imporant question is whether it is possible to construct low-dimensional
relevant modes inside the distributed open system so as to escape
the growth of its Hilbert space dimension.

\textit{Few-mode approximations of non-Markovian environments. }It
turns out that each moment of time only a finite number $m$ of degrees
of freedom of the environment are significantly coupled to the future
motion. Moreover, the significance of the neglected degrees of freedom
decreases exponentially fast. This opens interesting perspective on
the few-mode approximations of the environments with memory. 

\textit{Renormalization group methods. }We also obtain a novel perpective
for the renormalization group methods. Traditionally these methods
are based on the iterative application of the scale and coarsegraining
tranformations. Here we go beyond this paradigm. Any evolution can
be considered as a certain RG flow provided it is accompanied by the
continuous emergence of degrees of freedom which are not entangled
to the future evolution. These degrees of freedom are irrelevant and
can be iteratively traced out. Then the relevant subspace of such
an RG procedure is not the conventional infrared limit, but the subspace
of degrees of freedom which are significantly coupled to the future
evolution. We illustrate this point (down to the resulting numerical
schemes) for the case of real-time flow for open system. To extend
this ideas to other many-body problems is a subject of future research. 

\textit{Emergence of classical records. }There is a interesting interrelationship
between the entanglement, RG, and the models of continuous measurement
\citep{Megier2020}. The irrelevant degrees of freedom (which are
not entangled to the future), can be collapsed \citep{Adler2007,Carlesso2018}
to a classical measurement signal as soon as they emerge. Computationally,
this leads to a stochastic RG along a quantum trajectory. The numerical
benefit over the deterministic RG for density matrices is that the
latter scales as $N^{2}$ for $N$ being the dimension of the relevant
Fock subspace, whereas the former scales as $N$.

\textit{Quantum trajectories.} In this work we promote a constructive
viewpoint on quantum trajectories \citep{Strunz1999a,Strunz1999,Gambetta2004,Breuer2004a,Luoma2011,Durr2004}:
we propose to define the quantum trajectory as a model of how the
quantum complexity decays into the classical complexity of a stochastic
ensemble. 

These results are illustrated with a number of calculations for the
model of a two-level system (qubit), with and without driving, which
is coulped to the environments with subohmic/Ohmic/superohmic and
semicircle spectral densities. In particular, in the cases considered
only 4-3 relevant modes were enough to achieve the numerical convergence. 

Of couse, we illustrate these ideas on a single specific model: the
spin in the bosonic bath. However we believe the picture presented
in the Introduction, and the concept of irreversible decoupling as
negligible average intensity of future interactions (sec. \ref{sec:HOW-THE-QUANTUM}),
are rather physically transparent and general. They can be applied
to a varienty of other models and situations, although specific formal
implementation may differ.
\begin{acknowledgments}
The work was carried out in the framework of the Roadmap for Quantum
computing in Russia
\end{acknowledgments}

\appendix

\section{\label{sec:SOLVING-EQUATIONS-IN}SOLVING EQUATIONS IN THE TRUNCATED
FOCK SPACE}

The total Hilbert space for the procedure in sec. \ref{subsec:Propagation-of-density}
has the basis elements
\begin{equation}
\left|q\right\rangle _{\textrm{s}}\otimes\left|n_{1}\right\rangle _{1}\otimes\left|n_{2}\right\rangle _{2}\otimes\ldots\left|n_{m+1}\right\rangle _{m+1},\label{eq:basis_elements}
\end{equation}
where $\left|q\right\rangle _{\textrm{s}}$ is some basis in the Hilbert
space $\mathcal{H}_{\textrm{s}}$ of the open system; $\left|n_{i}\right\rangle _{i}$
is the state of $n_{i}$ quanta in mode $i$. The modes $i=1\ldots m$
are our relevant modes. The mode $m+1$ becomes coupled only after
$t=t_{m}$ and alternatively plays the role of incoming and outgoing
mode. 

We truncate the Hilbert space at certain number $n$ of quanta: only
the basis states with $\sum_{i}n_{i}\leq n$ are being kept. Let us
assume that $N$ is the total number of remaining basis elements.
All the basis elements (\ref{eq:basis_elements}) are numbered: 
\begin{equation}
\left|k\right\rangle =\left|q^{\left(k\right)}\right\rangle _{\textrm{s}}\otimes\left|n_{1}^{\left(k\right)}\right\rangle _{1}\otimes\left|n_{2}^{\left(k\right)}\right\rangle _{2}\otimes\ldots\left|n_{m+1}^{\left(k\right)}\right\rangle _{m+1}
\end{equation}
for $k=1\ldots N$. Then the joint many-particle wavefunction $\left|\Psi\right\rangle $
of open system and the modes is represented as $\left|\Psi\right\rangle =\sum_{k=1}^{N}\Psi_{k}\left|k\right\rangle $
with $\Psi_{k}$ being the $1\times N$ array of complex numbers.
The density matrix is represented as $\widehat{\rho}=\sum_{kl=1}^{N}\rho_{kl}\left|k\right\rangle \left\langle l\right|$,
with $\rho_{kl}$ being the $N\times N$ Hermitean matrix of complex
numbers.

The operators of open system, the creation/annihilation operators
for modes are represented as sparce matrices in the basis of $\left|k\right\rangle $.
Therefore, all the operations (\ref{eq:hamiltonian_propagation_of_density_matrix})-(\ref{eq:density_matrix_project_away})
are implemented in terms of space matrix multiplications. 

Please observe that at the time $t_{p}$ the mode $\textrm{min}\left(m+1,p\right)$
play the role of incoming mode before the disentangling step (\ref{eq:density_matrix_disentangler}).
After the disentangling step the mode $\textrm{min}\left(m+1,p\right)$
plays the role of outgoing mode. The oblivion step (\ref{eq:density_matrix_project_away})
is implemented by setting to zero all elements of $\rho_{kl}$ which
correspond to non-zero occupation of the mode $\textrm{min}\left(m+1,p\right)$.
After the oblivion step, at the next time moment, $p\to p+1$, the
mode $\textrm{min}\left(m+1,p\right)$ becomes again the incoming
mode.

\section{\label{sec:COMPUTING-THE-PAIRING}COMPUTING THE PAIRING UPDATE}

Our propagation procedure is permitted to have a local error $O\left(dt^{3}\right)$.
Then it is sufficient to compute the pairing update (\ref{eq:nonMarkovain_jump})
up to the error $O\left(dt^{3}\right)$. Observe how the pairing update
(\ref{eq:nonMarkovain_jump}) is acting: if we consider the Taylor
series of the pairing function, then at the order $n$ it annihilates
at least $n$ quanta in the incoming mode, either from the left (via
$\widehat{\psi}_{p}$) or from the right (via $\widehat{\psi}_{p}^{\dagger}$)
of the density matrix $\widehat{\rho}_{\textrm{rel+i}}^{\left(1\right)}\left(t_{p+1}\right)$.
However, the amplitude to find $n$ quanta in the incoming mode scales
as $dt^{n}$. Indeed, looking at the propagation (\ref{eq:hamiltonian_propagation_of_density_matrix})
and the Hamiltonian $\widehat{H}_{\textrm{st}}\left(\tau_{p}\right)$,
eq. (\ref{eq:midpoint_Hamiltonian-1-1-1-1}), we see that the new
quanta are created via $\propto dt\widehat{\psi}_{p}^{\dagger}$.
Therefore, with the second order accuracy we compute the pairing update
(\ref{eq:nonMarkovain_jump}) as
\begin{equation}
\widehat{\rho}_{\textrm{rel+i}}^{\left(2\right)}\left(t_{p+1}\right)=\widehat{\rho}_{0}+\widehat{\rho}_{1}+\frac{1}{2}\widehat{\rho}_{2},
\end{equation}
where 
\begin{equation}
\widehat{\rho}_{0}=\widehat{\rho}_{\textrm{rel+i}}^{\left(1\right)}\left(t_{p+1}\right),
\end{equation}
and the recurrent relation between $\widehat{\rho}_{i}$ and $\widehat{\rho}_{i-1}$,
\begin{multline}
\widehat{\rho}_{i}=M\left(0\right)\widehat{\psi}_{p}\widehat{\rho}_{i-1}\widehat{\psi}_{p}^{\dagger}\\
+\sum_{i}M_{i}\left(p\right)\widehat{\phi}_{i}\widehat{\rho}_{i-1}\widehat{\psi}_{p}^{\dagger}\\
+\sum_{i}M_{i}^{*}\left(p\right)\widehat{\psi}_{p}\widehat{\rho}_{i-1}\widehat{\phi}_{i}^{\dagger},
\end{multline}
for $i=1,2$.

\section{\label{sec:LANCZOS-INSPIRED-ALGORITHM-FOR} COMPUTING THE DISENTANGLER
TRANSFOM}

Here we present the algorithm we use in order to apply the disentangler
transforms (\ref{eq:density_matrix_disentangler}) and (\ref{eq:wavefunction_disentangler}).
The latter are defined as a unitary Bogoliubov transform, eq. (\ref{eq:many_particle_disentangler-1-1}).
The Bogoliubov transform has the general form 
\begin{equation}
\widehat{W}=\exp\left(i\sum_{kl}\widehat{\psi}_{k}^{\dagger}h_{kl}\widehat{\psi}_{l}\right)\equiv\exp\left(i\widehat{H}_{W}\right)
\end{equation}
with  suitably defined $\widehat{\psi}_{l},\,\widehat{\psi}_{l}^{\dagger}$
and a hermitean matrix $h_{kl}$. Here we introduce the effective
``Hamiltonian'' $\widehat{H}_{W}$ which generates $\widehat{W}$.
We need to apply $\widehat{W}$ to some wavefunction $\left|\Psi\right\rangle $,
\begin{equation}
\left|\Psi_{W}\right\rangle =\widehat{W}\left|\Psi\right\rangle .\label{eq:bogoliubov_transform}
\end{equation}
In general, $h_{kl}$ is a dense matrix (as it turns out to be in
eq. (\ref{eq:many_particle_disentangler-1-1})). Therefore, each application
of $\widehat{H}_{W}$ is expensive if $\left|\Psi\right\rangle $
is many-particle. We need to devise an algorithm which minimizes the
number of applications of$\widehat{H}_{W}$.

Suppose that our budget is $M$ applications of $\widehat{H}_{W}$.
Then we construct the orthonormal basis $\left|\Phi_{0}\right\rangle \ldots\left|\Phi_{M}\right\rangle $
from $M+1$ vectors $\left|\Psi\right\rangle ,\widehat{H}_{W}\left|\Psi\right\rangle ,\ldots,\widehat{H}_{W}^{M}\left|\Psi\right\rangle $.
The basis is constructed recurrently. We begin as 
\begin{equation}
\left|\Phi_{0}\right\rangle =\left|\Psi\right\rangle /\left\Vert \Psi\right\Vert ,
\end{equation}
where we normalize $\left|\Psi\right\rangle $ since it is normalizated
differently in the tape model, see eq. (\ref{eq:normalization_in_the_time_domain}).
Now suppose we know $\left|\Phi_{k}\right\rangle $. We apply $\widehat{H}_{W}$
and find new unnormalized basis element $\left|\widetilde{\Phi}_{k+1}\right\rangle $:
\begin{multline}
\left|\Phi_{k}^{H}\right\rangle =\widehat{H}_{W}\left|\Phi_{k}\right\rangle ,\,\,\beta_{k}=\left\langle \Phi_{k}\left|\Phi_{k}^{H}\right.\right\rangle ,\\
\left|\widetilde{\Phi}_{k+1}\right\rangle =\left|\Phi_{k}^{H}\right\rangle -\beta_{k}\left|\Phi_{k}\right\rangle -\alpha_{k-1}\left|\Phi_{k-1}\right\rangle ,\\
\alpha_{k}=\left\Vert \widetilde{\Phi}_{k}\right\Vert ,
\end{multline}
were we set $\alpha_{-1}=0$, $\left|\Phi_{-1}\right\rangle =0$.
Starting from $k=1$, we additionally orthogonalize $\left|\widetilde{\Phi}_{k+1}\right\rangle $
to all the previous basis elements: 
\begin{equation}
\left|\Phi^{\perp}{}_{k+1}\right\rangle =\left(1-\left|\Phi_{k}\right\rangle \left\langle \Phi_{k}\right|\right)\ldots\left(1-\left|\Phi_{0}\right\rangle \left\langle \Phi_{0}\right|\right)\left|\widetilde{\Phi}_{k+1}\right\rangle .
\end{equation}
Otherwise the procedure will be numerically unstable. Finally, we
normalize the basis element:
\begin{equation}
\left|\Phi{}_{k+1}\right\rangle =\left|\Phi^{\perp}{}_{k+1}\right\rangle /\left\Vert \left|\Phi^{\perp}{}_{k+1}\right\rangle \right\Vert .
\end{equation}
As a result, $\widehat{H}_{W}$ is represented as a tridiagonal matrix
\begin{equation}
H_{\textrm{ch}}=\left[\begin{array}{cccc}
\beta_{0} & \alpha_{0} & 0 & 0\\
\alpha_{0} & \beta_{1} & \alpha_{1} & 0\\
0 & \alpha_{1} & \ddots & \vdots\\
\vdots & \ddots & \ddots & \alpha_{M-1}\\
0 & 0 & \alpha_{M-1} & \beta_{M}
\end{array}\right].\label{eq:tridiagonal}
\end{equation}
We see from this form that the Hamiltonian $\widehat{H}_{W}$ is represented
as a semiinfinite chain attached to $\left|\Phi_{0}\right\rangle $.
The hopping between the sites of the chain is given by $\alpha_{k}$.
Therefore, we obtain the first criterion for truncation: if $\alpha_{k}^{2}$
is below some treshold, e.g. $10^{-5}$, then the remaining sites
are effectively decoupled, and we can stop at $k$ applications of
$\widehat{H}_{W}$ and set $M=k$. The second criterion is the norm
$\left\Vert \left|\Phi^{\perp}{}_{k+1}\right\rangle \right\Vert ^{2}$:
if it is below some treshold e.g. $10^{-4}$, then the new basis elements
are essentially linerarly dependent, so we stop at $k$ applications
of $\widehat{H}_{W}$ and set $M=k$. 

Finally, the $M\times M$ tridiagonal matrix $H_{\textrm{ch}}$, eq.
(\ref{eq:tridiagonal}), is diagonalized: 
\begin{equation}
H_{\textrm{ch}}=U\left[\begin{array}{ccc}
\omega_{1} & \dots & 0\\
\vdots & \ddots & \vdots\\
0 & \dots & \omega_{M}
\end{array}\right]U^{\dagger},
\end{equation}
and the approximation $\left|\Psi_{W}^{M}\right\rangle $ to $\left|\Psi_{W}\right\rangle $
is computed as
\begin{equation}
\left|\Psi_{W}^{M}\right\rangle =\sum_{k,l,l^{\prime}=1}^{M}\exp\left(-i\omega_{k}\right)U_{lk}\left|\Phi{}_{l}\right\rangle U_{l^{\prime}k}^{*}\left\langle \Phi_{l^{\prime}}\right|\left.\Psi\right\rangle .
\end{equation}
If $\alpha_{k}^{2}$ and $\left\Vert \left|\Phi^{\perp}{}_{k+1}\right\rangle \right\Vert ^{2}$
are still above their tresholds, then the last, fourth criterion,
is that we stop adding basis elements when $\left\Vert \left|\Psi_{W}^{M}\right\rangle -\left|\Psi_{W}^{M+1}\right\rangle \right\Vert ^{2}$
is below some treshold e.g. $10^{-3}$. Then we set $\left|\Psi_{W}\right\rangle \approx\left|\Psi_{W}^{M}\right\rangle $.
For the numerical computations in the paper we obtain that $M$ lies
in the range $3\ldots16$.

\section{\label{sec:IMPORTANCE-SAMPLING}IMPORTANCE SAMPLING}

The noise shift leads to an additional Bogoliubov transform in the
state vector (\ref{eq:dressed_state}):
\begin{equation}
S\left(\Delta\xi\right)=e^{\sum_{s=0}^{\infty}\Delta\xi_{s}^{*}\widehat{\psi}_{s}}.\label{eq:shift_noise_bogoliubov}
\end{equation}
Then the restriction on $\Delta\xi$ is that $S\left(\Delta\xi\right)$
should not couple to the outgoing modes which have appeared by the
time moment $t_{p}$. Otherwise the entanglement structure will be
spoiled and the oblivion step (\ref{eq:wavefunction_disentangler})
will become invalid. This restricts the shifts to
\begin{equation}
\Delta\xi_{r}=\begin{cases}
\textrm{arbitrary} & \textrm{for}\,\,r\geq p,\\
\textrm{const}\times M_{pr} & \textrm{for}\,\,r<p.
\end{cases}
\end{equation}
Apart from this, there is no restrictions on the choice of importance
sampling technique.

In this work we derive the importance sampling technique from the
evolution of the probability distribution $Q\left(\xi,t_{p}\right)$
of the classical noise trajectories $\xi$ at a time $t_{p}$. At
$t_{0}=0$ this distribution coincides with the probability distribution
of the vacuum fluctuations
\begin{equation}
Q\left(\xi,0\right)\propto\exp\left(-\sum_{rs=0}^{\infty}\xi_{r}^{*}\left[M^{-1}\right]_{rs}\xi_{s}\right).
\end{equation}
However at later time moments it becomes weighted: 
\begin{equation}
Q\left(\xi,t_{p}\right)=\left\Vert \ps{}{\textrm{t}}{\left\langle 0\right.}\left|\Phi\left(\xi,t_{p}\right)\right\rangle _{\textrm{st}}\right\Vert ^{2}Q\left(\xi,0\right),\label{eq:target_probability}
\end{equation}
which follows from the partial trace relation eq. (\ref{eq:partial_trace_as_stochastic_average}).
This probability distribution is properly normalized in the sense
that 
\begin{equation}
\overline{\left[\frac{Q\left(\xi,t_{p}\right)}{Q\left(\xi,0\right)}\right]}_{\xi}=1,\label{eq:ratio}
\end{equation}
which follows from (\ref{eq:partial_trace_as_stochastic_average})
and $\textrm{Tr}\widehat{\rho}_{\textrm{s}}\left(t_{p}\right)=1$.
The importance sampling is required when the ratio on the left of
eq. (\ref{eq:ratio}) starts to fluctuate violently. 

Let us find the master equation for $Q\left(\xi,t_{p}\right)$ \citep{Polyakov2019}.
From the tape model Hamiltonian (\ref{eq:discrete_time_hamiltonian})
we find: 
\begin{multline}
\partial_{t}Q\left(\xi,t_{p}\right)\\
=-i\textrm{Tr}_{\textrm{s}}\left\{ \left|\xi\right\rangle _{\textrm{t}}\ps{}{\textrm{t}}{\left\langle \xi\right|}\left[\widehat{H}_{\textrm{st}}\left(\tau_{p}\right)\left|\Phi\left(t_{p}\right)\right\rangle _{\textrm{st}}\ps{}{\textrm{st}}{\left\langle \Phi\left(t_{p}\right)\right|}-\textrm{c.c.}\right]\right\} \\
=-i\textrm{Tr}_{\textrm{s}}\left\{ \left[\left|\xi\right\rangle _{\textrm{t}}\ps{}{\textrm{t}}{\left\langle \xi\right|}\left(\widehat{s}\widehat{\psi}_{p}^{\dagger}+\widehat{s}^{\dagger}\sum_{r=0}^{\infty}M_{pr}\widehat{\psi}_{r}\right)-\textrm{c.c.}\right]\right.\\
\left.\vphantom{\left[\left|\xi\right\rangle _{\textrm{t}}\ps{}{\textrm{t}}{\left\langle \xi\right|}\left(\widehat{s}\widehat{\psi}_{p}^{\dagger}+\widehat{s}^{\dagger}\sum_{r=0}^{\infty}M_{pr}\widehat{\psi}_{r}\right)-\textrm{c.c.}\right]}\times\left|\Phi\left(t_{p}\right)\right\rangle _{\textrm{st}}\ps{}{\textrm{st}}{\left\langle \Phi\left(t_{p}\right)\right|}\right\} ,\label{eq:Q_derivative}
\end{multline}
where we introduce the time-cell coherent states 
\begin{equation}
\left|\xi\right\rangle _{\textrm{t}}=Q^{\frac{1}{2}}\left(\xi,0\right)e^{\sum_{s=0}^{\infty}\xi_{s}\widehat{\psi}_{s}^{\dagger}}\left|0\right\rangle _{\textrm{t}}.
\end{equation}
In the second line of eq. (\ref{eq:Q_derivative}) we have employed
the cyclic trace property and the fact that $\widehat{H}_{\textrm{s}}$
commutes with the time-cell coherent state projections. Observe that
we extend the summation to infinity in the annihilation term on the
second line of eq. (\ref{eq:Q_derivative}). This is possible since
the incoming modes are in vacuum. It turns out that the equation (\ref{eq:Q_derivative})
can be reduced to the convection (drift) form

\begin{multline}
\partial_{t}Q\left(\xi,t_{p}\right)=\sum_{r}\partial_{\xi_{r}}\left\{ \mathcal{A}_{r}\left(t_{p}\right)Q\left(\xi,t_{p}\right)\right\} \\
+\sum_{r}\partial_{\xi_{r}^{*}}\left\{ \mathcal{A}_{r}^{*}\left(t_{p}\right)Q\left(\xi,t_{p}\right)\right\} .
\end{multline}
In order to demonstrate this, we employ the fact that the creation
$\widehat{\psi}_{s}^{\dagger}$ and annihilation $\widehat{\psi}_{s}$
operators act as differential operators on the coherent state projections:
\begin{equation}
\left|\xi\right\rangle _{\textrm{t}}\ps{}{\textrm{t}}{\left\langle \xi\right|}\widehat{\psi}_{p}^{\dagger}=\xi_{p}^{*}\left|\xi\right\rangle _{\textrm{t}}\ps{}{\textrm{t}}{\left\langle \xi\right|},
\end{equation}
\begin{equation}
\widehat{\psi}_{p}\left|\xi\right\rangle _{\textrm{t}}\ps{}{\textrm{t}}{\left\langle \xi\right|}=\xi_{p}\left|\xi\right\rangle _{\textrm{t}}\ps{}{\textrm{t}}{\left\langle \xi\right|},
\end{equation}
\begin{equation}
\left|\xi\right\rangle _{\textrm{t}}\ps{}{\textrm{t}}{\left\langle \xi\right|}\sum_{r=0}^{\infty}M_{pr}\widehat{\psi}_{r}=\left\{ \sum_{r=0}^{\infty}M_{pr}\partial_{\xi_{r}^{*}}+\xi_{p}\right\} \left|\xi\right\rangle _{\textrm{t}}\ps{}{\textrm{t}}{\left\langle \xi\right|},
\end{equation}
\begin{equation}
\sum_{r=0}^{\infty}M_{pr}^{*}\widehat{\psi}_{r}^{\dagger}\left|\xi\right\rangle _{\textrm{t}}\ps{}{\textrm{t}}{\left\langle \xi\right|}=\left\{ \sum_{r=0}^{\infty}M_{pr}^{*}\partial_{\xi_{r}}+\xi_{p}^{*}\right\} \left|\xi\right\rangle _{\textrm{t}}\ps{}{\textrm{t}}{\left\langle \xi\right|}.
\end{equation}
Substituting these into the eq. (\ref{eq:Q_derivative}), we find
\begin{equation}
\mathcal{A}_{r}\left(t_{p}\right)=i\overline{s}\left(t_{p};\xi\right)\sum_{r=0}^{\infty}M_{pr}^{*},\label{eq:velocity}
\end{equation}
where the conditional average $\overline{s}\left(t_{p};\xi\right)$
of the coupling operator $\widehat{s}$ is defined as
\begin{equation}
\overline{s}\left(t_{p};\xi\right)=\frac{\textrm{Tr}\left\{ \widehat{s}\times\ps{}{\textrm{t}}{\left\langle 0\right.}\left|\Phi\left(\xi,t_{p}\right)\right\rangle _{\textrm{st}}\ps{}{\textrm{st}}{\left\langle \Phi\left(\xi,t_{p}\right)\right|\left.0\right\rangle _{\textrm{t}}}\right\} }{\left\Vert \ps{}{\textrm{t}}{\left\langle 0\right.}\left|\Phi\left(\xi,t_{p}\right)\right\rangle _{\textrm{st}}\right\Vert ^{2}}.
\end{equation}
Switching to the frame of incoming/relevant/outgoing modes and employing
the relevant wavefunction, we reexpress this equation as 
\begin{equation}
\overline{s}\left(t_{p};\xi\right)=\frac{\textrm{Tr}\left\{ \widehat{s}\times\ps{}{\textrm{rel}}{\left\langle 0\right.}\left|\Phi\left(\xi,t_{p}\right)\right\rangle _{\textrm{rel}}\ps{}{\textrm{rel}}{\left\langle \Phi\left(\xi,t_{p}\right)\right|\left.0\right\rangle _{\textrm{rel}}}\right\} }{\left\Vert \ps{}{\textrm{rel}}{\left\langle 0\right.}\left|\Phi\left(\xi,t_{p}\right)\right\rangle _{\textrm{rel}}\right\Vert ^{2}},
\end{equation}
which is the eq. (\ref{eq:conditional_average}). The velocity (\ref{eq:velocity})
means that the noise trajectory $\xi$ is shifted with time as 
\begin{equation}
\xi\left(\tau;t_{p}\right)=\xi\left(\tau\right)-i\intop_{0}^{t_{p}}dt\overline{s}\left(t;\xi\right)M^{*}\left(t-\tau\right).\label{eq:shifted_noise}
\end{equation}
Observe that here we switch back and forth between the continuous
and the discrete-time relations. This is permissible because we have
chosen the discrete propagator in sec. \ref{subsec:The-Hamiltonian-for}
so that it converges to the continous evolution with the global error
$O\left(dt^{2}\right)$. 

The importance sampling procedure is to shift the noise $\xi$ continuously
in time according to the relation (\ref{eq:shifted_noise}). As we
have mentioned in eq. (\ref{eq:shift_noise_bogoliubov}), the infinitesimal
shift of noise 
\begin{equation}
\xi_{r}\left(t_{p+1}\right)=\xi_{r}\left(t_{p}\right)-idtM_{pr}^{*}\overline{s}\left(\tau_{p};\xi\right)
\end{equation}
 will generate the infinitesimal Bogoliubov transform 
\begin{equation}
S\left(d\xi\right)=1+idt\overline{s}^{*}\left(\tau_{p};\xi\right)\sum_{r=0}^{p-1}M_{pr}\widehat{\psi}_{r},
\end{equation}
which should be incorporated into the Hamiltonian. The latter for
the continuously shifting noise becomes
\begin{multline}
\widehat{H}_{\textrm{st}}^{\prime}\left(\xi,\tau_{p}\right)=\widehat{H}_{\textrm{s}}+\widehat{s}\left\{ \widehat{\psi}_{p}^{\dagger}+\xi_{p}^{*}+f_{p}^{*}\right\} \\
+\left(\widehat{s}^{\dagger}-\overline{s}^{*}\left(\tau_{p};\xi\right)\right)\left\{ M\left(0\right)\widehat{\psi}_{p}+\sum_{i=1}^{\textrm{min}\left(p,m\right)}M_{i}\left(p\right)\widehat{\phi}_{i}\right\} ,\label{eq:shifting_noise_hamiltonian}
\end{multline}
which is the eq. (\ref{eq:shifting_noise_hamiltonian-1}). Here $\xi_{p}^{*}$
are sampled from the vacuum probability distribution, and the self-consistent
displacement is
\begin{equation}
f_{p}=-i\intop_{0}^{t_{p}}dt\overline{s}\left(t;\xi\right)M\left(\tau_{p}-t\right).
\end{equation}
The consistent midpoint discretization of $f_{p}$ leads to equation
(\ref{eq:midpoint_shift}). The Hamiltonian $\widehat{H}_{\textrm{st}}^{\prime}\left(\xi,\tau_{p}\right)$
should be employed in the entangling step eq. (\ref{eq:rg_pure_hamiltonian}).
The rest of RG procedure is the same except that the average of any
system observable $\widehat{o}$ should be computed as
\begin{equation}
\left\langle \widehat{o}\left(t_{p}\right)\right\rangle =\overline{\left[\frac{\textrm{Tr}\left\{ \widehat{o}\times\ps{}{\textrm{rel}}{\left\langle 0\right.}\left|\Phi\left(\xi,t_{p}\right)\right\rangle _{\textrm{rel}}\ps{}{\textrm{rel}}{\left\langle \Phi\left(\xi,t_{p}\right)\right|\left.0\right\rangle _{\textrm{rel}}}\right\} }{\left\Vert \ps{}{\textrm{rel}}{\left\langle 0\right.}\left|\Phi\left(\xi,t_{p}\right)\right\rangle _{\textrm{rel}}\right\Vert ^{2}}\right]}_{\xi}.
\end{equation}
The reason is that now we directly sample the probability distribution
(\ref{eq:target_probability}), hence the denominator.


\begin{thebibliography}{89}%
\makeatletter
\providecommand \@ifxundefined [1]{%
 \@ifx{#1\undefined}
}%
\providecommand \@ifnum [1]{%
 \ifnum #1\expandafter \@firstoftwo
 \else \expandafter \@secondoftwo
 \fi
}%
\providecommand \@ifx [1]{%
 \ifx #1\expandafter \@firstoftwo
 \else \expandafter \@secondoftwo
 \fi
}%
\providecommand \natexlab [1]{#1}%
\providecommand \enquote  [1]{``#1''}%
\providecommand \bibnamefont  [1]{#1}%
\providecommand \bibfnamefont [1]{#1}%
\providecommand \citenamefont [1]{#1}%
\providecommand \href@noop [0]{\@secondoftwo}%
\providecommand \href [0]{\begingroup \@sanitize@url \@href}%
\providecommand \@href[1]{\@@startlink{#1}\@@href}%
\providecommand \@@href[1]{\endgroup#1\@@endlink}%
\providecommand \@sanitize@url [0]{\catcode `\\12\catcode `\$12\catcode
  `\&12\catcode `\#12\catcode `\^12\catcode `\_12\catcode `\%12\relax}%
\providecommand \@@startlink[1]{}%
\providecommand \@@endlink[0]{}%
\providecommand \url  [0]{\begingroup\@sanitize@url \@url }%
\providecommand \@url [1]{\endgroup\@href {#1}{\urlprefix }}%
\providecommand \urlprefix  [0]{URL }%
\providecommand \Eprint [0]{\href }%
\providecommand \doibase [0]{https://doi.org/}%
\providecommand \selectlanguage [0]{\@gobble}%
\providecommand \bibinfo  [0]{\@secondoftwo}%
\providecommand \bibfield  [0]{\@secondoftwo}%
\providecommand \translation [1]{[#1]}%
\providecommand \BibitemOpen [0]{}%
\providecommand \bibitemStop [0]{}%
\providecommand \bibitemNoStop [0]{.\EOS\space}%
\providecommand \EOS [0]{\spacefactor3000\relax}%
\providecommand \BibitemShut  [1]{\csname bibitem#1\endcsname}%
\let\auto@bib@innerbib\@empty
\bibitem [{\citenamefont {Feynman}\ \emph {et~al.}(2010)\citenamefont
  {Feynman}, \citenamefont {Hibbs},\ and\ \citenamefont {Styer}}]{Feynman2010}%
  \BibitemOpen
  \bibfield  {author} {\bibinfo {author} {\bibfnamefont {R.~P.}\ \bibnamefont
  {Feynman}}, \bibinfo {author} {\bibfnamefont {A.~R.}\ \bibnamefont {Hibbs}},\
  and\ \bibinfo {author} {\bibfnamefont {D.~F.}\ \bibnamefont {Styer}},\
  }\href@noop {} {\emph {\bibinfo {title} {Quantum Mechanics and Path
  Integrals: Emended Edition}}}\ (\bibinfo  {publisher} {Dover Publications},\
  \bibinfo {year} {2010})\BibitemShut {NoStop}%
\bibitem [{\citenamefont {Dirac}(1982)}]{Dirac1982}%
  \BibitemOpen
  \bibfield  {author} {\bibinfo {author} {\bibfnamefont {P.~A.~M.}\
  \bibnamefont {Dirac}},\ }\href@noop {} {\emph {\bibinfo {title} {The
  Principles of Quantum Mechanics}}}\ (\bibinfo  {publisher} {Clarendon
  Press},\ \bibinfo {year} {1982})\BibitemShut {NoStop}%
\bibitem [{\citenamefont {Poulin}\ \emph {et~al.}(2011)\citenamefont {Poulin},
  \citenamefont {Quarry}, \citenamefont {Somma},\ and\ \citenamefont
  {Verstraete}}]{Poulin2011}%
  \BibitemOpen
  \bibfield  {author} {\bibinfo {author} {\bibfnamefont {D.}~\bibnamefont
  {Poulin}}, \bibinfo {author} {\bibfnamefont {A.}~\bibnamefont {Quarry}},
  \bibinfo {author} {\bibfnamefont {R.}~\bibnamefont {Somma}},\ and\ \bibinfo
  {author} {\bibfnamefont {F.}~\bibnamefont {Verstraete}},\ }\bibfield  {title}
  {\bibinfo {title} {Quantum simulation of time-dependent hamiltonians and the
  convenient illusion of hilbert space},\ }\href@noop {} {\bibfield  {journal}
  {\bibinfo  {journal} {Phys. Rev. Lett.}\ }\textbf {\bibinfo {volume} {106}},\
  \bibinfo {pages} {170501} (\bibinfo {year} {2011})}\BibitemShut {NoStop}%
\bibitem [{\citenamefont {Brandao}\ \emph {et~al.}(2019)\citenamefont
  {Brandao}, \citenamefont {Chemissany}, \citenamefont {Hunter-Jones},
  \citenamefont {Kueng},\ and\ \citenamefont {Preskill}}]{Brandao2019}%
  \BibitemOpen
  \bibfield  {author} {\bibinfo {author} {\bibfnamefont {F.~G.}\ \bibnamefont
  {Brandao}}, \bibinfo {author} {\bibfnamefont {W.}~\bibnamefont {Chemissany}},
  \bibinfo {author} {\bibfnamefont {N.}~\bibnamefont {Hunter-Jones}}, \bibinfo
  {author} {\bibfnamefont {R.}~\bibnamefont {Kueng}},\ and\ \bibinfo {author}
  {\bibfnamefont {J.}~\bibnamefont {Preskill}},\ }\href@noop {} {\bibinfo
  {title} {Models of quantum complexity growth}} (\bibinfo {year} {2019}),\
  \Eprint {https://arxiv.org/abs/1912.04297} {arXiv:1912.04297 [hep-th]}
  \BibitemShut {NoStop}%
\bibitem [{\citenamefont {Lin}\ \emph {et~al.}(2021)\citenamefont {Lin},
  \citenamefont {Dilip}, \citenamefont {Green}, \citenamefont {Smith},\ and\
  \citenamefont {Pollmann}}]{Lin2021}%
  \BibitemOpen
  \bibfield  {author} {\bibinfo {author} {\bibfnamefont {S.-H.}\ \bibnamefont
  {Lin}}, \bibinfo {author} {\bibfnamefont {R.}~\bibnamefont {Dilip}}, \bibinfo
  {author} {\bibfnamefont {A.~G.}\ \bibnamefont {Green}}, \bibinfo {author}
  {\bibfnamefont {A.}~\bibnamefont {Smith}},\ and\ \bibinfo {author}
  {\bibfnamefont {F.}~\bibnamefont {Pollmann}},\ }\bibfield  {title} {\bibinfo
  {title} {Real- and imaginary-time evolution with compressed quantum
  circuits},\ }\href@noop {} {\bibfield  {journal} {\bibinfo  {journal} {PRX
  Quantum}\ }\textbf {\bibinfo {volume} {2}},\ \bibinfo {pages} {010342}
  (\bibinfo {year} {2021})}\BibitemShut {NoStop}%
\bibitem [{\citenamefont {Zhou}\ \emph {et~al.}(2020)\citenamefont {Zhou},
  \citenamefont {Stoundenmire},\ and\ \citenamefont {Waintal}}]{Zhou2020}%
  \BibitemOpen
  \bibfield  {author} {\bibinfo {author} {\bibfnamefont {Y.}~\bibnamefont
  {Zhou}}, \bibinfo {author} {\bibfnamefont {E.~M.}\ \bibnamefont
  {Stoundenmire}},\ and\ \bibinfo {author} {\bibfnamefont {X.}~\bibnamefont
  {Waintal}},\ }\bibfield  {title} {\bibinfo {title} {What limits the
  simulation of quantum computers?},\ }\href@noop {} {\bibfield  {journal}
  {\bibinfo  {journal} {Phys. Rev. X}\ }\textbf {\bibinfo {volume} {10}},\
  \bibinfo {pages} {041038} (\bibinfo {year} {2020})}\BibitemShut {NoStop}%
\bibitem [{\citenamefont {Franca}\ and\ \citenamefont
  {Garcia-Patron}(2020)}]{Franca2020}%
  \BibitemOpen
  \bibfield  {author} {\bibinfo {author} {\bibfnamefont {D.~S.}\ \bibnamefont
  {Franca}}\ and\ \bibinfo {author} {\bibfnamefont {R.}~\bibnamefont
  {Garcia-Patron}},\ }\href@noop {} {\bibinfo {title} {Limitations of
  optimization algorithms on noisy quantum devices}} (\bibinfo {year} {2020}),\
  \Eprint {https://arxiv.org/abs/2009.05532} {arXiv:2009.05532 [quant-ph]}
  \BibitemShut {NoStop}%
\bibitem [{\citenamefont {Vidal}(2007)}]{Vidal2007}%
  \BibitemOpen
  \bibfield  {author} {\bibinfo {author} {\bibfnamefont {G.}~\bibnamefont
  {Vidal}},\ }\bibfield  {title} {\bibinfo {title} {Entanglement
  renormalizatiom},\ }\href@noop {} {\bibfield  {journal} {\bibinfo  {journal}
  {Phys. Rev. Lett.}\ }\textbf {\bibinfo {volume} {99}},\ \bibinfo {pages}
  {220405} (\bibinfo {year} {2007})}\BibitemShut {NoStop}%
\bibitem [{\citenamefont {Vidal}(2008)}]{Vidal2008}%
  \BibitemOpen
  \bibfield  {author} {\bibinfo {author} {\bibfnamefont {G.}~\bibnamefont
  {Vidal}},\ }\bibfield  {title} {\bibinfo {title} {Class of quantum many-body
  states that can be efficiently simulated},\ }\href@noop {} {\bibfield
  {journal} {\bibinfo  {journal} {Phys. Rev. Lett.}\ }\textbf {\bibinfo
  {volume} {101}},\ \bibinfo {pages} {110501} (\bibinfo {year}
  {2008})}\BibitemShut {NoStop}%
\bibitem [{\citenamefont {Acoleyen}\ \emph {et~al.}(2020)\citenamefont
  {Acoleyen}, \citenamefont {Hallam}, \citenamefont {Bal}, \citenamefont
  {Hauru}, \citenamefont {Haegeman},\ and\ \citenamefont
  {Verstraete}}]{Acoleyen2020}%
  \BibitemOpen
  \bibfield  {author} {\bibinfo {author} {\bibfnamefont {K.~V.}\ \bibnamefont
  {Acoleyen}}, \bibinfo {author} {\bibfnamefont {A.}~\bibnamefont {Hallam}},
  \bibinfo {author} {\bibfnamefont {M.}~\bibnamefont {Bal}}, \bibinfo {author}
  {\bibfnamefont {M.}~\bibnamefont {Hauru}}, \bibinfo {author} {\bibfnamefont
  {J.}~\bibnamefont {Haegeman}},\ and\ \bibinfo {author} {\bibfnamefont
  {F.}~\bibnamefont {Verstraete}},\ }\bibfield  {title} {\bibinfo {title}
  {Entanglement compression in scale space: From the multiscale entanglement
  renormalization ansatz to matrix product operators},\ }\href@noop {}
  {\bibfield  {journal} {\bibinfo  {journal} {Phys. Rev. B}\ }\textbf {\bibinfo
  {volume} {102}},\ \bibinfo {pages} {165131} (\bibinfo {year}
  {2020})}\BibitemShut {NoStop}%
\bibitem [{\citenamefont {Haegeman}\ \emph {et~al.}(2013)\citenamefont
  {Haegeman}, \citenamefont {Osborne}, \citenamefont {Verschelde},\ and\
  \citenamefont {Verstraete}}]{Haegeman2013}%
  \BibitemOpen
  \bibfield  {author} {\bibinfo {author} {\bibfnamefont {J.}~\bibnamefont
  {Haegeman}}, \bibinfo {author} {\bibfnamefont {T.~J.}\ \bibnamefont
  {Osborne}}, \bibinfo {author} {\bibfnamefont {H.}~\bibnamefont
  {Verschelde}},\ and\ \bibinfo {author} {\bibfnamefont {F.}~\bibnamefont
  {Verstraete}},\ }\bibfield  {title} {\bibinfo {title} {Entanglement
  renormalization for quantum fields in real space},\ }\href@noop {} {\bibfield
   {journal} {\bibinfo  {journal} {Phys. Rev. Lett.}\ }\textbf {\bibinfo
  {volume} {110}},\ \bibinfo {pages} {100402} (\bibinfo {year}
  {2013})}\BibitemShut {NoStop}%
\bibitem [{\citenamefont {White}(1992)}]{White1992}%
  \BibitemOpen
  \bibfield  {author} {\bibinfo {author} {\bibfnamefont {S.~R.}\ \bibnamefont
  {White}},\ }\bibfield  {title} {\bibinfo {title} {Density matrix formulation
  for quantum renormalization groups},\ }\href@noop {} {\bibfield  {journal}
  {\bibinfo  {journal} {Phys. Rev. Lett.}\ }\textbf {\bibinfo {volume} {69}},\
  \bibinfo {pages} {2863} (\bibinfo {year} {1992})}\BibitemShut {NoStop}%
\bibitem [{\citenamefont {Schollwock}(2011)}]{Schollwock2011}%
  \BibitemOpen
  \bibfield  {author} {\bibinfo {author} {\bibfnamefont {U.}~\bibnamefont
  {Schollwock}},\ }\bibfield  {title} {\bibinfo {title} {The density-matrix
  renormalization group in the age of matrix product states},\ }\href@noop {}
  {\bibfield  {journal} {\bibinfo  {journal} {Ann. Phys.}\ }\textbf {\bibinfo
  {volume} {326}},\ \bibinfo {pages} {96} (\bibinfo {year} {2011})}\BibitemShut
  {NoStop}%
\bibitem [{\citenamefont {Vega}\ and\ \citenamefont {Alonso}(2017)}]{Vega2017}%
  \BibitemOpen
  \bibfield  {author} {\bibinfo {author} {\bibfnamefont {I.~d.}\ \bibnamefont
  {Vega}}\ and\ \bibinfo {author} {\bibfnamefont {D.}~\bibnamefont {Alonso}},\
  }\bibfield  {title} {\bibinfo {title} {Dynamics of non-markovian open quantum
  systems},\ }\href@noop {} {\bibfield  {journal} {\bibinfo  {journal} {Rev.
  Mod. Phys.}\ }\textbf {\bibinfo {volume} {89}},\ \bibinfo {pages} {015001}
  (\bibinfo {year} {2017})}\BibitemShut {NoStop}%
\bibitem [{\citenamefont {Brandes}(2010)}]{Brandes2010}%
  \BibitemOpen
  \bibfield  {author} {\bibinfo {author} {\bibfnamefont {T.}~\bibnamefont
  {Brandes}},\ }\bibfield  {title} {\bibinfo {title} {Feedback control of
  quantum transport},\ }\href@noop {} {\bibfield  {journal} {\bibinfo
  {journal} {Phys. Rev. Lett.}\ }\textbf {\bibinfo {volume} {105}},\ \bibinfo
  {pages} {060602} (\bibinfo {year} {2010})}\BibitemShut {NoStop}%
\bibitem [{\citenamefont {Kiesslich}\ \emph {et~al.}(2012)\citenamefont
  {Kiesslich}, \citenamefont {Emary}, \citenamefont {Schaller},\ and\
  \citenamefont {Brandes}}]{Kiesslich2012}%
  \BibitemOpen
  \bibfield  {author} {\bibinfo {author} {\bibfnamefont {G.}~\bibnamefont
  {Kiesslich}}, \bibinfo {author} {\bibfnamefont {C.}~\bibnamefont {Emary}},
  \bibinfo {author} {\bibfnamefont {G.}~\bibnamefont {Schaller}},\ and\
  \bibinfo {author} {\bibfnamefont {T.}~\bibnamefont {Brandes}},\ }\bibfield
  {title} {\bibinfo {title} {Reverse quantum state engineering using electronic
  feedback loops},\ }\href@noop {} {\bibfield  {journal} {\bibinfo  {journal}
  {New J. Phys.}\ }\textbf {\bibinfo {volume} {14}},\ \bibinfo {pages} {123036}
  (\bibinfo {year} {2012})}\BibitemShut {NoStop}%
\bibitem [{\citenamefont {Gough}(2014)}]{Gough2014}%
  \BibitemOpen
  \bibfield  {author} {\bibinfo {author} {\bibfnamefont {J.}~\bibnamefont
  {Gough}},\ }\bibfield  {title} {\bibinfo {title} {Feedback network models for
  quantum transport},\ }\href@noop {} {\bibfield  {journal} {\bibinfo
  {journal} {Phys. Rev. E}\ }\textbf {\bibinfo {volume} {90}},\ \bibinfo
  {pages} {062109} (\bibinfo {year} {2014})}\BibitemShut {NoStop}%
\bibitem [{\citenamefont {Brandes}\ and\ \citenamefont
  {Emary}(2016)}]{Brandes2016}%
  \BibitemOpen
  \bibfield  {author} {\bibinfo {author} {\bibfnamefont {T.}~\bibnamefont
  {Brandes}}\ and\ \bibinfo {author} {\bibfnamefont {C.}~\bibnamefont
  {Emary}},\ }\bibfield  {title} {\bibinfo {title} {Feedback control of waiting
  times},\ }\href@noop {} {\bibfield  {journal} {\bibinfo  {journal} {Phys.
  Rev. E}\ }\textbf {\bibinfo {volume} {93}},\ \bibinfo {pages} {042103}
  (\bibinfo {year} {2016})}\BibitemShut {NoStop}%
\bibitem [{\citenamefont {Luo}\ \emph {et~al.}(2016)\citenamefont {Luo},
  \citenamefont {Jin}, \citenamefont {Wang}, \citenamefont {Hu}, \citenamefont
  {Huang},\ and\ \citenamefont {He}}]{Luo2016}%
  \BibitemOpen
  \bibfield  {author} {\bibinfo {author} {\bibfnamefont {J.~Y.}\ \bibnamefont
  {Luo}}, \bibinfo {author} {\bibfnamefont {J.}~\bibnamefont {Jin}}, \bibinfo
  {author} {\bibfnamefont {S.-K.}\ \bibnamefont {Wang}}, \bibinfo {author}
  {\bibfnamefont {J.}~\bibnamefont {Hu}}, \bibinfo {author} {\bibfnamefont
  {Y.}~\bibnamefont {Huang}},\ and\ \bibinfo {author} {\bibfnamefont {X.-L.}\
  \bibnamefont {He}},\ }\bibfield  {title} {\bibinfo {title} {Unraveling of a
  detailed-balance-preserved quantum master equation and continuous feedback
  control of a measured qubit},\ }\href@noop {} {\bibfield  {journal} {\bibinfo
   {journal} {Phys. Rev. B}\ }\textbf {\bibinfo {volume} {93}},\ \bibinfo
  {pages} {125122} (\bibinfo {year} {2016})}\BibitemShut {NoStop}%
\bibitem [{\citenamefont {Wagner}\ \emph {et~al.}(2017)\citenamefont {Wagner},
  \citenamefont {Strasberg}, \citenamefont {Bayer}, \citenamefont
  {Rugeramigabo}, \citenamefont {Brandes},\ and\ \citenamefont
  {Haug}}]{Wagner2016}%
  \BibitemOpen
  \bibfield  {author} {\bibinfo {author} {\bibfnamefont {T.}~\bibnamefont
  {Wagner}}, \bibinfo {author} {\bibfnamefont {P.}~\bibnamefont {Strasberg}},
  \bibinfo {author} {\bibfnamefont {J.~C.}\ \bibnamefont {Bayer}}, \bibinfo
  {author} {\bibfnamefont {E.~P.}\ \bibnamefont {Rugeramigabo}}, \bibinfo
  {author} {\bibfnamefont {T.}~\bibnamefont {Brandes}},\ and\ \bibinfo {author}
  {\bibfnamefont {R.~J.}\ \bibnamefont {Haug}},\ }\bibfield  {title} {\bibinfo
  {title} {Strong suppression of shot noise in a feedback-controlled
  single-electron transistor},\ }\href@noop {} {\bibfield  {journal} {\bibinfo
  {journal} {Nat. Nanotechnol.}\ }\textbf {\bibinfo {volume} {12}},\ \bibinfo
  {pages} {218} (\bibinfo {year} {2017})}\BibitemShut {NoStop}%
\bibitem [{\citenamefont {Kondo}(1984)}]{Kondo1984}%
  \BibitemOpen
  \bibfield  {author} {\bibinfo {author} {\bibfnamefont {J.}~\bibnamefont
  {Kondo}},\ }\bibfield  {title} {\bibinfo {title} {Diffusion of light
  interstitials in metals},\ }\href@noop {} {\bibfield  {journal} {\bibinfo
  {journal} {Physica B+C}\ }\textbf {\bibinfo {volume} {125}},\ \bibinfo
  {pages} {279} (\bibinfo {year} {1984})}\BibitemShut {NoStop}%
\bibitem [{\citenamefont {Wilson}(1975)}]{Wilson1975}%
  \BibitemOpen
  \bibfield  {author} {\bibinfo {author} {\bibfnamefont {K.~G.}\ \bibnamefont
  {Wilson}},\ }\bibfield  {title} {\bibinfo {title} {The renormalization group:
  Critical phenomena and the kondo problem},\ }\href@noop {} {\bibfield
  {journal} {\bibinfo  {journal} {Rev. Mod. Phys.}\ }\textbf {\bibinfo {volume}
  {47}},\ \bibinfo {pages} {773} (\bibinfo {year} {1975})}\BibitemShut
  {NoStop}%
\bibitem [{\citenamefont {Bulla}\ and\ \citenamefont
  {Vojta}(2003)}]{Bulla2003}%
  \BibitemOpen
  \bibfield  {author} {\bibinfo {author} {\bibfnamefont {L.}~\bibnamefont
  {Bulla}}\ and\ \bibinfo {author} {\bibfnamefont {M.}~\bibnamefont {Vojta}},\
  }\bibinfo {title} {Quantum phase transitions in models of magnetic
  impurities},\ in\ \href@noop {} {\emph {\bibinfo {booktitle} {Concepts in
  Electron Correlations. NATO Science Series (Series II: Mathematics, Physics
  and Chemistry)}}},\ Vol.\ \bibinfo {volume} {110},\ \bibinfo {editor} {edited
  by\ \bibinfo {editor} {\bibfnamefont {A.~C.}\ \bibnamefont {Hewson}}\ and\
  \bibinfo {editor} {\bibfnamefont {V.}~\bibnamefont {Zlatic}}}\ (\bibinfo
  {publisher} {Springer, Dordrecht},\ \bibinfo {year} {2003})\ pp.\ \bibinfo
  {pages} {209--217}\BibitemShut {NoStop}%
\bibitem [{\citenamefont {Bulla}\ \emph {et~al.}(2008)\citenamefont {Bulla},
  \citenamefont {Costi},\ and\ \citenamefont {Pruschke}}]{Bulla2008}%
  \BibitemOpen
  \bibfield  {author} {\bibinfo {author} {\bibfnamefont {R.}~\bibnamefont
  {Bulla}}, \bibinfo {author} {\bibfnamefont {T.~A.}\ \bibnamefont {Costi}},\
  and\ \bibinfo {author} {\bibfnamefont {T.}~\bibnamefont {Pruschke}},\
  }\bibfield  {title} {\bibinfo {title} {Numerical renormalization group method
  for quantum impurity systems},\ }\href@noop {} {\bibfield  {journal}
  {\bibinfo  {journal} {Rev. Mod. Phys.}\ }\textbf {\bibinfo {volume} {80}},\
  \bibinfo {pages} {395} (\bibinfo {year} {2008})}\BibitemShut {NoStop}%
\bibitem [{\citenamefont {Lazarou}\ \emph {et~al.}(2012)\citenamefont
  {Lazarou}, \citenamefont {Luoma}, \citenamefont {Maniscalco}, \citenamefont
  {Piilo},\ and\ \citenamefont {Garraway}}]{Lazarou2012}%
  \BibitemOpen
  \bibfield  {author} {\bibinfo {author} {\bibfnamefont {C.}~\bibnamefont
  {Lazarou}}, \bibinfo {author} {\bibfnamefont {K.}~\bibnamefont {Luoma}},
  \bibinfo {author} {\bibfnamefont {S.}~\bibnamefont {Maniscalco}}, \bibinfo
  {author} {\bibfnamefont {J.}~\bibnamefont {Piilo}},\ and\ \bibinfo {author}
  {\bibfnamefont {B.~M.}\ \bibnamefont {Garraway}},\ }\bibfield  {title}
  {\bibinfo {title} {Entanglement trapping in a nonstationary structured
  reservoir},\ }\href@noop {} {\bibfield  {journal} {\bibinfo  {journal} {Phys.
  Rev. A}\ }\textbf {\bibinfo {volume} {86}},\ \bibinfo {pages} {012331}
  (\bibinfo {year} {2012})}\BibitemShut {NoStop}%
\bibitem [{\citenamefont {Rams}\ and\ \citenamefont {Zwolak}(2020)}]{Rams2020}%
  \BibitemOpen
  \bibfield  {author} {\bibinfo {author} {\bibfnamefont {M.~M.}\ \bibnamefont
  {Rams}}\ and\ \bibinfo {author} {\bibfnamefont {M.}~\bibnamefont {Zwolak}},\
  }\bibfield  {title} {\bibinfo {title} {Breaking the entanglement barrier:
  Tensor network simulation of quantum transport},\ }\href@noop {} {\bibfield
  {journal} {\bibinfo  {journal} {Phys. Rev. Lett.}\ }\textbf {\bibinfo
  {volume} {127}},\ \bibinfo {pages} {137701} (\bibinfo {year}
  {2020})}\BibitemShut {NoStop}%
\bibitem [{\citenamefont {Polyakov}\ and\ \citenamefont
  {Rubtsov}(2019)}]{Polyakov2019}%
  \BibitemOpen
  \bibfield  {author} {\bibinfo {author} {\bibfnamefont {E.~A.}\ \bibnamefont
  {Polyakov}}\ and\ \bibinfo {author} {\bibfnamefont {A.~N.}\ \bibnamefont
  {Rubtsov}},\ }\bibfield  {title} {\bibinfo {title} {Dressed quantum
  trajectories: novel approach to the non-markovian dynamics of open quantum
  systems on a wide time scale},\ }\href@noop {} {\bibfield  {journal}
  {\bibinfo  {journal} {New. J. Phys.}\ }\textbf {\bibinfo {volume} {21}},\
  \bibinfo {pages} {063004} (\bibinfo {year} {2019})}\BibitemShut {NoStop}%
\bibitem [{\citenamefont {Strathearn}\ \emph
  {et~al.}(2018{\natexlab{a}})\citenamefont {Strathearn}, \citenamefont
  {Kirton}, \citenamefont {Kilda}, \citenamefont {Keeling},\ and\ \citenamefont
  {Lovett}}]{Strathearn2018}%
  \BibitemOpen
  \bibfield  {author} {\bibinfo {author} {\bibfnamefont {A.}~\bibnamefont
  {Strathearn}}, \bibinfo {author} {\bibfnamefont {P.}~\bibnamefont {Kirton}},
  \bibinfo {author} {\bibfnamefont {D.}~\bibnamefont {Kilda}}, \bibinfo
  {author} {\bibfnamefont {J.}~\bibnamefont {Keeling}},\ and\ \bibinfo {author}
  {\bibfnamefont {B.~W.}\ \bibnamefont {Lovett}},\ }\bibfield  {title}
  {\bibinfo {title} {Efficient non-markovian quantum dynamics using
  time-evolving matrix product operators},\ }\href@noop {} {\bibfield
  {journal} {\bibinfo  {journal} {Nat. Commun.}\ }\textbf {\bibinfo {volume}
  {9}},\ \bibinfo {pages} {3322} (\bibinfo {year}
  {2018}{\natexlab{a}})}\BibitemShut {NoStop}%
\bibitem [{\citenamefont {Eisert}\ \emph {et~al.}(2010)\citenamefont {Eisert},
  \citenamefont {Cramer},\ and\ \citenamefont {Plenio}}]{Eisert2010}%
  \BibitemOpen
  \bibfield  {author} {\bibinfo {author} {\bibfnamefont {J.}~\bibnamefont
  {Eisert}}, \bibinfo {author} {\bibfnamefont {M.}~\bibnamefont {Cramer}},\
  and\ \bibinfo {author} {\bibfnamefont {M.~B.}\ \bibnamefont {Plenio}},\
  }\bibfield  {title} {\bibinfo {title} {Colloquium: Area laws for the
  entanglement entropy},\ }\href@noop {} {\bibfield  {journal} {\bibinfo
  {journal} {Rev. Mod. Phys.}\ }\textbf {\bibinfo {volume} {82}},\ \bibinfo
  {pages} {277} (\bibinfo {year} {2010})}\BibitemShut {NoStop}%
\bibitem [{\citenamefont {Breuer}\ and\ \citenamefont
  {Petruccione}(2007)}]{Breuer2011}%
  \BibitemOpen
  \bibfield  {author} {\bibinfo {author} {\bibfnamefont {H.-P.}\ \bibnamefont
  {Breuer}}\ and\ \bibinfo {author} {\bibfnamefont {F.}~\bibnamefont
  {Petruccione}},\ }\href@noop {} {\emph {\bibinfo {title} {The Theory of Open
  Quantum Systems}}}\ (\bibinfo  {publisher} {Oxford University Press},\
  \bibinfo {address} {Oxford},\ \bibinfo {year} {2007})\BibitemShut {NoStop}%
\bibitem [{\citenamefont {Percival}(1999)}]{Percival1999}%
  \BibitemOpen
  \bibfield  {author} {\bibinfo {author} {\bibfnamefont {I.}~\bibnamefont
  {Percival}},\ }\href@noop {} {\emph {\bibinfo {title} {Quantum State
  Diffusion}}}\ (\bibinfo  {publisher} {Cambridge University Press},\ \bibinfo
  {address} {Cambridge},\ \bibinfo {year} {1999})\BibitemShut {NoStop}%
\bibitem [{\citenamefont {Wiseman}(1996)}]{Wiseman1996}%
  \BibitemOpen
  \bibfield  {author} {\bibinfo {author} {\bibfnamefont {H.~M.}\ \bibnamefont
  {Wiseman}},\ }\bibfield  {title} {\bibinfo {title} {Quantum trajectories and
  quantum measurement theory},\ }\href@noop {} {\bibfield  {journal} {\bibinfo
  {journal} {Quantum Semiclass. Opt.}\ }\textbf {\bibinfo {volume} {8}},\
  \bibinfo {pages} {205} (\bibinfo {year} {1996})}\BibitemShut {NoStop}%
\bibitem [{\citenamefont {Wiseman}\ and\ \citenamefont
  {Milburn}(1993)}]{Wiseman1993}%
  \BibitemOpen
  \bibfield  {author} {\bibinfo {author} {\bibfnamefont {H.~M.}\ \bibnamefont
  {Wiseman}}\ and\ \bibinfo {author} {\bibfnamefont {G.~J.}\ \bibnamefont
  {Milburn}},\ }\bibfield  {title} {\bibinfo {title} {Quantum theory of
  field-quadrature measurements},\ }\href@noop {} {\bibfield  {journal}
  {\bibinfo  {journal} {Phys. Rev. A}\ }\textbf {\bibinfo {volume} {47}},\
  \bibinfo {pages} {642} (\bibinfo {year} {1993})}\BibitemShut {NoStop}%
\bibitem [{\citenamefont {Daley}(2014)}]{Daley2014}%
  \BibitemOpen
  \bibfield  {author} {\bibinfo {author} {\bibfnamefont {A.~J.}\ \bibnamefont
  {Daley}},\ }\bibfield  {title} {\bibinfo {title} {Quantum trajectories and
  open many-body quantum systems},\ }\href@noop {} {\bibfield  {journal}
  {\bibinfo  {journal} {Adv. in Phys.}\ }\textbf {\bibinfo {volume} {63}},\
  \bibinfo {pages} {77} (\bibinfo {year} {2014})}\BibitemShut {NoStop}%
\bibitem [{\citenamefont {Segal}\ \emph {et~al.}(2010)\citenamefont {Segal},
  \citenamefont {Millis},\ and\ \citenamefont {Reichman}}]{Segal2010}%
  \BibitemOpen
  \bibfield  {author} {\bibinfo {author} {\bibfnamefont {D.}~\bibnamefont
  {Segal}}, \bibinfo {author} {\bibfnamefont {A.~J.}\ \bibnamefont {Millis}},\
  and\ \bibinfo {author} {\bibfnamefont {D.~R.}\ \bibnamefont {Reichman}},\
  }\bibfield  {title} {\bibinfo {title} {Numerically exact path-integral
  simulation of nonequilibrium quantum transport and dissipation},\ }\href
  {https://doi.org/10.1103/PhysRevB.82.205323} {\bibfield  {journal} {\bibinfo
  {journal} {Phys. Rev. B}\ }\textbf {\bibinfo {volume} {82}},\ \bibinfo
  {pages} {205323} (\bibinfo {year} {2010})}\BibitemShut {NoStop}%
\bibitem [{\citenamefont {Makri}(1995)}]{Makri1995}%
  \BibitemOpen
  \bibfield  {author} {\bibinfo {author} {\bibfnamefont {N.}~\bibnamefont
  {Makri}},\ }\bibfield  {title} {\bibinfo {title} {Numerical path integral
  techniques for long time dynamics of quantum dissipative systems},\ }\href
  {https://doi.org/10.1063/1.531046} {\bibfield  {journal} {\bibinfo  {journal}
  {Journal of Mathematical Physics}\ }\textbf {\bibinfo {volume} {36}},\
  \bibinfo {pages} {2430} (\bibinfo {year} {1995})},\ \Eprint
  {https://arxiv.org/abs/https://doi.org/10.1063/1.531046}
  {https://doi.org/10.1063/1.531046} \BibitemShut {NoStop}%
\bibitem [{\citenamefont {Makri}\ and\ \citenamefont
  {Makarov}(1995)}]{Makri1995a}%
  \BibitemOpen
  \bibfield  {author} {\bibinfo {author} {\bibfnamefont {N.}~\bibnamefont
  {Makri}}\ and\ \bibinfo {author} {\bibfnamefont {D.~E.}\ \bibnamefont
  {Makarov}},\ }\bibfield  {title} {\bibinfo {title} {Tensor propagator for
  iterative quantum time evolution of reduced density matrices. i. theory},\
  }\href {https://doi.org/10.1063/1.469508} {\bibfield  {journal} {\bibinfo
  {journal} {The Journal of Chemical Physics}\ }\textbf {\bibinfo {volume}
  {102}},\ \bibinfo {pages} {4600} (\bibinfo {year} {1995})},\ \Eprint
  {https://arxiv.org/abs/https://doi.org/10.1063/1.469508}
  {https://doi.org/10.1063/1.469508} \BibitemShut {NoStop}%
\bibitem [{\citenamefont {Makri}\ \emph {et~al.}(1996)\citenamefont {Makri},
  \citenamefont {Sim}, \citenamefont {Makarov},\ and\ \citenamefont
  {Topaler}}]{Makri1996}%
  \BibitemOpen
  \bibfield  {author} {\bibinfo {author} {\bibfnamefont {N.}~\bibnamefont
  {Makri}}, \bibinfo {author} {\bibfnamefont {E.}~\bibnamefont {Sim}}, \bibinfo
  {author} {\bibfnamefont {D.~E.}\ \bibnamefont {Makarov}},\ and\ \bibinfo
  {author} {\bibfnamefont {M.}~\bibnamefont {Topaler}},\ }\bibfield  {title}
  {\bibinfo {title} {Long-time quantum simulation of the primary charge
  separation in bacterial photosynthesis},\ }\href
  {http://www.pnas.org/content/93/9/3926.abstract} {\bibfield  {journal}
  {\bibinfo  {journal} {Proceedings of the National Academy of Sciences}\
  }\textbf {\bibinfo {volume} {93}},\ \bibinfo {pages} {3926} (\bibinfo {year}
  {1996})},\ \Eprint
  {https://arxiv.org/abs/http://www.pnas.org/content/93/9/3926.full.pdf}
  {http://www.pnas.org/content/93/9/3926.full.pdf} \BibitemShut {NoStop}%
\bibitem [{\citenamefont {Makarov}\ and\ \citenamefont
  {Makri}(1994)}]{Makarov1994}%
  \BibitemOpen
  \bibfield  {author} {\bibinfo {author} {\bibfnamefont {D.~E.}\ \bibnamefont
  {Makarov}}\ and\ \bibinfo {author} {\bibfnamefont {N.}~\bibnamefont
  {Makri}},\ }\bibfield  {title} {\bibinfo {title} {Path integrals for
  dissipative systems by tensor multiplication. condensed phase quantum
  dynamics for arbitrarily long time},\ }\href
  {https://doi.org/https://doi.org/10.1016/0009-2614(94)00275-4} {\bibfield
  {journal} {\bibinfo  {journal} {Chemical Physics Letters}\ }\textbf {\bibinfo
  {volume} {221}},\ \bibinfo {pages} {482 } (\bibinfo {year}
  {1994})}\BibitemShut {NoStop}%
\bibitem [{\citenamefont {Richter}\ and\ \citenamefont
  {Fingerhut}(2017)}]{Richter2017}%
  \BibitemOpen
  \bibfield  {author} {\bibinfo {author} {\bibfnamefont {M.}~\bibnamefont
  {Richter}}\ and\ \bibinfo {author} {\bibfnamefont {B.~P.}\ \bibnamefont
  {Fingerhut}},\ }\bibfield  {title} {\bibinfo {title} {Coarse-grained
  representation of the quasi-adiabatic propagator path integral for the
  treatment of non-markovian long-time bath memory},\ }\href@noop {} {\bibfield
   {journal} {\bibinfo  {journal} {J. Chem. Phys.}\ }\textbf {\bibinfo {volume}
  {146}},\ \bibinfo {pages} {214101} (\bibinfo {year} {2017})}\BibitemShut
  {NoStop}%
\bibitem [{\citenamefont {Woods}\ \emph {et~al.}(2014)\citenamefont {Woods},
  \citenamefont {Groux}, \citenamefont {Chin}, \citenamefont {Huelga},\ and\
  \citenamefont {Plenio}}]{Woods2014}%
  \BibitemOpen
  \bibfield  {author} {\bibinfo {author} {\bibfnamefont {M.~P.}\ \bibnamefont
  {Woods}}, \bibinfo {author} {\bibfnamefont {R.}~\bibnamefont {Groux}},
  \bibinfo {author} {\bibfnamefont {A.~W.}\ \bibnamefont {Chin}}, \bibinfo
  {author} {\bibfnamefont {S.~F.}\ \bibnamefont {Huelga}},\ and\ \bibinfo
  {author} {\bibfnamefont {M.~B.}\ \bibnamefont {Plenio}},\ }\bibfield  {title}
  {\bibinfo {title} {Mappings of open quantum systems onto chain
  representations and markovian embeddings},\ }\href@noop {} {\bibfield
  {journal} {\bibinfo  {journal} {J. Math. Phys.}\ }\textbf {\bibinfo {volume}
  {55}},\ \bibinfo {pages} {032101} (\bibinfo {year} {2014})}\BibitemShut
  {NoStop}%
\bibitem [{\citenamefont {Mazzola}\ \emph {et~al.}(2009)\citenamefont
  {Mazzola}, \citenamefont {Maniscalco}, \citenamefont {Piilo}, \citenamefont
  {Suominen},\ and\ \citenamefont {Garraway}}]{Mazzola2009}%
  \BibitemOpen
  \bibfield  {author} {\bibinfo {author} {\bibfnamefont {L.}~\bibnamefont
  {Mazzola}}, \bibinfo {author} {\bibfnamefont {S.}~\bibnamefont {Maniscalco}},
  \bibinfo {author} {\bibfnamefont {J.}~\bibnamefont {Piilo}}, \bibinfo
  {author} {\bibfnamefont {K.-A.}\ \bibnamefont {Suominen}},\ and\ \bibinfo
  {author} {\bibfnamefont {B.~M.}\ \bibnamefont {Garraway}},\ }\bibfield
  {title} {\bibinfo {title} {Pseudomodes as an effective description of memory:
  Non-markovian dynamics of two-state systems in structured reservoirs},\
  }\href@noop {} {\bibfield  {journal} {\bibinfo  {journal} {Phys. Rev. A}\
  }\textbf {\bibinfo {volume} {80}},\ \bibinfo {pages} {012104} (\bibinfo
  {year} {2009})}\BibitemShut {NoStop}%
\bibitem [{\citenamefont {Lambert}\ \emph {et~al.}(2019)\citenamefont
  {Lambert}, \citenamefont {Ahmed}, \citenamefont {Cirio},\ and\ \citenamefont
  {Nori}}]{Lambert2019}%
  \BibitemOpen
  \bibfield  {author} {\bibinfo {author} {\bibfnamefont {N.}~\bibnamefont
  {Lambert}}, \bibinfo {author} {\bibfnamefont {S.}~\bibnamefont {Ahmed}},
  \bibinfo {author} {\bibfnamefont {M.}~\bibnamefont {Cirio}},\ and\ \bibinfo
  {author} {\bibfnamefont {F.}~\bibnamefont {Nori}},\ }\bibfield  {title}
  {\bibinfo {title} {Modelling the ultra-strongly coupled spin-boson model with
  unphysical modes},\ }\href@noop {} {\bibfield  {journal} {\bibinfo  {journal}
  {Nat. Commun.}\ }\textbf {\bibinfo {volume} {10}},\ \bibinfo {pages} {3721}
  (\bibinfo {year} {2019})}\BibitemShut {NoStop}%
\bibitem [{\citenamefont {Tamascelli}\ \emph {et~al.}(2018)\citenamefont
  {Tamascelli}, \citenamefont {Smirne}, \citenamefont {Huelga},\ and\
  \citenamefont {Plenio}}]{Tamascelli2018}%
  \BibitemOpen
  \bibfield  {author} {\bibinfo {author} {\bibfnamefont {D.}~\bibnamefont
  {Tamascelli}}, \bibinfo {author} {\bibfnamefont {A.}~\bibnamefont {Smirne}},
  \bibinfo {author} {\bibfnamefont {S.~F.}\ \bibnamefont {Huelga}},\ and\
  \bibinfo {author} {\bibfnamefont {M.~B.}\ \bibnamefont {Plenio}},\ }\bibfield
   {title} {\bibinfo {title} {Nonperturbative treatment of non-markovian
  dynamics of open quantum systems},\ }\href@noop {} {\bibfield  {journal}
  {\bibinfo  {journal} {Phys. Rev. Lett.}\ }\textbf {\bibinfo {volume} {120}},\
  \bibinfo {pages} {030402} (\bibinfo {year} {2018})}\BibitemShut {NoStop}%
\bibitem [{\citenamefont {Jorgensen}\ and\ \citenamefont
  {Pollock}(2019)}]{Jorgensen2019}%
  \BibitemOpen
  \bibfield  {author} {\bibinfo {author} {\bibfnamefont {M.~R.}\ \bibnamefont
  {Jorgensen}}\ and\ \bibinfo {author} {\bibfnamefont {F.~A.}\ \bibnamefont
  {Pollock}},\ }\bibfield  {title} {\bibinfo {title} {Exploiting the causal
  tensor network structure of quantum processes to efficiently simulate
  non-markovian path integrals},\ }\href@noop {} {\bibfield  {journal}
  {\bibinfo  {journal} {Phys. Rev. Lett.}\ }\textbf {\bibinfo {volume} {123}},\
  \bibinfo {pages} {240602} (\bibinfo {year} {2019})}\BibitemShut {NoStop}%
\bibitem [{\citenamefont {Luchnikov}\ \emph {et~al.}(2019)\citenamefont
  {Luchnikov}, \citenamefont {Vintskevich}, \citenamefont {Querdane},\ and\
  \citenamefont {Filippov}}]{Luchnikov2019}%
  \BibitemOpen
  \bibfield  {author} {\bibinfo {author} {\bibfnamefont {I.~A.}\ \bibnamefont
  {Luchnikov}}, \bibinfo {author} {\bibfnamefont {S.~V.}\ \bibnamefont
  {Vintskevich}}, \bibinfo {author} {\bibfnamefont {H.}~\bibnamefont
  {Querdane}},\ and\ \bibinfo {author} {\bibfnamefont {S.~N.}\ \bibnamefont
  {Filippov}},\ }\bibfield  {title} {\bibinfo {title} {Simulation complexity of
  open quantum dynamics: Connection with tensor networks},\ }\href@noop {}
  {\bibfield  {journal} {\bibinfo  {journal} {Phys. Rev. Lett.}\ }\textbf
  {\bibinfo {volume} {122}},\ \bibinfo {pages} {160401} (\bibinfo {year}
  {2019})}\BibitemShut {NoStop}%
\bibitem [{\citenamefont {Somoza}\ \emph {et~al.}(2019)\citenamefont {Somoza},
  \citenamefont {Marty}, \citenamefont {Lin}, \citenamefont {Huelga},\ and\
  \citenamefont {Plenio}}]{Somoza2019}%
  \BibitemOpen
  \bibfield  {author} {\bibinfo {author} {\bibfnamefont {A.~D.}\ \bibnamefont
  {Somoza}}, \bibinfo {author} {\bibfnamefont {O.}~\bibnamefont {Marty}},
  \bibinfo {author} {\bibfnamefont {J.}~\bibnamefont {Lin}}, \bibinfo {author}
  {\bibfnamefont {S.~F.}\ \bibnamefont {Huelga}},\ and\ \bibinfo {author}
  {\bibfnamefont {M.~B.}\ \bibnamefont {Plenio}},\ }\bibfield  {title}
  {\bibinfo {title} {Dissipation-assisted matrix product factorization},\
  }\href@noop {} {\bibfield  {journal} {\bibinfo  {journal} {Phys. Rev. Lett.}\
  }\textbf {\bibinfo {volume} {123}},\ \bibinfo {pages} {100502} (\bibinfo
  {year} {2019})}\BibitemShut {NoStop}%
\bibitem [{\citenamefont {Zhu}\ \emph {et~al.}(2020)\citenamefont {Zhu},
  \citenamefont {Huang}, \citenamefont {He},\ and\ \citenamefont
  {Wen}}]{Zhu2020}%
  \BibitemOpen
  \bibfield  {author} {\bibinfo {author} {\bibfnamefont {W.}~\bibnamefont
  {Zhu}}, \bibinfo {author} {\bibfnamefont {Z.}~\bibnamefont {Huang}}, \bibinfo
  {author} {\bibfnamefont {Y.-C.}\ \bibnamefont {He}},\ and\ \bibinfo {author}
  {\bibfnamefont {X.}~\bibnamefont {Wen}},\ }\bibfield  {title} {\bibinfo
  {title} {Entanglement hamiltonian of many-body dynamics in strongly
  correlated systems},\ }\href@noop {} {\bibfield  {journal} {\bibinfo
  {journal} {Phys. Rev. Lett.}\ }\textbf {\bibinfo {volume} {2020}},\ \bibinfo
  {pages} {124} (\bibinfo {year} {2020})}\BibitemShut {NoStop}%
\bibitem [{\citenamefont {Diosi}\ \emph {et~al.}(1998)\citenamefont {Diosi},
  \citenamefont {Gisin},\ and\ \citenamefont {Strunz}}]{Diosi1998}%
  \BibitemOpen
  \bibfield  {author} {\bibinfo {author} {\bibfnamefont {L.}~\bibnamefont
  {Diosi}}, \bibinfo {author} {\bibfnamefont {N.}~\bibnamefont {Gisin}},\ and\
  \bibinfo {author} {\bibfnamefont {W.~T.}\ \bibnamefont {Strunz}},\ }\bibfield
   {title} {\bibinfo {title} {Non-markovian quantum state diffusion},\
  }\href@noop {} {\bibfield  {journal} {\bibinfo  {journal} {Phys.l Rev. A}\
  }\textbf {\bibinfo {volume} {58}},\ \bibinfo {pages} {1699} (\bibinfo {year}
  {1998})}\BibitemShut {NoStop}%
\bibitem [{\citenamefont {Diosi}\ and\ \citenamefont
  {Strunz}(1997)}]{Diosi1997}%
  \BibitemOpen
  \bibfield  {author} {\bibinfo {author} {\bibfnamefont {L.}~\bibnamefont
  {Diosi}}\ and\ \bibinfo {author} {\bibfnamefont {W.~T.}\ \bibnamefont
  {Strunz}},\ }\bibfield  {title} {\bibinfo {title} {The non-markovian
  stochastic schrijdinger equation for open systems},\ }\href@noop {}
  {\bibfield  {journal} {\bibinfo  {journal} {Phys. Lett. A}\ }\textbf
  {\bibinfo {volume} {235}},\ \bibinfo {pages} {569} (\bibinfo {year}
  {1997})}\BibitemShut {NoStop}%
\bibitem [{\citenamefont {Zhou}\ and\ \citenamefont {Shao}(2008)}]{Shao2008}%
  \BibitemOpen
  \bibfield  {author} {\bibinfo {author} {\bibfnamefont {Y.}~\bibnamefont
  {Zhou}}\ and\ \bibinfo {author} {\bibfnamefont {J.}~\bibnamefont {Shao}},\
  }\bibfield  {title} {\bibinfo {title} {Solving the spin-boson model of strong
  dissipation with flexible random-deterministic scheme},\ }\href@noop {}
  {\bibfield  {journal} {\bibinfo  {journal} {J. Chem. Phys.}\ }\textbf
  {\bibinfo {volume} {128}},\ \bibinfo {pages} {034106} (\bibinfo {year}
  {2008})}\BibitemShut {NoStop}%
\bibitem [{\citenamefont {Yan}\ and\ \citenamefont {Shao}(2016)}]{Yan2016}%
  \BibitemOpen
  \bibfield  {author} {\bibinfo {author} {\bibfnamefont {Y.-A.}\ \bibnamefont
  {Yan}}\ and\ \bibinfo {author} {\bibfnamefont {J.}~\bibnamefont {Shao}},\
  }\bibfield  {title} {\bibinfo {title} {Stochastic description of quantum
  brownian dynamics},\ }\href@noop {} {\bibfield  {journal} {\bibinfo
  {journal} {Front. Phys.}\ }\textbf {\bibinfo {volume} {11}},\ \bibinfo
  {pages} {110309} (\bibinfo {year} {2016})}\BibitemShut {NoStop}%
\bibitem [{\citenamefont {Shao}(2004)}]{Shao2004}%
  \BibitemOpen
  \bibfield  {author} {\bibinfo {author} {\bibfnamefont {J.}~\bibnamefont
  {Shao}},\ }\bibfield  {title} {\bibinfo {title} {Decoupling quantum
  dissipation via stochastic fields},\ }\href@noop {} {\bibfield  {journal}
  {\bibinfo  {journal} {J. Chem. Phys.}\ }\textbf {\bibinfo {volume} {120}},\
  \bibinfo {pages} {5053} (\bibinfo {year} {2004})}\BibitemShut {NoStop}%
\bibitem [{\citenamefont {Han}\ \emph {et~al.}(2019)\citenamefont {Han},
  \citenamefont {Chemyak}, \citenamefont {Yan}, \citenamefont {Zheng},\ and\
  \citenamefont {Yan}}]{Han2019}%
  \BibitemOpen
  \bibfield  {author} {\bibinfo {author} {\bibfnamefont {L.}~\bibnamefont
  {Han}}, \bibinfo {author} {\bibfnamefont {V.}~\bibnamefont {Chemyak}},
  \bibinfo {author} {\bibfnamefont {Y.-A.}\ \bibnamefont {Yan}}, \bibinfo
  {author} {\bibfnamefont {X.}~\bibnamefont {Zheng}},\ and\ \bibinfo {author}
  {\bibfnamefont {Y.}~\bibnamefont {Yan}},\ }\bibfield  {title} {\bibinfo
  {title} {Stochastic representation of non-markovian fermionic quantum
  dissipation},\ }\href@noop {} {\bibfield  {journal} {\bibinfo  {journal}
  {Phys. Rev. Lett.}\ }\textbf {\bibinfo {volume} {123}},\ \bibinfo {pages}
  {050601} (\bibinfo {year} {2019})}\BibitemShut {NoStop}%
\bibitem [{\citenamefont {Suess}\ \emph {et~al.}(2014)\citenamefont {Suess},
  \citenamefont {Eisfeld},\ and\ \citenamefont {Strunz}}]{Suess2014}%
  \BibitemOpen
  \bibfield  {author} {\bibinfo {author} {\bibfnamefont {D.}~\bibnamefont
  {Suess}}, \bibinfo {author} {\bibfnamefont {A.}~\bibnamefont {Eisfeld}},\
  and\ \bibinfo {author} {\bibfnamefont {W.~T.}\ \bibnamefont {Strunz}},\
  }\bibfield  {title} {\bibinfo {title} {Hierarchy of stochastic pure states
  for open quantum system dynamics},\ }\href@noop {} {\bibfield  {journal}
  {\bibinfo  {journal} {Phys. Rev. Lett.}\ }\textbf {\bibinfo {volume} {113}},\
  \bibinfo {pages} {150403} (\bibinfo {year} {2014})}\BibitemShut {NoStop}%
\bibitem [{\citenamefont {Hartmann}\ and\ \citenamefont
  {Strunz}(2017)}]{Hartmann2017}%
  \BibitemOpen
  \bibfield  {author} {\bibinfo {author} {\bibfnamefont {R.}~\bibnamefont
  {Hartmann}}\ and\ \bibinfo {author} {\bibfnamefont {W.~T.}\ \bibnamefont
  {Strunz}},\ }\bibfield  {title} {\bibinfo {title} {Exact open quantum system
  dynamics using the hierarchy of pure states (hops)},\ }\href@noop {}
  {\bibfield  {journal} {\bibinfo  {journal} {J. Chem. Theor. Comput.}\
  }\textbf {\bibinfo {volume} {13}},\ \bibinfo {pages} {5834} (\bibinfo {year}
  {2017})}\BibitemShut {NoStop}%
\bibitem [{\citenamefont {Moix}\ and\ \citenamefont {Cao}(2013)}]{Moix2013}%
  \BibitemOpen
  \bibfield  {author} {\bibinfo {author} {\bibfnamefont {J.~M.}\ \bibnamefont
  {Moix}}\ and\ \bibinfo {author} {\bibfnamefont {J.}~\bibnamefont {Cao}},\
  }\bibfield  {title} {\bibinfo {title} {A hybrid stochastic hierarchy
  equations of motion approach to treat the low temperature dynamics of
  non-markovian open quantum systems},\ }\href@noop {} {\bibfield  {journal}
  {\bibinfo  {journal} {J. Chem. Phys}\ }\textbf {\bibinfo {volume} {139}},\
  \bibinfo {pages} {134106} (\bibinfo {year} {2013})}\BibitemShut {NoStop}%
\bibitem [{\citenamefont {Strümpfer}\ and\ \citenamefont
  {Schulten}(2012)}]{Strumpfer2012}%
  \BibitemOpen
  \bibfield  {author} {\bibinfo {author} {\bibfnamefont {J.}~\bibnamefont
  {Strümpfer}}\ and\ \bibinfo {author} {\bibfnamefont {K.}~\bibnamefont
  {Schulten}},\ }\bibfield  {title} {\bibinfo {title} {Open quantum dynamics
  calculations with the hierarchy equations of motion on parallel computers},\
  }\href {https://doi.org/10.1021/ct3003833} {\bibfield  {journal} {\bibinfo
  {journal} {Journal of Chemical Theory and Computation}\ }\textbf {\bibinfo
  {volume} {8}},\ \bibinfo {pages} {2808} (\bibinfo {year} {2012})},\ \bibinfo
  {note} {pMID: 23105920},\ \Eprint
  {https://arxiv.org/abs/http://dx.doi.org/10.1021/ct3003833}
  {http://dx.doi.org/10.1021/ct3003833} \BibitemShut {NoStop}%
\bibitem [{\citenamefont {Ishizaki}\ and\ \citenamefont
  {Fleming}(2009)}]{Ishizaki2009}%
  \BibitemOpen
  \bibfield  {author} {\bibinfo {author} {\bibfnamefont {A.}~\bibnamefont
  {Ishizaki}}\ and\ \bibinfo {author} {\bibfnamefont {G.~R.}\ \bibnamefont
  {Fleming}},\ }\bibfield  {title} {\bibinfo {title} {Unified treatment of
  quantum coherent and incoherent hopping dynamics in electronic energy
  transfer: Reduced hierarchy equation approach},\ }\href
  {https://doi.org/10.1063/1.3155372} {\bibfield  {journal} {\bibinfo
  {journal} {The Journal of Chemical Physics}\ }\textbf {\bibinfo {volume}
  {130}},\ \bibinfo {pages} {234111} (\bibinfo {year} {2009})},\ \Eprint
  {https://arxiv.org/abs/https://doi.org/10.1063/1.3155372}
  {https://doi.org/10.1063/1.3155372} \BibitemShut {NoStop}%
\bibitem [{\citenamefont {Duan}\ \emph {et~al.}(2017)\citenamefont {Duan},
  \citenamefont {Tang}, \citenamefont {Cao},\ and\ \citenamefont
  {Wu}}]{Duan2017}%
  \BibitemOpen
  \bibfield  {author} {\bibinfo {author} {\bibfnamefont {C.}~\bibnamefont
  {Duan}}, \bibinfo {author} {\bibfnamefont {Z.}~\bibnamefont {Tang}}, \bibinfo
  {author} {\bibfnamefont {J.}~\bibnamefont {Cao}},\ and\ \bibinfo {author}
  {\bibfnamefont {J.}~\bibnamefont {Wu}},\ }\bibfield  {title} {\bibinfo
  {title} {Zero-temperature localization in a sub-ohmic spin-boson model
  investigated by an extended hierarchy equation of motion},\ }\href@noop {}
  {\bibfield  {journal} {\bibinfo  {journal} {Phys. Rev. B}\ }\textbf {\bibinfo
  {volume} {95}},\ \bibinfo {pages} {214308} (\bibinfo {year}
  {2017})}\BibitemShut {NoStop}%
\bibitem [{\citenamefont {Rahman}\ and\ \citenamefont
  {Kleinekathofer}(2019)}]{Rahman2019}%
  \BibitemOpen
  \bibfield  {author} {\bibinfo {author} {\bibfnamefont {H.}~\bibnamefont
  {Rahman}}\ and\ \bibinfo {author} {\bibfnamefont {U.}~\bibnamefont
  {Kleinekathofer}},\ }\bibfield  {title} {\bibinfo {title} {Chebyshev
  hierarchical equations of motion for systems with arbitrary spectral
  densities and temperatures},\ }\href@noop {} {\bibfield  {journal} {\bibinfo
  {journal} {J. Chem. Phys.}\ }\textbf {\bibinfo {volume} {150}},\ \bibinfo
  {pages} {244104} (\bibinfo {year} {2019})}\BibitemShut {NoStop}%
\bibitem [{\citenamefont {Polyakov}\ and\ \citenamefont
  {Rubtsov}(2018{\natexlab{a}})}]{Polyakov2017a}%
  \BibitemOpen
  \bibfield  {author} {\bibinfo {author} {\bibfnamefont {E.~A.}\ \bibnamefont
  {Polyakov}}\ and\ \bibinfo {author} {\bibfnamefont {A.~N.}\ \bibnamefont
  {Rubtsov}},\ }\bibfield  {title} {\bibinfo {title} {Stochastic wave-function
  simulation of irreversible emission processes for open quantum systems in a
  non-markovian environment},\ }\href@noop {} {\bibfield  {journal} {\bibinfo
  {journal} {AIP Conf. Proc.}\ }\textbf {\bibinfo {volume} {1936}},\ \bibinfo
  {pages} {020028} (\bibinfo {year} {2018}{\natexlab{a}})}\BibitemShut
  {NoStop}%
\bibitem [{\citenamefont {Polyakov}\ and\ \citenamefont
  {Rubtsov}(2018{\natexlab{b}})}]{Polyakov2018b}%
  \BibitemOpen
  \bibfield  {author} {\bibinfo {author} {\bibfnamefont {E.~A.}\ \bibnamefont
  {Polyakov}}\ and\ \bibinfo {author} {\bibfnamefont {A.~N.}\ \bibnamefont
  {Rubtsov}},\ }\bibfield  {title} {\bibinfo {title} {Information loss pathways
  in a nuremically exact simulation of a non-markovian open quantum system}}
  (\bibinfo {year} {2018}{\natexlab{b}}),\ \bibinfo {note} {to be
  published}\BibitemShut {NoStop}%
\bibitem [{\citenamefont {Diosi}(2012)}]{Diosi2012}%
  \BibitemOpen
  \bibfield  {author} {\bibinfo {author} {\bibfnamefont {L.}~\bibnamefont
  {Diosi}},\ }\bibfield  {title} {\bibinfo {title} {Non-markovian open quantum
  systems: Input-output fields, memory, and monitoring},\ }\href@noop {}
  {\bibfield  {journal} {\bibinfo  {journal} {Phys. Rev. A}\ }\textbf {\bibinfo
  {volume} {85}},\ \bibinfo {pages} {034101} (\bibinfo {year}
  {2012})}\BibitemShut {NoStop}%
\bibitem [{\citenamefont {Atland}\ and\ \citenamefont
  {Simons}(2010)}]{Atland2010}%
  \BibitemOpen
  \bibfield  {author} {\bibinfo {author} {\bibfnamefont {A.}~\bibnamefont
  {Atland}}\ and\ \bibinfo {author} {\bibfnamefont {B.~D.}\ \bibnamefont
  {Simons}},\ }\href@noop {} {\emph {\bibinfo {title} {Condensed Matter Field
  Theory}}}\ (\bibinfo  {publisher} {Cambridge University Press},\ \bibinfo
  {address} {New York},\ \bibinfo {year} {2010})\BibitemShut {NoStop}%
\bibitem [{\citenamefont {Plenio}\ and\ \citenamefont
  {Knight}(1998)}]{Plenio1998}%
  \BibitemOpen
  \bibfield  {author} {\bibinfo {author} {\bibfnamefont {M.~B.}\ \bibnamefont
  {Plenio}}\ and\ \bibinfo {author} {\bibfnamefont {P.~L.}\ \bibnamefont
  {Knight}},\ }\bibfield  {title} {\bibinfo {title} {The quantum-jump approach
  to dissipative dynamics in quantum optics},\ }\href@noop {} {\bibfield
  {journal} {\bibinfo  {journal} {Rev. Mod. Phys.}\ }\textbf {\bibinfo {volume}
  {70}},\ \bibinfo {pages} {101} (\bibinfo {year} {1998})}\BibitemShut
  {NoStop}%
\bibitem [{\citenamefont {Aharonov}\ and\ \citenamefont
  {Vaidman}(2008)}]{Aharonov2008}%
  \BibitemOpen
  \bibfield  {author} {\bibinfo {author} {\bibfnamefont {Y.}~\bibnamefont
  {Aharonov}}\ and\ \bibinfo {author} {\bibfnamefont {L.}~\bibnamefont
  {Vaidman}},\ }\bibfield  {title} {\bibinfo {title} {The two-state vector
  formalism: An updated review},\ }in\ \href@noop {} {\emph {\bibinfo
  {booktitle} {Time in Quantum Mechanics}}},\ \bibinfo {series} {Lecture Notes
  in Physics}, Vol.\ \bibinfo {volume} {734},\ \bibinfo {editor} {edited by\
  \bibinfo {editor} {\bibfnamefont {J.}~\bibnamefont {Muga}}, \bibinfo {editor}
  {\bibfnamefont {R.~S.}\ \bibnamefont {Mayato}},\ and\ \bibinfo {editor}
  {\bibfnamefont {I.}~\bibnamefont {Egusquiza}}}\ (\bibinfo  {publisher}
  {Springer},\ \bibinfo {address} {Berlin},\ \bibinfo {year} {2008})\
  Chap.~\bibinfo {chapter} {13}, pp.\ \bibinfo {pages} {399--447}\BibitemShut
  {NoStop}%
\bibitem [{Aha(2014)}]{Aharonov2014}%
  \BibitemOpen
  \href@noop {} {\emph {\bibinfo {title} {Quantum Theory: A Two-Time Success
  Story}}}\ (\bibinfo  {publisher} {Springer, Milano},\ \bibinfo {year}
  {2014})\BibitemShut {NoStop}%
\bibitem [{\citenamefont {Aharonov}\ \emph {et~al.}(2017)\citenamefont
  {Aharonov}, \citenamefont {Cohen},\ and\ \citenamefont
  {Landsberger}}]{Aharonov2017}%
  \BibitemOpen
  \bibfield  {author} {\bibinfo {author} {\bibfnamefont {Y.}~\bibnamefont
  {Aharonov}}, \bibinfo {author} {\bibfnamefont {E.}~\bibnamefont {Cohen}},\
  and\ \bibinfo {author} {\bibfnamefont {T.}~\bibnamefont {Landsberger}},\
  }\bibfield  {title} {\bibinfo {title} {The two-time interpretation and
  macroscopic time-reversibility},\ }\href@noop {} {\bibfield  {journal}
  {\bibinfo  {journal} {Entropy}\ }\textbf {\bibinfo {volume} {19}},\ \bibinfo
  {pages} {111} (\bibinfo {year} {2017})}\BibitemShut {NoStop}%
\bibitem [{\citenamefont {Gardiner}\ and\ \citenamefont
  {Zoller}(2004)}]{Gardiner2004}%
  \BibitemOpen
  \bibfield  {author} {\bibinfo {author} {\bibfnamefont {C.~W.}\ \bibnamefont
  {Gardiner}}\ and\ \bibinfo {author} {\bibfnamefont {P.}~\bibnamefont
  {Zoller}},\ }\href@noop {} {\emph {\bibinfo {title} {Quantum Noise}}},\
  \bibinfo {edition} {3rd}\ ed.\ (\bibinfo  {publisher} {Springer-Verlag Berlin
  Heidelberg},\ \bibinfo {year} {2004})\BibitemShut {NoStop}%
\bibitem [{\citenamefont {Plenio}\ and\ \citenamefont
  {Virmani}(2007)}]{Plenio2007}%
  \BibitemOpen
  \bibfield  {author} {\bibinfo {author} {\bibfnamefont {M.~B.}\ \bibnamefont
  {Plenio}}\ and\ \bibinfo {author} {\bibfnamefont {S.}~\bibnamefont
  {Virmani}},\ }\bibfield  {title} {\bibinfo {title} {An introduction to the
  entanglement measures},\ }\href@noop {} {\bibfield  {journal} {\bibinfo
  {journal} {Quantum Info. Comput.}\ }\textbf {\bibinfo {volume} {7}},\
  \bibinfo {pages} {1} (\bibinfo {year} {2007})}\BibitemShut {NoStop}%
\bibitem [{\citenamefont {Chin}\ \emph {et~al.}(2010)\citenamefont {Chin},
  \citenamefont {Rivas}, \citenamefont {Huelga},\ and\ \citenamefont
  {Plenio}}]{Chin2010}%
  \BibitemOpen
  \bibfield  {author} {\bibinfo {author} {\bibfnamefont {A.~W.}\ \bibnamefont
  {Chin}}, \bibinfo {author} {\bibfnamefont {A.}~\bibnamefont {Rivas}},
  \bibinfo {author} {\bibfnamefont {S.~F.}\ \bibnamefont {Huelga}},\ and\
  \bibinfo {author} {\bibfnamefont {M.~B.}\ \bibnamefont {Plenio}},\ }\bibfield
   {title} {\bibinfo {title} {Exact mapping between system-reservoir quantum
  models and semi-infinite discrete chains using orthogonal polynomials},\
  }\href@noop {} {\bibfield  {journal} {\bibinfo  {journal} {J. Math. Phys.}\
  }\textbf {\bibinfo {volume} {51}},\ \bibinfo {pages} {092109} (\bibinfo
  {year} {2010})}\BibitemShut {NoStop}%
\bibitem [{\citenamefont {Breuer}\ \emph {et~al.}(2009)\citenamefont {Breuer},
  \citenamefont {Laine},\ and\ \citenamefont {Piilo}}]{Breuer2009}%
  \BibitemOpen
  \bibfield  {author} {\bibinfo {author} {\bibfnamefont {H.-P.}\ \bibnamefont
  {Breuer}}, \bibinfo {author} {\bibfnamefont {E.-M.}\ \bibnamefont {Laine}},\
  and\ \bibinfo {author} {\bibfnamefont {J.}~\bibnamefont {Piilo}},\ }\bibfield
   {title} {\bibinfo {title} {Measure for the degree of non-markoviani
  behaviour of quantum processes in open systems},\ }\href@noop {} {\bibfield
  {journal} {\bibinfo  {journal} {Phys. Rev. Lett.}\ }\textbf {\bibinfo
  {volume} {103}},\ \bibinfo {pages} {210401} (\bibinfo {year}
  {2009})}\BibitemShut {NoStop}%
\bibitem [{\citenamefont {Bylicka}\ \emph {et~al.}(2017)\citenamefont
  {Bylicka}, \citenamefont {Johansson},\ and\ \citenamefont
  {Acin}}]{Bylicka2017}%
  \BibitemOpen
  \bibfield  {author} {\bibinfo {author} {\bibfnamefont {B.}~\bibnamefont
  {Bylicka}}, \bibinfo {author} {\bibfnamefont {M.}~\bibnamefont {Johansson}},\
  and\ \bibinfo {author} {\bibfnamefont {A.}~\bibnamefont {Acin}},\ }\bibfield
  {title} {\bibinfo {title} {Constructive method for detecting the information
  backflow of non-markovian dynamics},\ }\href@noop {} {\bibfield  {journal}
  {\bibinfo  {journal} {Phys. Rev. Lett}\ }\textbf {\bibinfo {volume} {118}},\
  \bibinfo {pages} {120501} (\bibinfo {year} {2017})}\BibitemShut {NoStop}%
\bibitem [{\citenamefont {Buscemi}\ and\ \citenamefont
  {Datta}(2016)}]{Buscemi2016}%
  \BibitemOpen
  \bibfield  {author} {\bibinfo {author} {\bibfnamefont {F.}~\bibnamefont
  {Buscemi}}\ and\ \bibinfo {author} {\bibfnamefont {N.}~\bibnamefont
  {Datta}},\ }\bibfield  {title} {\bibinfo {title} {Equivalence between
  divisibility and monotonic decrease of information in classical and quantum
  stochastic processes},\ }\href@noop {} {\bibfield  {journal} {\bibinfo
  {journal} {Phys. Rev. A}\ }\textbf {\bibinfo {volume} {93}},\ \bibinfo
  {pages} {012101} (\bibinfo {year} {2016})}\BibitemShut {NoStop}%
\bibitem [{\citenamefont {Ruskai}(2009)}]{Ruskai1994}%
  \BibitemOpen
  \bibfield  {author} {\bibinfo {author} {\bibfnamefont {M.~B.}\ \bibnamefont
  {Ruskai}},\ }\bibfield  {title} {\bibinfo {title} {Beyond strong
  subadditivity? improved bounds on the contraction of generalized relative
  entropy},\ }\href@noop {} {\bibfield  {journal} {\bibinfo  {journal} {Rev.
  Math. Phys.}\ }\textbf {\bibinfo {volume} {6}},\ \bibinfo {pages} {1147}
  (\bibinfo {year} {2009})}\BibitemShut {NoStop}%
\bibitem [{\citenamefont {Popovic}\ \emph {et~al.}(2021)\citenamefont
  {Popovic}, \citenamefont {Mitchison}, \citenamefont {Strathearn},
  \citenamefont {Lovett}, \citenamefont {Goold},\ and\ \citenamefont
  {Eastham}}]{Popovic2021}%
  \BibitemOpen
  \bibfield  {author} {\bibinfo {author} {\bibfnamefont {M.}~\bibnamefont
  {Popovic}}, \bibinfo {author} {\bibfnamefont {M.~T.}\ \bibnamefont
  {Mitchison}}, \bibinfo {author} {\bibfnamefont {A.}~\bibnamefont
  {Strathearn}}, \bibinfo {author} {\bibfnamefont {B.~W.}\ \bibnamefont
  {Lovett}}, \bibinfo {author} {\bibfnamefont {J.}~\bibnamefont {Goold}},\ and\
  \bibinfo {author} {\bibfnamefont {P.~R.}\ \bibnamefont {Eastham}},\
  }\bibfield  {title} {\bibinfo {title} {Quantum heat statistics with
  time-evolving matrix product operators},\ }\href@noop {} {\bibfield
  {journal} {\bibinfo  {journal} {PRX Quantum}\ }\textbf {\bibinfo {volume}
  {2}},\ \bibinfo {pages} {020338} (\bibinfo {year} {2021})}\BibitemShut
  {NoStop}%
\bibitem [{\citenamefont {Strathearn}\ \emph
  {et~al.}(2018{\natexlab{b}})\citenamefont {Strathearn}, \citenamefont
  {Kirton}, \citenamefont {Kilda}, \citenamefont {Keeling},\ and\ \citenamefont
  {Lovett}}]{Strathearn2018b}%
  \BibitemOpen
  \bibfield  {author} {\bibinfo {author} {\bibfnamefont {A.}~\bibnamefont
  {Strathearn}}, \bibinfo {author} {\bibfnamefont {P.}~\bibnamefont {Kirton}},
  \bibinfo {author} {\bibfnamefont {D.}~\bibnamefont {Kilda}}, \bibinfo
  {author} {\bibfnamefont {J.}~\bibnamefont {Keeling}},\ and\ \bibinfo {author}
  {\bibfnamefont {B.~W.}\ \bibnamefont {Lovett}},\ }\href@noop {} {\bibinfo
  {title} {Efficient non-markovian quantum dynamics using time-evolving matrix
  product operators}},\ \bibinfo {howpublished}
  {\url{https://doi.org/10.5281/zenodo.1322407}} (\bibinfo {year}
  {2018}{\natexlab{b}})\BibitemShut {NoStop}%
\bibitem [{\citenamefont {Ali}\ \emph {et~al.}(2019)\citenamefont {Ali},
  \citenamefont {Bhattacharyya}, \citenamefont {Haque}, \citenamefont {Kim},\
  and\ \citenamefont {Moynihan}}]{Ali2019}%
  \BibitemOpen
  \bibfield  {author} {\bibinfo {author} {\bibfnamefont {T.}~\bibnamefont
  {Ali}}, \bibinfo {author} {\bibfnamefont {A.}~\bibnamefont {Bhattacharyya}},
  \bibinfo {author} {\bibfnamefont {S.~S.}\ \bibnamefont {Haque}}, \bibinfo
  {author} {\bibfnamefont {E.~H.}\ \bibnamefont {Kim}},\ and\ \bibinfo {author}
  {\bibfnamefont {N.}~\bibnamefont {Moynihan}},\ }\bibfield  {title} {\bibinfo
  {title} {Time evolution of complexity: a critique of three methods},\
  }\href@noop {} {\bibfield  {journal} {\bibinfo  {journal} {JHEP}\ }\textbf
  {\bibinfo {volume} {2019}},\ \bibinfo {pages} {87}}\BibitemShut {NoStop}%
\bibitem [{\citenamefont {Bhattacharyya}\ \emph {et~al.}(2020)\citenamefont
  {Bhattacharyya}, \citenamefont {Nandy},\ and\ \citenamefont
  {Sinha}}]{Bhattacharyya2020}%
  \BibitemOpen
  \bibfield  {author} {\bibinfo {author} {\bibfnamefont {A.}~\bibnamefont
  {Bhattacharyya}}, \bibinfo {author} {\bibfnamefont {P.}~\bibnamefont
  {Nandy}},\ and\ \bibinfo {author} {\bibfnamefont {A.}~\bibnamefont {Sinha}},\
  }\bibfield  {title} {\bibinfo {title} {Renormalized circuit complexity},\
  }\href@noop {} {\bibfield  {journal} {\bibinfo  {journal} {Phys. Rev. Lett.}\
  }\textbf {\bibinfo {volume} {124}},\ \bibinfo {pages} {101602} (\bibinfo
  {year} {2020})}\BibitemShut {NoStop}%
\bibitem [{\citenamefont {Megier}\ \emph {et~al.}(2020)\citenamefont {Megier},
  \citenamefont {Strunz},\ and\ \citenamefont {Luoma}}]{Megier2020}%
  \BibitemOpen
  \bibfield  {author} {\bibinfo {author} {\bibfnamefont {N.}~\bibnamefont
  {Megier}}, \bibinfo {author} {\bibfnamefont {W.~N.}\ \bibnamefont {Strunz}},\
  and\ \bibinfo {author} {\bibfnamefont {K.}~\bibnamefont {Luoma}},\
  }\href@noop {} {\bibinfo {title} {Continuous quantum measurement for general
  gaussian unravelings can exist}} (\bibinfo {year} {2020}),\ \Eprint
  {https://arxiv.org/abs/1912.08662} {arXiv:1912.08662 [quant-ph]} \BibitemShut
  {NoStop}%
\bibitem [{\citenamefont {Adler}\ and\ \citenamefont
  {Bassi}(2007)}]{Adler2007}%
  \BibitemOpen
  \bibfield  {author} {\bibinfo {author} {\bibfnamefont {S.~L.}\ \bibnamefont
  {Adler}}\ and\ \bibinfo {author} {\bibfnamefont {A.}~\bibnamefont {Bassi}},\
  }\bibfield  {title} {\bibinfo {title} {Collapse models with non-white
  noises},\ }\href@noop {} {\bibfield  {journal} {\bibinfo  {journal} {J. Phys.
  A: Math. Theor.}\ }\textbf {\bibinfo {volume} {40}},\ \bibinfo {pages}
  {15083} (\bibinfo {year} {2007})}\BibitemShut {NoStop}%
\bibitem [{\citenamefont {Carlesso}\ \emph {et~al.}(2018)\citenamefont
  {Carlesso}, \citenamefont {Ferialdi},\ and\ \citenamefont
  {Bassi}}]{Carlesso2018}%
  \BibitemOpen
  \bibfield  {author} {\bibinfo {author} {\bibfnamefont {M.}~\bibnamefont
  {Carlesso}}, \bibinfo {author} {\bibfnamefont {L.}~\bibnamefont {Ferialdi}},\
  and\ \bibinfo {author} {\bibfnamefont {A.}~\bibnamefont {Bassi}},\ }\bibfield
   {title} {\bibinfo {title} {Colored collapse models from the
  non-interferometric perspective},\ }\href@noop {} {\bibfield  {journal}
  {\bibinfo  {journal} {EPJ D}\ }\textbf {\bibinfo {volume} {72}},\ \bibinfo
  {pages} {159} (\bibinfo {year} {2018})}\BibitemShut {NoStop}%
\bibitem [{\citenamefont {Strunz}\ \emph
  {et~al.}(1999{\natexlab{a}})\citenamefont {Strunz}, \citenamefont {Diosi},
  \citenamefont {Gisin},\ and\ \citenamefont {Yu}}]{Strunz1999a}%
  \BibitemOpen
  \bibfield  {author} {\bibinfo {author} {\bibfnamefont {W.~T.}\ \bibnamefont
  {Strunz}}, \bibinfo {author} {\bibfnamefont {L.}~\bibnamefont {Diosi}},
  \bibinfo {author} {\bibfnamefont {N.}~\bibnamefont {Gisin}},\ and\ \bibinfo
  {author} {\bibfnamefont {T.}~\bibnamefont {Yu}},\ }\bibfield  {title}
  {\bibinfo {title} {Quantum trajectories for brownian motion},\ }\href@noop {}
  {\bibfield  {journal} {\bibinfo  {journal} {Phys. Rev. Lett.}\ }\textbf
  {\bibinfo {volume} {83}},\ \bibinfo {pages} {4909} (\bibinfo {year}
  {1999}{\natexlab{a}})}\BibitemShut {NoStop}%
\bibitem [{\citenamefont {Strunz}\ \emph
  {et~al.}(1999{\natexlab{b}})\citenamefont {Strunz}, \citenamefont {Diosi},\
  and\ \citenamefont {Gisin}}]{Strunz1999}%
  \BibitemOpen
  \bibfield  {author} {\bibinfo {author} {\bibfnamefont {W.~T.}\ \bibnamefont
  {Strunz}}, \bibinfo {author} {\bibfnamefont {L.}~\bibnamefont {Diosi}},\ and\
  \bibinfo {author} {\bibfnamefont {N.}~\bibnamefont {Gisin}},\ }\bibfield
  {title} {\bibinfo {title} {Open system dynamics with non-markovian quantum
  trajectories},\ }\href@noop {} {\bibfield  {journal} {\bibinfo  {journal}
  {Phys. Rev. A}\ }\textbf {\bibinfo {volume} {82}},\ \bibinfo {pages} {1801}
  (\bibinfo {year} {1999}{\natexlab{b}})}\BibitemShut {NoStop}%
\bibitem [{\citenamefont {Gambetta}\ \emph {et~al.}(2004)\citenamefont
  {Gambetta}, \citenamefont {Askerud},\ and\ \citenamefont
  {Wiseman}}]{Gambetta2004}%
  \BibitemOpen
  \bibfield  {author} {\bibinfo {author} {\bibfnamefont {J.}~\bibnamefont
  {Gambetta}}, \bibinfo {author} {\bibfnamefont {T.}~\bibnamefont {Askerud}},\
  and\ \bibinfo {author} {\bibfnamefont {H.~M.}\ \bibnamefont {Wiseman}},\
  }\bibfield  {title} {\bibinfo {title} {Jumplike unravelings for non-markovian
  open quantum systems},\ }\href@noop {} {\bibfield  {journal} {\bibinfo
  {journal} {Phys. Rev. A}\ }\textbf {\bibinfo {volume} {69}},\ \bibinfo
  {pages} {052104} (\bibinfo {year} {2004})}\BibitemShut {NoStop}%
\bibitem [{\citenamefont {Breuer}(2004)}]{Breuer2004a}%
  \BibitemOpen
  \bibfield  {author} {\bibinfo {author} {\bibfnamefont {H.-P.}\ \bibnamefont
  {Breuer}},\ }\bibfield  {title} {\bibinfo {title} {Genuine quantum
  trajectories for non-markovian processes},\ }\href@noop {} {\bibfield
  {journal} {\bibinfo  {journal} {Phys. Rev. A}\ }\textbf {\bibinfo {volume}
  {70}},\ \bibinfo {pages} {012106} (\bibinfo {year} {2004})}\BibitemShut
  {NoStop}%
\bibitem [{\citenamefont {Luoma}\ \emph {et~al.}(2011)\citenamefont {Luoma},
  \citenamefont {Suominen},\ and\ \citenamefont {Piilo}}]{Luoma2011}%
  \BibitemOpen
  \bibfield  {author} {\bibinfo {author} {\bibfnamefont {K.}~\bibnamefont
  {Luoma}}, \bibinfo {author} {\bibfnamefont {K.-A.}\ \bibnamefont
  {Suominen}},\ and\ \bibinfo {author} {\bibfnamefont {J.}~\bibnamefont
  {Piilo}},\ }\bibfield  {title} {\bibinfo {title} {Connecting two jumplike
  unravelings for non-markovian open quantum systems},\ }\href@noop {}
  {\bibfield  {journal} {\bibinfo  {journal} {Phys. Rev. A}\ }\textbf {\bibinfo
  {volume} {84}},\ \bibinfo {pages} {032113} (\bibinfo {year}
  {2011})}\BibitemShut {NoStop}%
\bibitem [{\citenamefont {Durr}\ \emph {et~al.}(2004)\citenamefont {Durr},
  \citenamefont {Goldstein}, \citenamefont {Tumulka},\ and\ \citenamefont
  {Zanghi}}]{Durr2004}%
  \BibitemOpen
  \bibfield  {author} {\bibinfo {author} {\bibfnamefont {D.}~\bibnamefont
  {Durr}}, \bibinfo {author} {\bibfnamefont {S.}~\bibnamefont {Goldstein}},
  \bibinfo {author} {\bibfnamefont {R.}~\bibnamefont {Tumulka}},\ and\ \bibinfo
  {author} {\bibfnamefont {N.}~\bibnamefont {Zanghi}},\ }\bibfield  {title}
  {\bibinfo {title} {Bohmian mechanics and quantum field theory},\ }\href@noop
  {} {\bibfield  {journal} {\bibinfo  {journal} {Phys. Rev. Lett.}\ }\textbf
  {\bibinfo {volume} {93}},\ \bibinfo {pages} {090402} (\bibinfo {year}
  {2004})}\BibitemShut {NoStop}%
\end{thebibliography}
\end{document}